\pdfoutput=1
\documentclass[draft]{agujournal}
\draftfalse

\usepackage[ampersand]{easylist}

\usepackage{epstopdf}

\journalname{Reviews of Geophysics}

\begin{document}
\title{The physics of sediment transport initiation, cessation, and entrainment across aeolian and fluvial environments}
\authors{Thomas P\"ahtz\affil{1,2}, Abram H. Clark\affil{3}, Manousos Valyrakis\affil{4}, and Orencio Dur\'an\affil{5}}

\affiliation{1}{Institute of Port, Coastal and Offshore Engineering, Ocean College, Zhejiang University, 866 Yu Hang Tang Road, 310058 Hangzhou, China}
\affiliation{2}{State Key Laboratory of Satellite Ocean Environment Dynamics, Second Institute of Oceanography, 36 North Baochu Road, 310012 Hangzhou, China}
\affiliation{3}{Naval Postgraduate School, Department of Physics, Monterey, CA, 93943 USA}
\affiliation{4}{Infrastructure and Environment Research Division, School of Engineering, University of Glasgow, Glasgow, UK}
\affiliation{5}{Department of Ocean Engineering, Texas A\&M University, College Station, Texas 77843-3136, USA}

\correspondingauthor{Thomas P\"ahtz}{0012136@zju.edu.cn}

\begin{keypoints}
 \item The physics of sediment transport initiation, cessation, and entrainment across aeolian and fluvial environments is reviewed
 \item The focus lies on the simplest physical systems: mildly sloped, nearly monodisperse sediment beds without complexities such as vegetation
 \item A large part of the review concerns consensus-changing developments in the field within the last two decades
\end{keypoints}

\begin{abstract}
Predicting the morphodynamics of sedimentary landscapes due to fluvial and aeolian flows requires answering the following questions: Is the flow strong enough to initiate sediment transport, is the flow strong enough to sustain sediment transport once initiated, and how much sediment is transported by the flow in the saturated state (i.e., what is the transport capacity)? In the geomorphological and related literature, the widespread consensus has been that the initiation, cessation, and capacity of fluvial transport, and the initiation of aeolian transport, are controlled by fluid entrainment of bed sediment caused by flow forces overcoming local resisting forces, whereas aeolian transport cessation and capacity are controlled by impact entrainment caused by the impacts of transported particles with the bed. Here the physics of sediment transport initiation, cessation, and capacity is reviewed with emphasis on recent consensus-challenging developments in sediment transport experiments, two-phase flow modeling, and the incorporation of granular physics' concepts. Highlighted are the similarities between dense granular flows and sediment transport, such as a superslow granular motion known as creeping (which occurs for arbitrarily weak driving flows) and system-spanning force networks that resist bed sediment entrainment; the roles of the magnitude and duration of turbulent fluctuation events in fluid entrainment; the traditionally overlooked role of particle-bed impacts in triggering entrainment events in fluvial transport; and the common physical underpinning of transport thresholds across aeolian and fluvial environments. This sheds a new light on the well-known Shields diagram, where measurements of fluid entrainment thresholds could actually correspond to entrainment-independent cessation thresholds.
\end{abstract}

\section*{Plane Language Summary}
Loose sediment grains can be transported by blowing wind (aeolian) or water flowing in a riverbed (fluvial). These processes are responsible for shaping much of the natural world, but they involve the combination of several very complex physical systems, like turbulent fluid flow near a rough boundary and the mechanical behavior of granular materials. Thus, there is no consensus about the minimum wind or water speeds required to initiate and sustain sediment transport. Additionally, wind and water-driven sediment transport are obviously similar, suggesting that it should be possible to capture both under one description. Recent advances in experiments and computer simulations have helped scientists to answer some key questions about why sediment transport is initiated and sustained. This article reviews many of these recent discoveries, focusing on three key topics: (1) the mechanical behavior of granular materials; (2) how turbulence in the fluid helps to move grains; and (3) the role of inertia of mobile grains. We show that a deeper understanding of these topics helps to resolve some major inconsistencies in our understanding of why sediment transport is initiated and sustained and may help to unify sediment transport by wind and water under a single theoretical description.

\section{Introduction} \label{Introduction}
When an erodible sediment bed is subjected to a shearing flow of a Newtonian fluid, such as air or water, bed particles may be \textit{entrained} (i.e., set into motion) by the action of flow forces and then transported by the flow, initiating a process known as \textit{sediment transport}. The critical conditions that are required for the initiation of sediment transport have been studied for more than two centuries~\citep[e.g.,][]{Brahms57}. Dating back to the pioneering studies for water-driven transport by \citet{Shields36} and for wind-driven transport by \citet{Bagnold36,Bagnold37,Bagnold38} (summarized in his book~\citep{Bagnold41}), the initiation of sediment transport in both cases has been commonly described by threshold values of the time-averaged shear stress $\tau$ that the flow applies onto the bed~\citep[see reviews by][and references therein]{Duranetal11,Koketal12,Merrison12,DeyAli18,DeyAli19,Yangetal19}. The idea of a threshold value of $\tau$ is natural, since a necessary condition for flow-driven entrainment (or \textit{fluid entrainment}) is that flow forces and/or flow-induced torques acting on bed surface particles must overcome resisting forces and/or torques. Consistently, for wall-bounded flows (to which sediment transport belongs) at a given shear Reynolds number $\mathrm{Re}_\ast\equiv u_\ast d/\nu_f$, the shear velocity $u_\ast\equiv\sqrt{\tau/\rho_f}$ controls the near-surface profile of the streamwise flow velocity when averaged over the entire spectrum of turbulent fluctuations~\citep[see review by][and references therein]{Smitsetal11}, where $\rho_f$ is the fluid density, $\nu_f$ the kinematic fluid viscosity, and $d$ a particle diameter characteristic for bed particles. As forces resisting entrainment of a bed particle scale with the submerged gravity force ($\propto(\rho_s-\rho_f)gd^3$), where $\rho_s$ is the particle density and $g$ the gravity constant, it has been common among geomorphologists to nondimenionalize $\tau$ via $\Theta\equiv\tau/[(\rho_p-\rho_f)gd]$~\citep{Shields36}, which is known as the \textit{Shields number} or \textit{Shields parameter}. In the aeolian research community, the \textit{threshold parameter} $\sqrt{\Theta}$~\citep[][p.~86]{Bagnold41} is also often used. \citet{Shields36} and numerous researchers after him have measured transport thresholds for water-driven transport~\citep[see reviews by][and references therein]{Milleretal77,BuffingtonMontgomery97,Paphitis01,DeyPapanicolaou08,DeyAli19,Yangetal19}. These measurements are usually summarized in a diagram showing the threshold Shields number $\Theta_t$ as a function of $\mathrm{Re}_\ast$ (the \textit{Shields curve} $\Theta_t(\mathrm{Re}_\ast)$), which is known as the \textit{Shields diagram}.

However, the concept of a threshold shear stress for \textit{incipient motion} (i.e., for the initiation of sediment transport by fluid entrainment) has had several consistency problems. First, for wind-driven transport, the most widely used incipient motion models~\citep{IversenWhite82,ShaoLu00}, when applied to Martian atmospheric conditions, predict threshold shear stresses for fine sand particles that are so large that transport should occur only during rare strong Mars storms~\citep{SullivanKok17}. However, this prediction is contradicted by modern observations indicating widespread and persistent sediment activity~\citep{Bridgesetal12a,Bridgesetal12b,Silvestroetal13,Chojnackietal15}, even of very coarse sand~\citep{Bakeretal18}.

A second inconsistency, which has long been known, concerns water-driven sediment transport and is tacitly acknowledged whenever the concept of an incipient motion shear stress is applied: the sediment transport rate $Q$ (i.e., the average particle momentum per unit bed area) seems to never truly vanish for nearly any $\Theta>0$ in water flume experiments because of occasional strong turbulent fluctuation events causing entrainment by bursts of much-larger-than-average flow forces. That is why measurements of $\Theta_t$ have relied either on indirect extrapolation methods or on vague criteria defining the value of $Q$ (or a proxy of $Q$) at which transport is critical~\citep{BuffingtonMontgomery97}. Such criteria had been introduced even before Shields~\citep{Gilbert14,Kramer35}. In particular, the experiments by \citet{Paintal71} suggest a power law relationship between $Q$, appropriately nondimensionalized, and $\Theta$ for weak flows over gravel beds: $Q_\ast\equiv Q/[\rho_pd\sqrt{(\rho_p/\rho_f-1)gd}]\propto\Theta^{16}$ (it was necessary to measure $Q$ over tens of hours for the weakest flows), which describes a dramatic but not infinitely rapid decrease of $Q_\ast$ with decreasing $\Theta$. Qualitatively similar observations were reported by \citet{HellandHansenetal74}. Largely because of Paintal's experiments, \citet{LavelleMofjeld87} strongly argued in favor of stochastic sediment transport models that do not contain a threshold shear stress~\citep[e.g.,][]{Einstein50} in a highly cited paper with the title, ``Do Critical Stresses for Incipient Motion and Erosion Really Exist?'' Despite the fact that many researchers have been well aware of this inconsistency, the concept of a threshold shear stress has remained alive and never been truly questioned by the majority of scientists working on water-driven sediment transport~\citep{DeyAli18,DeyAli19,Yangetal19}. There are two main reasons for the trust in this concept. First, above a value of $Q_\ast$ that roughly coincides with typical criteria defining critical transport ($Q_\ast\approx0.007$), the relationship between $Q_\ast$ and $\Theta$ turns into a much milder power law~\citep{Paintal71}: $Q_\ast\propto\Theta^{2.5}$, suggesting a clear physical meaning of the threshold Shields number associated with this transition ($\Theta_t\approx0.05$). Second, descriptions of water-driven sediment transport that are based on a threshold shear stress (i.e., expressions $Q_\ast(\Theta)$ with $Q_\ast(\Theta\leq\Theta_t)=0$) have been quite successful in reproducing transport rate measurements for well-controlled conditions when using very similar values of $\Theta_t$. For example, the scaling $Q_\ast\propto(\Theta-\Theta_t)^{3/2}$ by \citet{MeyerPeterMuller48} with $\Theta_t\approx0.05$ is one of the most widely used expressions in hydraulic engineering for gravel transport driven by water~\citep{WongParker06}. However, if this value of $\Theta_t$ has a real physical meaning, what is it? Does it truly describe incipient motion, which has always been the predominant interpretation~\citep[see reviews by][and references therein]{Milleretal77,BuffingtonMontgomery97,Paphitis01,DeyPapanicolaou08,DeyAli18,DeyAli19,Yangetal19}, despite the fact that $Q_\ast(\Theta\leq\Theta_t)>0$ (in Paintal's experiments, $Q_\ast>0$ even for $\Theta\approx0.007\ll\Theta_t$)?

A third inconsistency in the concept of an incipient motion shear stress, which also concerns water-driven sediment transport, is also old but much less well known, perhaps because one of the key papers~\citep{GrafPazis77} is published in French language. Graf's and Pazis' measurements show that increasing the shear stress on the bed due to the water flow from zero up to a certain value $\tau$ (a transport initiation protocol) results in smaller transport rates $Q$ than decreasing the shear stress from a larger value down to $\tau$ (a transport cessation protocol). This clearly indicates an important role of particle inertia in sustaining water-driven sediment transport. Hence, any measurement of $\Theta_t$ is affected by particle inertia because, regardless of whether an initiation or cessation protocol is used, particles are already transported when $\Theta$ approaches $\Theta_t$ (see the second inconsistency discussed above). Hence, $\Theta_t$ is not, or at least not only, associated with fluid entrainment and thus incipient motion. The importance of particle inertia was proposed and indirectly shown even earlier, in a largely ignored study (only eight citations indexed by Web of Science today, half a century after publication) by \citet{Ward69}. In this study, \citet{Ward69} measured smaller values of $\Theta_t$ for a larger particle-fluid-density ratio $s\equiv\rho_p/\rho_f$ (which is a measure for particle inertia) at the same shear Reynolds number $\mathrm{Re}_\ast$. A slight downward trend of $\Theta_t$ with $s$ even existed in the pioneering experiments by \citet{Shields36}. Interestingly, a particle inertia effect in water-driven sediment transport has actually been studied. It is well known, although often not considered to be crucial in the context of transport thresholds, that the flow strength at which a transported particle can come to rest at the bed surface is weaker than the one at which it can reenter transport~\citep[e.g.,][]{Francis73,Reidetal85,Drakeetal88,Anceyetal02}. In contrast, another potentially important effect of particle inertia in water-driven sediment transport has not received the same attention: the interaction between particles that are already in transport and particles of the bed surface (e.g., particle-bed impacts) may support bed particle entrainment or even be predominantly responsible for it (\textit{impact entrainment}).

Particle inertia and particularly impact entrainment have been widely recognized as crucial for sustaining wind-driven sediment transport since the pioneering studies by \citet{Bagnold41}. Yet, in contrast to water-driven transport, there seems to be a clear-cut shear stress threshold when applying an initiation protocol in wind tunnel experiments~\citep[e.g.,][]{Bagnold41}. This rather curious difference between wind-driven and water-driven transport is usually not discussed in the context of incipient motion. Why is it necessary to define critical transport rates for measuring an incipient motion shear stress threshold in water-driven transport but not in wind-driven transport? A complete description of incipient motion should be generally applicable and not limited to a subset of possible sediment transport conditions, since there is no reason to believe that the physical mechanisms involved in the entrainment of a bed particle by a turbulent flow depend much on the nature of the flow. In fact, frameworks unifying sediment transport across driving fluids (not only in regard to transport thresholds) are scarce in general (e.g., apart from modern studies, only \citet{Bagnold56,Bagnold73} seems to have attempted unifying water-driven and wind-driven transport conditions).

One of the most desired aspects of a general framework of sediment transport would be its ability to reliably predict the general dependency of $Q_\ast$ on $\Theta$ and other dimensionless environmental parameters, such as the density ratio $s$. However, there is an obvious problem: since measured transport rates may depend on the experimental protocol for a given condition, as was the case in the experiments by \citet{GrafPazis77} (see third inconsistency), does the concept of a general relationship even make sense? The consensus is, yes, it does make sense when referring to transport \textit{capacity} (also known as transport \textit{saturation} in aeolian geomorphology), which loosely defines the maximal amount of sediment a given flow can carry without causing net sediment deposition at the bed. However, a precise definition of transport capacity is very tricky and controversial~\citep[see review by][and references therein]{Wainwrightetal15}. For example, the fact that equilibrium transport rates may depend on the experimental protocol for a given condition implies that not every equilibrium transport condition is equivalent to transport capacity and that transport capacity is in some way linked to particle inertia. In fact, that the latter may be the case was recognized by \citet{NinoGarcia98a}, who numerically modeled water-driven sediment transport as a continuous motion of particles hopping along a flat wall. In particular, these authors mentioned that the capacity relation obtained from their numerical simulations contains a threshold Shields number that may not be associated with fluid entrainment, demonstrating the necessity for a good understanding of transport capacity and its relationship to particle inertia in the context of sediment transport thresholds.

While this introduction has focused on introducing issues in our understanding of fluid entrainment, shear stress thresholds, particle inertia, transport capacity, and their mutual relationships from a historical perspective, there have been major developments in these topics in the last two decades, largely because of the emergence of novel experimental designs and modeling techniques. The purpose of this review is to draw the attention of the involved research communities to these developments that, if put together, resolve the above issues and provide a largely improved conceptual understanding of sediment entrainment and transport thresholds.

A large portion of recent developments in the field can be attributed to numerical studies modeling the particle phase using the discrete element method (DEM). In comparison to other methods modeling the particle phase (e.g., continuum models), this method has the big advantage that it approximates the laws of physics at a very basic level, namely, at the level of intergrain contacts. In fact, the force laws commonly used to model intergrain contacts are known to produce system results that match experiments extremely well~\citep[e.g.,][]{Stewartetal01,Latzeletal03,Clarketal16}. Additionally, granular continuum models are formulated using DEM simulations~\citep{DaCruzetal05} but reproduce complex experiments on granular flows often very accurately~\citep{Jopetal06}. In the context of sediment transport, the main uncertainty of DEM-based models lies therefore in the modeling of the coupling between the particle phase and the Newtonian fluid driving transport. However, many of the simulations that are described in this review show that the results are often insensitive to the details of how this coupling is treated. The authors of this review thus argue that new physics uncovered by DEM-based numerical simulations are on a relatively solid footing.

To limit the scope of this review, it focuses on studies of \textit{mildly sloped} beds of \textit{relatively uniform} sediments unless mentioned otherwise. Also, because of the focus on physical processes involving the bed surface, this review largely concerns \textit{nonsuspended} sediment transport (i.e., the fluid turbulence is unable to support the submerged particle weight), in which transported particles remain in regular contact with the bed surface (typical for particles of sand size and larger) and which is the relevant transport mode for the morphodynamics of planetary landscapes, riverscapes, and seascapes. In contrast, in suspended transport (typical for particles of silt or dust size and smaller), transported particles can remain out of contact with the bed surface for very long times (e.g., as atmospheric dust aerosols). In typical nonsuspended wind-driven (\textit{aeolian}) sediment transport, many particles move in large ballistic hops and the transport layer thickness $h$ is therefore much larger than the particle diameter $d$. In the aeolian geomorphology community, such hopping particles are said to move in \textit{saltation} and explicitly distinguished from particles rolling and sliding along the surface. However, this terminology is not used in this review. Instead, the term \textit{saltation transport} is used for general transport regimes with $h\gg d$, that is, it refers to all rather than a subset of transported particles. In typical nonsuspended liquid-driven transport (henceforth referred to as \textit{fluvial} transport for simplicity although this mode is not limited to fluvial environments), $h$ is of the order of $d$ because the largest particle hops are small. Following the fluvial geomorphology community, transport regimes with $h\sim d$ are termed \textit{bedload transport}.

This manuscript is organized into sections that focus on specific topics (sections~\ref{DenseGranularFlow}--\ref{ParticleInertia}) followed by a summary and outlook section (section~\ref{Summary}) and a Notation section describing the definitions of technical terms and mathematical symbols. It is noted that readers may find it useful to read section~\ref{Summary} first in order to organize the contents of the manuscript, and then consult sections~\ref{DenseGranularFlow}--\ref{ParticleInertia} for more detailed information on a particular topic. Section~\ref{DenseGranularFlow} reviews recent insights into the mechanics of beginning sediment motion and fluid entrainment gained from studying sediment transport as a dense granular flow phenomenon. For example, it has become increasingly clear that granular material can flow even when a macroscopic motion does not occur, such as for a collapsed pile of sand, because of a process known as \textit{creeping}, which describes an irreversible superslow granular motion associated with sporadic microscopic rearrangements. That is, it is crucial to clearly define what kind of motion one refers to when introducing sediment transport thresholds. Likewise, forces resisting the entrainment of a bed particle do not only depend on the local arrangement of bed particles but also on granular interactions with regions within the bed that are far away from the entrainment location (i.e., sediment entrainment is a nonlocal phenomenon). This is because of collective granular structures that particles can form. Section~\ref{FluidEntrainment} reviews insights gained from recent experimental and theoretical studies showing that the fluid shear stress applied onto the bed surface alone only poorly characterizes the critical conditions required for fluid entrainment by turbulent flows. These studies have provided more suitable criteria for sediment entrainment that take into account turbulent fluctuation events and, in particular, their durations. However, section~\ref{FluidEntrainment} also explains that a critical fluid shear stress for incipient motion does make sense when referring to the shear stress at which the fluid entrainment probability exceeds zero (which, for turbulent fluvial bedload transport, occurs much below the Shields curve~\citep{Paintal71}). For example, in wind tunnel studies (but not necessarily in the field), aeolian saltation transport is initiated at about this threshold. Finally, section~\ref{ParticleInertia} reviews studies on the role of particle inertia in sediment transport, a topic that has very recently undergone a dramatic change. In fact, while it is well established that impact entrainment is crucial for aeolian saltation transport~\citep[see reviews by][and references therein]{Duranetal11,Koketal12,Valanceetal15}, very recent experimental and theoretical studies revealed that it is also crucial for sustaining fluvial bedload transport. Likewise, a very old argument by \citet{Bagnold41}, which was forgotten or deemed unimportant, has recently been revived. \citet{Bagnold41} pointed out that, for aeolian saltation transport, a predominant role of impact entrainment requires that the flow is able to sustain the motion of transported particles. This is only possible if the energy loss of transported particles rebounding with the bed is compensated by their energy gain during their trajectories via fluid drag acceleration. Models that explicitly incorporate this requirement have been able to partially unify aeolian saltation and viscous and turbulent fluvial bedload transport. When combined, the insights from the studies reviewed in sections~\ref{DenseGranularFlow}--\ref{ParticleInertia} provide a conceptual picture free of inconsistencies, which is described in section~\ref{Summary}. For example, the shear stress threshold compiled in the Shields diagram seems to characterize the cessation of sediment bulk motion and an appropriately defined transport capacity rather than incipient motion. Section~\ref{Summary} also summarizes important open problems and provides a brief outlook into related problems that have not been discussed in this review, such as the effects of particle size heterogeneity on transport thresholds and bed sediment entrainment.

\section{Yield and Flow of Dense Granular Media in the Context of Sediment Transport} \label{DenseGranularFlow}
In theoretical considerations of a problem as complex as the mechanics of beginning sediment motion, simplifying assumptions must be made. This often means that the granular phase is treated extremely coarsely, as a continuum with a Coulomb-like friction coefficient~\citep{Terzaghi51,DruckerPrager52}, or very finely, where the pocket geometry of individual grains sets the bed strength~\citep{WibergSmith87}. However, recent advances in granular mechanics have shown that Coulomb-like behavior of granular materials is inherently \textit{nonlocal}, so it must be treated on intermediate length scales. This is due to the fact that the \textit{yielding} condition, defined as the minimum shear stress required to achieve permanent granular flow, is set by emergent, collective networks of grains. These networks can couple different sections of the material together over large distances. The purpose of this section is to provide an overview of recent work on yield and flow of dense granular materials in the context of sediment transport, with a particular focus on the nonlocal nature of granular yielding. To simplify the discussion, it is assumed throughout this section that the granular bed is subjected to a constant bed shear stress (like for laminar flows), in which case the existence of a fluid entrainment threshold associated with bed failure does make sense. However, this is no longer true for turbulent flows, as reviewed in section~\ref{FluidEntrainment}. For more information on dense granular flow, readers might consult recent reviews~\citep{ForterrePouliquen08,Jop15,Kamrin18} devoted exclusively to the topic of dense granular flow. For the connection between granular flow and sediment transport, the perspective and review by \citet{FreyChurch09,FreyChurch11} are also recommended.

\subsection{Yielding of Granular Media} \label{Yielding}
Surface grains sit in \textit{pockets} on top of the bed, and the geometry of the pocket determines the entrainment conditions for that particular grain via its protrusion (i.e., the grain height above surrounding grains) and friction angle. When the downstream drag force from the fluid overcomes resistive forces from gravity and from contact forces with the pocket, the grain will begin to move. This conceptually simple scenario appears in many theoretical studies~\citep[e.g.,][]{WibergSmith87,Ling95,Dey99,DeyPapanicolaou08,AliDey16}. However, this picture has several conceptual problems. For example, there are many different pocket geometries~\citep{Kirchneretal90,Buffingtonetal92} implying a distribution of entrainment thresholds. \citet{Kirchneretal90} made a similar argument, advocating for a statistical treatment of pocket geometries, where only the grains with the smallest entrainment thresholds would be relevant. Additionally, when transport thresholds are discussed, one typically does not include transient behavior, after the flow has pushed grains from less stable to more stable pockets. For example, an entrained grain that then restabilizes in a nearby pocket would not constitute sediment transport. After such a rearrangement, the resulting bed would have a different intergrain force and contact structure, which would be more suited to resisting the applied flow forces~\citep{MastellerFinnegan17}. Thus, determining the fluid entrainment threshold amounts to determining the strongest bed that can be formed by the grains, subject to the flow forces and dynamics. This process necessarily involves transient behavior, as grains search for stable configurations, and spatial correlations, since information about each grain's movement is transmitted through the intergrain force network. 

\begin{figure}[!htb]
\RaggedRight (a) \hspace{60mm} (b) \\
    \centering
    \includegraphics[width=1.0\columnwidth]{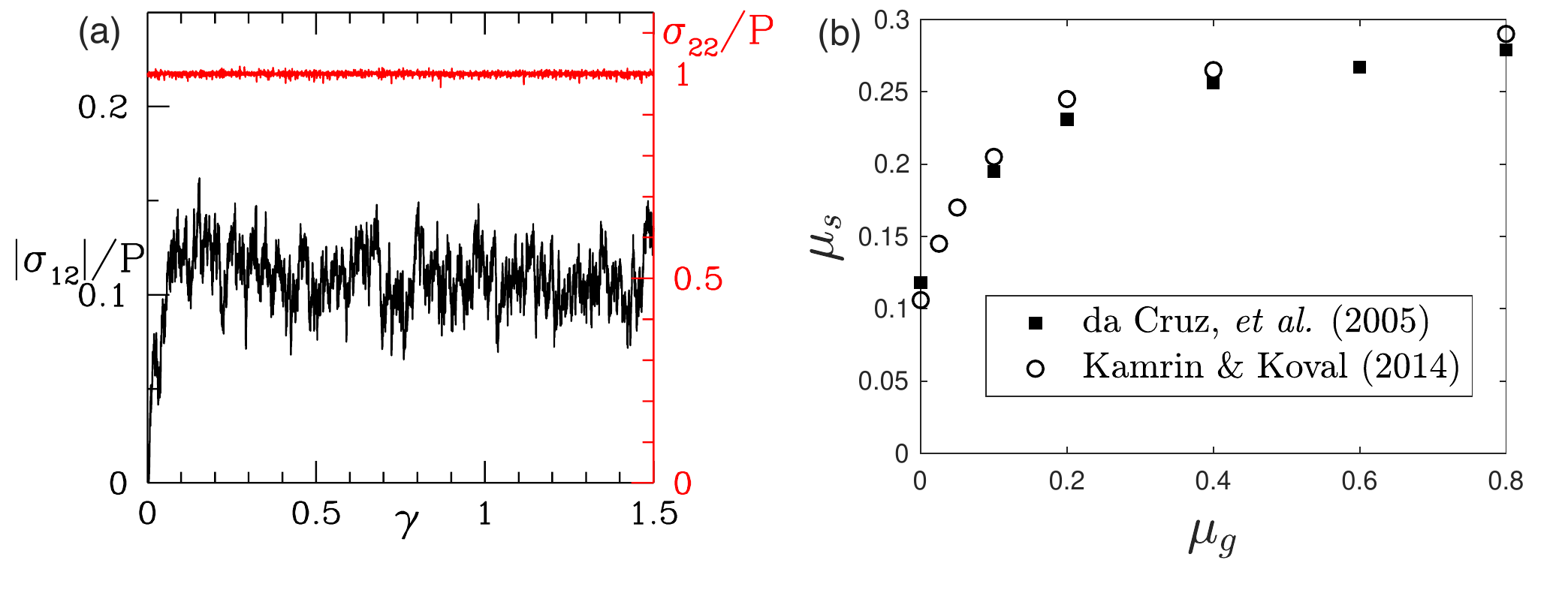}
    \caption{(a) From \citet{PeyneauRoux08a}, the normalized shear stress $\sigma_{12}/P$ is plotted as a function of strain $\gamma$. The shear stress builds up from zero, reaching its maximum value at $\gamma \approx 0.1$. Copyright 2008 American Physical Society. (b) Data adapted from \citet{DaCruzetal05} and \citet{KamrinKoval14} showing a measurement of the bulk static friction coefficient $\mu_s$ as a function of $\mu_g$, which is the static friction coefficient between the surfaces of two grains (simulated as two-dimensional disks).}
    \label{fig:friction-fig}
\end{figure}

While this represents a very challenging problem, it is exactly the picture that has emerged in recent years regarding the physical origin of frictional behavior in noncohesive soils or sediments. The yield criterion of granular materials is defined by the maximum internal shear stress that a granular material can achieve, but grains must rearrange to find this maximum stress, sometimes for a long time~\citep{Clarketal18,Srivistavaetal19}. The yield criterion has the form of a friction coefficient, where flow occurs only when $\mu \equiv \tau_p/P > \mu_s$, where $\tau_p$ and $P$ are the granular shear stress and pressure $P$, respectively, that arise from intergrain contacts, and $\mu_s$ is the static friction coefficient of the material. At first glance, this is not surprising, since the grains themselves have a surface friction coefficient $\mu_g$. However, $\mu_s$ is only weakly dependent on $\mu_g$~\citep{DaCruzetal05}, as shown in Figure~\ref{fig:friction-fig}.
Even \textit{frictionless} spheres have $\mu_s\approx 0.1$ \citep{PeyneauRoux08a,PeyneauRoux08b}, which arises from a preferred orientation for intergrain contacts that aligns with the compressive direction of the applied shear deformation. This effect is independent of whether the grains interact via linear spring forces~\citep{ThompsonClark19} or more realistic Hertzian interactions~\citep{PeyneauRoux08a}. Similar behavior is observed for grains with surface friction and irregular shape~\citep{Radjaietal98,AzemaRadjai10,AzemaRadjai14,Trulsson18}, but the maximum stress anisotropy is enhanced by these effects, since grain-grain contacts can have both normal and tangential components. This raises the yield stress slightly: frictional disks have $\mu_s \approx 0.2-0.3$~\citep{DaCruzetal05} and frictional spheres have $\mu_s \approx 0.3-0.4$~\citep{Jopetal06}, with only a weak dependence on $\mu_g$ for $\mu_g > 0.1$. Additionally, $\mu_s$ is nearly independent of polydispersity~\citep{Voivretetal09}. This picture assumes grains are slowly moving with persistent intergrain contacts, but $\mu_s$ can be lowered significantly for more energetic kinds of driving, like vibration~\citep{GaudelDeRichter19} or in aeolian saltation transport~\citep{Pahtzetal19}, probably because the tendency of the contact orientation to align with the compressive direction is somewhat suppressed~\citep{Pahtzetal19}. Thus, frictional behavior in granular media arises primarily from the anisotropic structure of force and contact networks, and grain-grain friction, shape, and polydispersity play secondary roles.

Here, $\mu$ is used to denote the local nondimensional shear stress in the granular material itself, while the Shields number $\Theta$ is the dimensionless shear stress applied to the granular bed surface, so the two quantities are not equivalent but are closely related. At the surface of the bed, $\mu\approx\Theta$ if lift forces are neglected. The existence of a maximum shear stress that can be supported by a granular material (which is independent of grain size) suggests that, for noncohesive sediments, there should be a theoretical upper limit to the threshold Shields number $\Theta_t$, $\Theta_t^{\rm max}\approx\mu_s$. This implies that the Shields curve must plateau at low values of the shear Reynolds number ${\rm Re}_\ast$ for laminar flows. This fact has been a subject of debate for many years, with some authors~\citep{Shields36,Mantz77,Milleretal77,YalinKarahan79,Govers87,BuffingtonMontgomery97,Dey99,Hongetal15} showing a trend where $\Theta_t$ continues to grow as ${\rm Re}_\ast$ gets smaller, while other studies~\citep{WibergSmith87,Paphitis01,PilottiMenduni01,Ouriemietal07} show a plateau at low ${\rm Re}_\ast$. Recent work by the present authors~\citep{Clarketal15a,Clarketal17,PahtzDuran18a} has investigated sediment transport thresholds over a wide range of ${\rm Re}_\ast$ and density ratio $s$ using simulations based on the DEM to model noncohesive grains that are coupled to fluid-driven shear forces. These studies all suggest that $\Theta_t$ is a constant at low ${\rm Re}_\ast$ and $s$, corresponding to the strongest possible state of the bed. It is noted that cohesive effects become important for very small grains, which can cause $\Theta_t$ to continue to grow for smaller ${\rm Re}_\ast$.

\subsubsection*{Open Problem: Value of Viscous Yield Stress $\Theta_t^{\rm max}$}
Measured values of the viscous yield stress $\Theta_t^{\rm max}$ vary substantially. For nearly monodisperse beds of spherical particles, most studies reported $\Theta_t^{\rm max}\approx0.12$~\citep{Charruetal04,Loiseleuxetal05,Ouriemietal07,Seizillesetal14,Houssaisetal15}, but larger values of up to about $0.37$ have also been reported~\citep{Lobkovskyetal08,Hongetal15}. Also, some measurements suggest that $\Theta_t^{\rm max}$ depends on the median grain size~\citep{Hongetal15}, in contradiction to the grain size independence of $\mu_s$, while other studies find no such dependence~\citep{Ouriemietal07}. To the authors' knowledge, there is currently no convincing explanation for these contradicting observations.  However, the scatter in the reported values for $\Theta_t^{\rm max}$ (between 0.12 and 0.37) is within the range reported for the yield stress of granular materials, ranging from low-friction spheres to rougher, more frictional particles. Thus, the yield stress of the bulk granular material may at least play some role in setting the scatter in $\Theta_t^{\rm max}$. In this context, it is worth noting that, for the entrainment of particles resting on an idealized substrate by a laminar flow, threshold Shields numbers range from zero to very large values depending on the packing arrangement~\citep{Agudoetal17,DeskosDiplas18,Topicetal19,ShihDiplas19}.

\subsection{Rheological Descriptions}
The existence of a yield stress is one piece of a rheological description, which is a constitutive law that mathematically connects the strain rate to the local stress at each point in a material. For granular materials, dissipation implies that more force is required for faster strain rates, so $\mu$ will increase with strain rate $\dot{\gamma}$. For the case of sediment transport, formulation of a constitutive law has obvious practical benefits, namely that it would allow an analytical prediction of transport rates $Q$ at varying Shields number $\Theta$ for transport conditions dominated by granular interactions. However, note that a bulk constitutive law may not be able to capture certain cases, particularly very near to the onset or cessation of fluvial bedload or aeolian saltation transport, where the transport layer is dominated by the isolated motion of a single grain along the bed (which is the typical situation in gravel-bed rivers~\citep{Parker78,PhillipsJerolmack16}). Despite the fact that the force and contact networks discussed above are spatially extended, some progress has been made by considering so-called \textit{local rheologies}. Based on dimensional analysis, \citet{DaCruzetal05} showed that $\mu$ for dry, uniform granular flows must depend on $\dot{\gamma}$ via a single dimensionless number, $I \equiv \dot{\gamma} d/\sqrt{P/\rho_p}$, where $I$ is called the inertial number, similar to the Savage~\citep{Savage84} or Coulomb~\citep{Anceyetal99} numbers. A functional form for $\mu(I)$ can then be measured from experiments or DEM simulations (a crude approximation is given by $\mu = \mu_s + c_II$, where $c_I$ is a constant parameter). If one then assumes that a three-dimensional, tensorial generalization of this law is \textit{locally} satisfied at each point in space in arbitrary geometries, then the equations of motion are closed and one can predict (at least numerically) flow in any arbitrary geometry where the forces and boundary conditions are known. Experimental measurements of rapid, dense flow in several geometries show good agreement with the local rheology~\citep{MiDi04,Jopetal05,Jopetal06}.

\subsubsection*{Open Problem: Rheology of Nonsuspended Sediment Transport}
There are many physical mechanisms that are relevant to nonsuspended sediment transport that are not included in the inertial number description, but recent work has suggested that appropriate dimensional analysis can be used to find a general rheological description that is relevant in all contexts. For example, viscous effects from the fluid can be included~\citep{Boyeretal11,Trulssonetal12,NessSun15,NessSun16,Houssaisetal16,Amarsidetal17,HoussaisJerolmack17,GuazzelliPouliquen18} by replacing the inertial number $I$ with the viscous number $J \equiv \rho_f\nu_f \dot{\gamma} / P$. This description is valid when the Stokes-like number $I^2/J$ is small, and the standard $\mu(I)$ rheology again takes over for large $I^2/J$. This crossover can be heuristically written in terms of a viscoinertial number $K\equiv J+c_KI^2$, where $c_K$ is an order-unity fit parameter~\citep{Trulssonetal12,NessSun15,NessSun16,Amarsidetal17}, and the rheology takes the form $\mu(K)$. 

The previous paragraph describes a unification of dry and wet, \textit{viscous} granular flows, but some situations, like turbulent bedload or aeolian saltation transport, do not fit neatly into this description. \citet{Maurinetal16} showed that, for intense turbulent bedload transport, the inertial number $I$ (used for dry flows) collapses the data best, but with a different $\mu(I)$ relation compared to dry flows. Additionally, the presence of more severe velocity fluctuations and grain-grain collisions can weaken the material, giving a $\mu$ that is smaller than would be predicted by a $\mu(I)$ or $\mu(K)$ rheology at a given shear rate~\citep{PahtzDuran18b}. Another option is to build a rheological description that explicitly accounts for these fluctuations and collisions via the P\'eclet number ${\rm Pe}\equiv\dot{\gamma} d / \sqrt{T} $~\citep{Pahtzetal19}, where the granular temperature $T$ equals the mean square of kinetic particle velocity fluctuations. The advantage of $\mathrm{Pe}$ is that it is applicable to a wide range of different granular flows (e.g., it unifies intense fluvial bedload and aeolian saltation transport), whereas $K$ is limited to relatively homogeneous flows. The disadvantage is that $\mathrm{Pe}$ involves another granular property ($T$) that requires modeling. 

\subsection{Creep and Nonlocal Rheologies} \label{Creeping}
\begin{figure}[!htb]
    \centering
    \includegraphics[width=1.0\columnwidth]{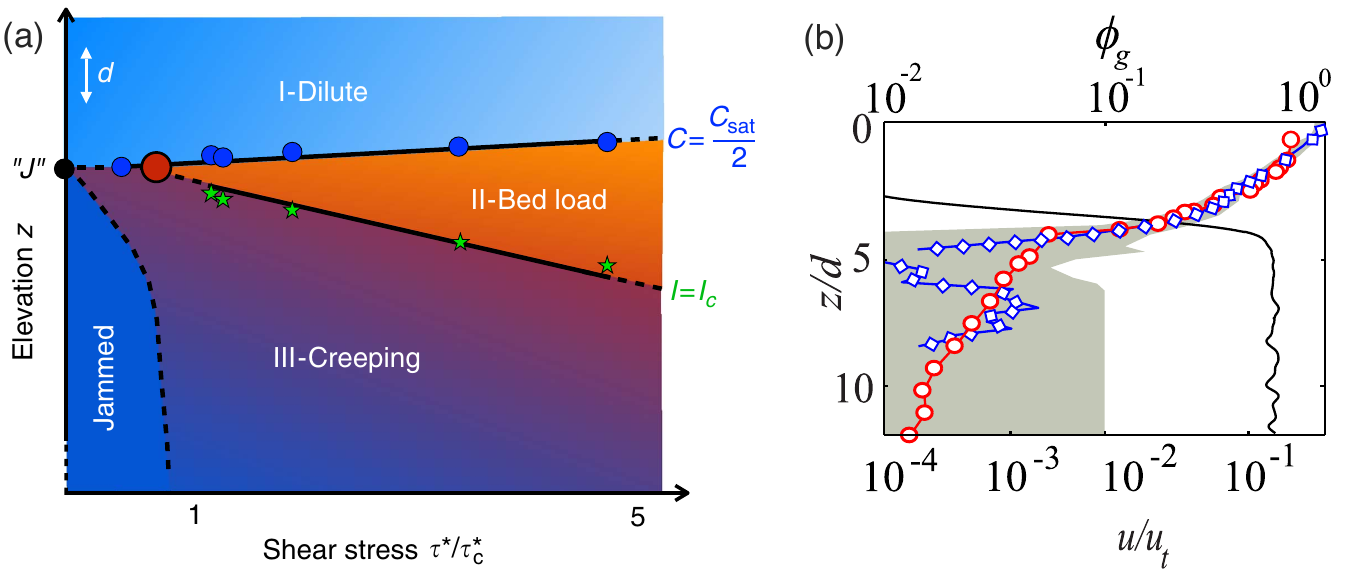}
    \caption{(a) From \citet{Houssaisetal15} (Creative Commons Attribution 4.0 International License), a proposed phase diagram for granular flow behavior as a function of elevation $z$ in the bed (vertical axis) and applied shear stress $\tau^\star$ from the overlying fluid flow (horizontal axis). Bedload transport triggers slow creeping flow below it, consistent with nonlocal rheological models that have recently been formulated for dry granular media, as described in the text. (b) From \citet{AllenKudrolli17}, normalized velocity profiles $u/u_t$ for the fluid (blue squares) and grains (red circles) are plotted as a function of height $z/d$. Also plotted is the packing fraction of the grains $\phi_g$ as a function of height. The top of the bed corresponds to the drop in $\phi_g$. Above the bed, grains move with the fluid. Below the bed, the grain velocity profile decays exponentially (a straight line on the semilogarithmic plot), which is a prediction of the nonlocal granular flow rheologies discussed in the text. Copyright 2017 American Physical Society.}
    \label{fig:creep-1}
\end{figure}

As discussed in section~\ref{Introduction}, some water flume experiments suggest that fluvial bedload transport never truly ceases for nearly any $\Theta>0$, which is usually attributed to turbulent fluctuations. However, as discussed in this section, the granular material itself may be partially responsible. In fact, it is well known that granular creep can be observed in a variety of observational geophysical contexts~\citep{BoultonHindmarsh87,Piersonetal87,Ferdowsietal18} as well as more idealized granular flows in a laboratory setting~\citep{Roeringetal01,Komatsuetal01,Nicholetal10,Moosavietal13,Amonetal13}, including sediment transport explicitly~\citep{Houssaisetal15,AllenKudrolli18}, as depicted in Figure~\ref{fig:creep-1}.
Generally, creeping refers to slow, typically intermittent flow (not limited to the bed surface) that occurs below a macroscopic yield criterion. 

One class of creeping flow involves systems where regions with $\mu>\mu_s$ and $\mu<\mu_s$ exist nearby each other, which often occurs in systems with stress gradients (e.g., due to gravity or curvature). In this case, creeping flow is observed in regions with $\mu<\mu_s$~\citep{FenisteinVanHecke03,MiDi04,Crassousetal08,Kovaletal09}. This creeping flow is not steady or continuous, but occurs in a series of intermittent, avalanche-like slips, which are triggered by the nearby steadily flowing region with $\mu>\mu_s$. The time-averaged shear rate profiles decay quasi-exponentially with spatial distance to the steadily flowing region. Various nonlocal theories have been proposed~\citep{Baranetal06,PouliquenForterre09} that include a spatial length scale $\xi$ over which flow can be triggered in this way. The most successful theories~\citep{KamrinKoval12,HenannKamrin13,KamrinHenann15,Bouzidetal13,Bouzidetal15} suggest that the cooperative length scale $\xi$ \textit{diverges at the yield stress} (i.e., $\xi \propto |\mu - \mu_s|^{-\nu}$, where $\nu\approx 0.5$). This means that, near the yield stress, flow events can be triggered over arbitrarily large distances; this point is revisited below. The grain-scale physical origin of the nonlocal models and associated spatial correlations~\citep{ZhangKamrin17} as well as how exactly to best mathematically formulate a nonlocal rheology~\citep{Bouzidetal17,LiHenann19} is still a subject of debate in the literature.

The creeping flow captured by these nonlocal models is also apparent in laboratory flumes used to model fluvial sediment transport. \citet{Houssaisetal15,Houssaisetal16} showed that sediment transport involves the coexistence of three regimes: a dilute suspension above the bed surface, the bedload layer at the bed surface, and creeping behavior below the surface. These regions are depicted in Figure~\ref{fig:creep-1}a. The shear rate profile in the creeping regime follows an exponential decay, which is consistent with the predictions of nonlocal models. Similar behavior was also observed by \citet{AllenKudrolli17}, shown in Figure~\ref{fig:creep-1}b, who also stressed that the apparent agreement with nonlocal models formulated for dry granular materials implies that the fluid stress is not playing a major role in the observed creeping behavior. In the creeping regime, $\mu<\mu_s$, but flow events are triggered via the bedload transport regime at the top of the bed via spatial correlations in the force network. These creeping events, although slow and intermittent, can lead to segregation effects over long times ($\sim$10--100~hr), where large particles are sorted to the top~\citep{Ferdowsietal17}. Thus, creep and nonlocal rheology may play a crucial role in armoring of gravel-bedded rivers, as opposed to size sorting in the transported layer. Additionally, recent computational work~\citep{PahtzDuran18b} has shown that sediment transport rheology is nonlocal even relatively far from the sediment transport threshold.

\begin{figure}[!htb]
    \centering
    \includegraphics[width=0.6\columnwidth]{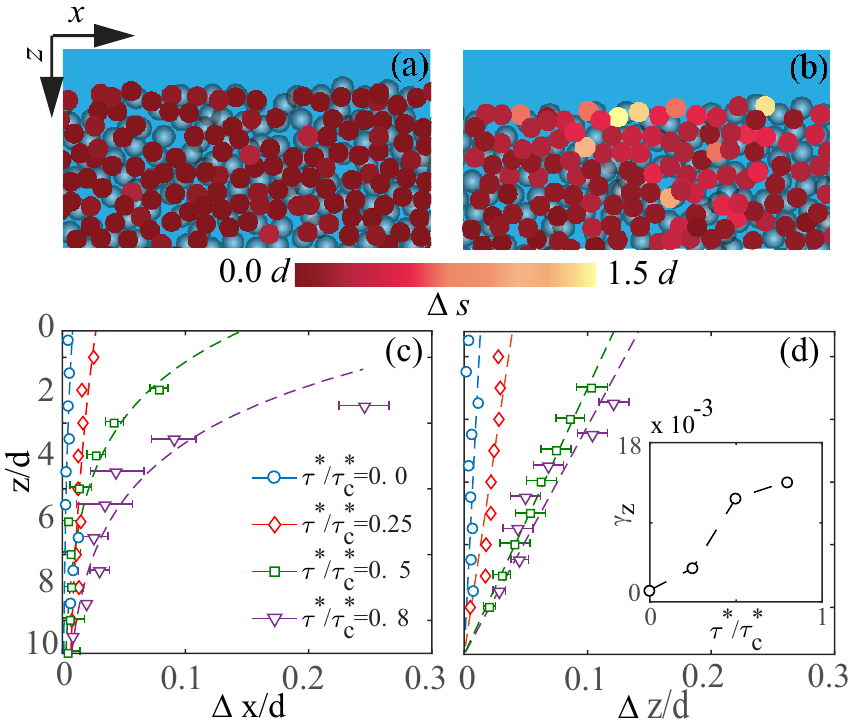}
    \caption{From \citet{AllenKudrolli18}, particle movement during 90 minutes with (a) no fluid flow and (b) fluid flow at 80\% of the critical flow rate (i.e., $\tau^\ast / \tau^\ast_c = 0.8$) to initiate particle transport (brighter colors indicate more particle movement). Movement is also plotted during times $t=30$--90~s in the (c) flow ($x$) direction and (d) gravity ($z$) direction. There is exponential behavior in the $x$~direction and a linear shift in the $z$~direction. The strain $\gamma_z$ is shown in the inset to (d), by fitting the slopes of the data in (d). Copyright 2018 American Physical Society.}
    \label{fig:creep-2}
\end{figure}

There is a second class of creeping flow, which is currently not explained by any rheological model. In the above discussion, creeping granular flow at $\mu<\mu_s$ was always induced by nearby regions with $\mu>\mu_s$. In some cases, creeping flow can be observed at $\mu<\mu_s$ without any apparent granular flow nearby at $\mu>\mu_s$~\citep{Amonetal13}. This class of creep is often accompanied by compaction of the bed. Slow shear and compaction interact in a complex way that is not fully understood but can be crucial in regulating slow (e.g., millimeters to meters per day) geophysical flows~\citep{MooreIverson02}. Similar behavior was also observed in laboratory sediment transport experiments by \citet{Houssaisetal15} and further studied by \citet{AllenKudrolli18}, as shown in Figure~\ref{fig:creep-2}.
The latter authors observed a granular bed with an overlying laminar shear flow and showed that slow (less than 0.1 grain diameters in 90~min) creeping flow persisted even for $\Theta \ll \Theta_t$ (meaning that $\mu<\mu_s$ everywhere in the granular bed). The grain motion in the direction of fluid flow followed an exponential decay with depth, similar to the creep described by nonlocal models. However, it was not induced by granular flow but somehow by the laminar fluid flow. Streamwise creep was also accompanied by compaction of the bed, which can strengthen the material and thus reduce creep. This second class of creep is therefore similar to compaction~\citep{Knightetal95,Ribietal05} and creep~\citep{Divouxetal08,CandelierDauchot09} that is induced by tapping or vibrations, despite the fact that no explicit vibrations were applied. The existence of this class of creep implies that sediment is likely \textit{always} transported (albeit slowly) for arbitrarily small values of $\Theta$, even in the absence of turbulence. Another recent experimental flume study~\citep{MastellerFinnegan17} showed a similar result, where conditioning a bed by applying weak fluid flow led to zero net transport but a smoother bed profile with fewer protruding grains. Then, when the fluid flow rate was increased to a value associated with significant transport for a conditioned bed, sediment transport rates were smaller when compared with an unconditioned bed. 

\subsubsection*{Open Problem: Physical Origin of Creeping Below Macroscopic Yield}
The physical mechanisms that lead to the second class of creep, where $\mu<\mu_s$ everywhere in the system, are not known. One possible mechanism is contact aging~\citep{Jiaetal11}, where the microscopic contact structure between two solid objects (i.e., grains) can evolve and weaken with time for reasons that are not fully understood~\citep{LiuSzlufarska12}. Additionally, \citet{Ponsetal16} showed that this second class of creep could be induced in dry granular flow by applying small pressure fluctuations to the interstitial air, with resulting shear rates of the order of 10$^{-7}$. Similar fluctuations likely always exist in natural systems. These two hypotheses are supported by the fact that, to the authors' knowledge, this class of creep does not occur in DEM simulations, which use a Cundall-Strack model~\citep{CundallStrack79} or similar Coulomb-like yield criterion for the frictional forces between grains, and fluctuating forces or slow variations in grain-grain friction are not included. Some DEM studies have observed creeping below a macroscopic yield criterion like the angle of response~\citep{Ferdowsietal18}, but the results from these studies seem to always include some region of $\mu>\mu_s$. 

\subsection{Critical Behavior and Weak Links} \label{CriticalBehavior}
Many experimental and computational studies~\citep{Carneiroetal11,Heymanetal13,Houssaisetal15} have observed that, near sediment transport thresholds (including the impact entrainment threshold, reviewed in section~\ref{ImpactEntrainmentContinuousTransport}), the time $t_{\rm conv}$ required for some system measurement (e.g., the sediment transport rate $Q$) to converge to its steady state value appears to grow very large. A common form~\citep{Clarketal15a} to capture these long time scales is $t_{\rm conv} \propto |\Theta - \Theta_t|^{-\beta}$, where $\beta$ is some positive exponent. A diverging time scale can arise in many ways, but one possibility is a \textit{critical phase transition}. The study of phase transitions, where a material abruptly changes as a control parameter is smoothly varied, originated in thermal physics (e.g., liquid-gas or ferromagnetic transition), but it has also been successful in describing many other kinds of systems where thermal physics is not applicable. The key feature of a \textit{critical} phase transition is a diverging correlation length, such that small changes near the critical point can have system-spanning effects that last for arbitrarily long times. The system is thus said to be \textit{scale-free} at the critical point, since there is no largest length or time scale that is affected by a perturbation. 

\subsubsection*{Open Question: Is Flow-Induced Bed Failure a Critical Phenomenon?}
Bed failure at the yield stress describes by definition a phase transition, but whether this transition is critical and how it arises from grain-grain and grain-fluid interactions remain open questions. However, there is a growing body of work~\citep{Clarketal18,Srivistavaetal19,ThompsonClark19} suggesting that the yielding transition for granular media is a critical transition. This is also suggested by the diverging correlation length $\xi \propto |\mu - \mu_s|^{-\nu}$ that is present in the nonlocal models discussed above~\citep{KamrinKoval12,Bouzidetal13}. In addition to describing creeping flow for $\mu<\mu_s$, nonlocal theories are also able to correctly predict other size-dependent effects, like strengthening of thin layers~\citep{MiDi04,KamrinHenann15}. The idea that yielding of granular media is a critical transition helps to explain certain experimentally observed behaviors in laboratory and computational models of sediment transport. For example, using a laboratory flume near the viscous limit, \citet{Houssaisetal15,Houssaisetal16} found a diverging time scale near the critical Shields number that is ``associated with the slowing down, and increasing variability, of the particle dynamics; it is unrelated to hydrodynamics.'' Evidence of scale-free channeling patterns~\citep{Aussillousetal16} was also observed during erosion of granular beds, which was attributed to the fact that the onset of erosion was behaving like a critical phase transition. 

When the physics controlling the onset of grain motion is no longer just the yield strength of the granular material itself, then the picture changes somewhat. For example, once particle inertia becomes important in sustaining nonsuspended sediment transport (see section~\ref{ParticleInertia}), the granular phase may not have a frictional state $\mu$ that is close to $\mu_s$, and thus it may be far from the critical point. For viscous bedload transport (small ${\rm Re}_\ast$), when particle inertia is not important, computational studies typically show that $t_{\rm conv}$ obeys system size dependence that is consistent with a critical phase transition~\citep{Yanetal16,Clarketal18}. However, under steady driving conditions, when grain inertia starts to play a role (e.g., for larger ${\rm Re}_\ast$), then $t_{\rm conv}$ still diverges, $t_{\rm conv} \propto |\Theta - \Theta_t|^{-\beta}$, but systems of different sizes will have the same $t_{\rm conv}$~\citep{Clarketal15a,Clarketal17}. Thus, $\Theta_t$ for inertial particles appears to be more similar to a dynamical instability rather than a true critical point.

However, nonlocal effects still likely play a role in the initiation of permanent bed failure. For example, if particle inertia plays a crucial role in sustaining sediment transport, as argued below in section~\ref{ParticleInertia}, then a bed could be above the threshold needed to sustain motion but not have any way to get started. Returning to the argument from \citet{Kirchneretal90} discussed above, if only the grains with the lowest entrainment thresholds are susceptible to being moved by the fluid, then these grains might be thought of as \textit{weak links} in the bed. Motion that is initiated by these weak links could trigger flow elsewhere in the system, via the redistribution of forces or by collision. \citet{Clarketal15a,Clarketal17} showed that the initiation of motion did indeed obey statistics consistent with a Weibullian weakest link scenario.

\subsection{Summary}
This section has described recent advances in the physics of sheared granular flows, with a focus on application to sediment transport. The main ideas are as follows. First, the yield condition for granular materials (e.g., a sediment bed) has the form of a static friction coefficient $\mu_s$, but it is not set directly by grain-grain friction. Instead, $\mu_s$ is an emergent property that arises from the maximum structural anisotropy that the grain-grain contact network can support. Friction plays a minor role in determining this maximum anisotropy, and grain shape and polydispersity also play minor roles. Second, although these contact networks are extended in space (and thus inherently nonlocal), local rheological descriptions (i.e., constitutive laws) can be very successful in many contexts. Recent advances suggest that a unified, local rheological description might be within reach. This rule could be used to model any context of wet or dry granular flow with appropriate boundary conditions. Such a description could be used to predict sediment transport rates and thresholds if the grain properties (i.e., size distribution, friction coefficient, grain shape, etc.) were known, even approximately. Third, the inherently nonlocal nature of yielding is dominant when the material is near its yield condition. This causes creeping behavior in regions where a local rheology would predict no flow, which complicates the search for a unified rheological description. However, the results described in Figures~\ref{fig:creep-1} and \ref{fig:creep-2} showed that creeping is similar in wet and dry flows, since it very slow and thus dominated by grain rearrangements (not fluid). This suggests that the nonlocal descriptions for wet and dry flows might also be unified in a relatively simple way. The underlying physics behind this nonlocal behavior is not fully understood, but there is mounting evidence that yielding of granular materials represents a kind of critical transition, where different parts of the system can be correlated over arbitrary distances. Remarkably, for sediment transport, creep seems to occur even much below the yield transition, that is, for seemingly arbitrarily small Shields numbers $\Theta$.

This section has considered only sediment beds sheared by nonfluctuating flows and usually neglected the effects of particle inertia in sustaining sediment transport. That is, except for the occurrence of creep, many of the results of this section do not apply to turbulent flows nor flows with significant particle inertia effects that are near the threshold for grain motion (occurring for sufficiently large ${\rm Re}_\ast$ and/or $s$, see section~\ref{ParticleInertia}). In particular, the average fluid shear stress at which turbulent flows are able to entrain bed particles is usually much below the yield stress of the granular phase. Nonetheless, both creep and the viscous yields stress $\Theta^{\rm max}_t$ will play crucial roles in the new conceptual picture of sediment transport thresholds and sediment entrainment that is presented in section~\ref{Summary}.

\section{Fluid Entrainment by Turbulent Flows} \label{FluidEntrainment}
This section reviews the state of the art on the entrainment of bed particles by a turbulent flow of Newtonian fluid. This process is not equivalent to the initiation of overall sediment motion, which occurs even in the absence of bed sediment entrainment because of creeping (see section~\ref{Creeping}). It is also not equivalent to the comparably simple physics of fluid entrainment by a nonfluctuating flow. For example, when a laminar flow of a Newtonian fluid shears a target particle resting on the sediment bed, there are critical values of the fluid shear stress $\tau$, which depend on the local bed arrangement, above which this particle begins to roll and slide, respectively~\citep{Agudoetal17,DeskosDiplas18}. Once motion begins, resisting forces weaken and, since the flow does not fluctuate, the particle will inevitably leave its bed pocket (i.e., become entrained). The entrained particle will travel along the bed until it comes to rest in another pocket in which it can resist the flow, provided such a pocket exists and is accessible (when the sediment bed has yielded, particles can no longer find stable resting place, see section~\ref{Yielding}). In contrast, in turbulent flows, even though resisting forces weaken when a bed particle becomes mobilized, such a mobilized particle may not find its way out of its initial bed pocket (i.e., incomplete entrainment). The prototype for this situation is a turbulent fluctuation of the flow that exerts a large force on the particle, but the fluctuation is too short-lived for it to become entrained. Hence, there are two important ingredients that need to be considered to accurately describe sediment entrainment by turbulent flows for a given pocket geometry: the magnitude and duration of turbulent fluctuations (evidence for this statement is briefly reviewed in section~\ref{RoleFluctuations}). Only entrainment criteria that account for both aspects are able to accurately describe fluid entrainment experiments (section~\ref{EntrainmentCriteria}). Shear stress-based criteria, in general, do not belong to this category. Yet one can still define the critical shear stress $\tau^{\rm In}_t$ above which the probability of fluid entrainment exceeds zero. This and related thresholds have received a lot of attention in studies on aeolian and planetary transport (section~\ref{ShearStressThreshold}). 

\subsection{The Role of Turbulent Fluctuations in Fluid Entrainment} \label{RoleFluctuations}
Turbulent fluctuations have been known to play a crucial role in fluid entrainment for a long time. For example, \citet{EinsteinElSamni49}, and later \citet{MollingerNieuwstadt96}, measured large fluctuating lift forces on a fixed rough surface induced by pressure gradient fluctuations of the order of the mean pressure gradient. These authors concluded that such pressure gradient fluctuations must be important also for the mobilization of bed sediment. In fact, numerous laboratory, field, and theoretical studies have advocated the viewpoint that the magnitude of peaks of the instantaneous flow force acting on a bed particle, consisting of both lift and drag forces, is a key aspect of fluid entrainment~\citep[e.g.,][]{Kalinske47,Sutherland67,Paintal71,HeathershawThorne85,ApperleyRaudkivi89,Kirchneretal90,Nelsonetal95,Papanicolaouetal01,Sumeretal03,Zanke03,Hoflandetal05,Schmeeckleetal07,VollmerKleinhans07,GimenezCurtoCorniero09,Dwivedietal10,Dwivedietal11,Cameronetal19,Cameronetal20}. However, while such force peaks explain certain observations, such as the episodic character of very weak turbulent bedload transport~\citep{Paintal71,HellandHansenetal74,Hofland05} or the strong increase of weak turbulent bedload transport in the presence of vegetation~\citep{YagerSchmeeckle13,YangNepf18,YangNepf19}, they do not explain all observations. In fact, experiments in which a target particle was placed on an idealized rough substrate and exposed to an electrodynamic force revealed that very high force pulses do not lead to entrainment if their duration is too short~\citep{Diplasetal08}. Likewise, moderate force pulses that only barely exceed resisting forces lead to entrainment if their duration is sufficiently long. That the duration of force peaks is as important as their magnitude has also been experimentally confirmed both for particles resting on idealized, fixed beds~\citep{Diplasetal08,Celiketal10,Celiketal13,Celiketal14,Valyrakisetal10,Valyrakisetal11,Valyrakisetal13,Valyrakis13} and natural erodible sediment beds~\citep{Salimetal17,Salimetal18}. However, note that, for sediment transport along erodible beds (with the exception of viscous bedload transport), the vast majority of entrainment events are triggered by particle-bed impacts, except for very weak transport conditions (see sections~\ref{ImpactEntrainmentBedload} and \ref{ImpactEntrainmentContinuousTransport}). In the following, criteria are reviewed that account for both the magnitude and duration of turbulent fluctuation events.

\subsection{Entrainment Criteria That Account for the Magnitude and Duration of Turbulent Fluctuation Events} \label{EntrainmentCriteria}
\subsubsection{Impulse Criterion} \label{ImpulseCriterion}
\begin{figure}[!htb]
 \begin{center}
  \includegraphics[width=1.0\columnwidth]{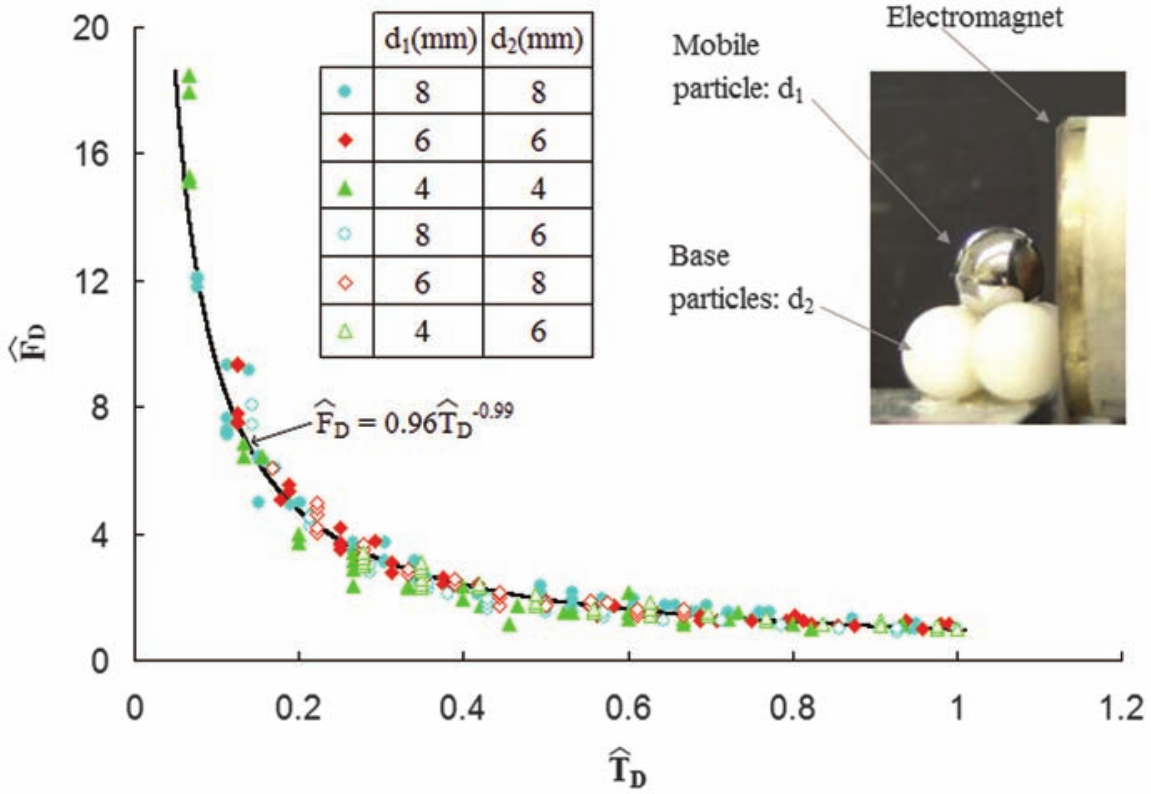}
 \end{center}
 \caption{From \citet{Diplasetal08} (M.V. is copyright holder), normalized magnitude $\hat F_D$ of the electrodynamic force pulse that is required for entrainment versus normalized duration $\hat T_D$ of the force pulse. Data correspond to the entrainment experiments that were carried out for various particle arrangements and varying sizes of the target ($d_1$) and base particles ($d_2$). The line corresponds to the prediction $\hat F_D=\hat T_D^{-1}$ associated with a constant impulse threshold.}
\label{ImpulseExperiments}
\end{figure}

The initiation of movement of a target particle resting in a pocket of the bed surface necessarily requires that the instantaneous flow forces (or torques) $F(t_0)$ acting on it at the instant $t_0$ of initial motion overcome resisting forces (or torques) $F_c$:
\begin{linenomath*}
\begin{equation}
 F(t_0)\geq F_c,
\end{equation}
\end{linenomath*}
However, this criterion is not sufficient for entrainment to occur as the target particle may merely move back to its initial resting place if $F(t)$ becomes subcritical for times $t$ too soon after $t_o$ so that its gained kinetic energy is insufficient to overcome the potential barrier of its bed pocket. For this reason, \citet{Diplasetal08} proposed that the fluid impulse $I_f$ associated with larger-than-critical flow forces must exceed a critical value:
\begin{linenomath*}
\begin{equation}
 I_f\equiv\int\limits_{t_0}^{t_0+T}F(t)\mathrm{d}t\geq I_{fc}\quad\text{with}\quad F(t)\geq F_c\quad\text{for}\quad t\in(t_0,t_0+T), \label{ImpulseDef}
\end{equation}
\end{linenomath*}
where $T$ is the duration of the impulse event (i.e., the duration of the particle acceleration phase of a turbulent fluctuation event). Note that $T$ can be much smaller than the time needed to leave the bed pocket as the latter also includes the particle deceleration phase. \citet{Diplasetal08} confirmed their hypothesis with idealized experiments in which they subjected an isolated target particle with a constant electrodynamic, horizontal force $F_D$ for a given time $T_D$, for which $I_f=F_DT_D$. In fact, their measured data of the force that is required for entrainment roughly obey the relation $\hat F_D\equiv F_D/F^{\mathrm{min}}_D=T^{\mathrm{max}}_D/T_D\equiv\hat T_D^{-1}$, where $F^{\mathrm{min}}_D$ is the minimal force required for measurable particle motion (but not necessarily entrainment) and $T^{\mathrm{max}}_D$ the associated time that is needed for $F^{\mathrm{min}}_D$ to cause entrainment (Figure~\ref{ImpulseExperiments}).

\begin{figure}[!htb]
 \begin{center}
  \includegraphics[width=0.5\columnwidth]{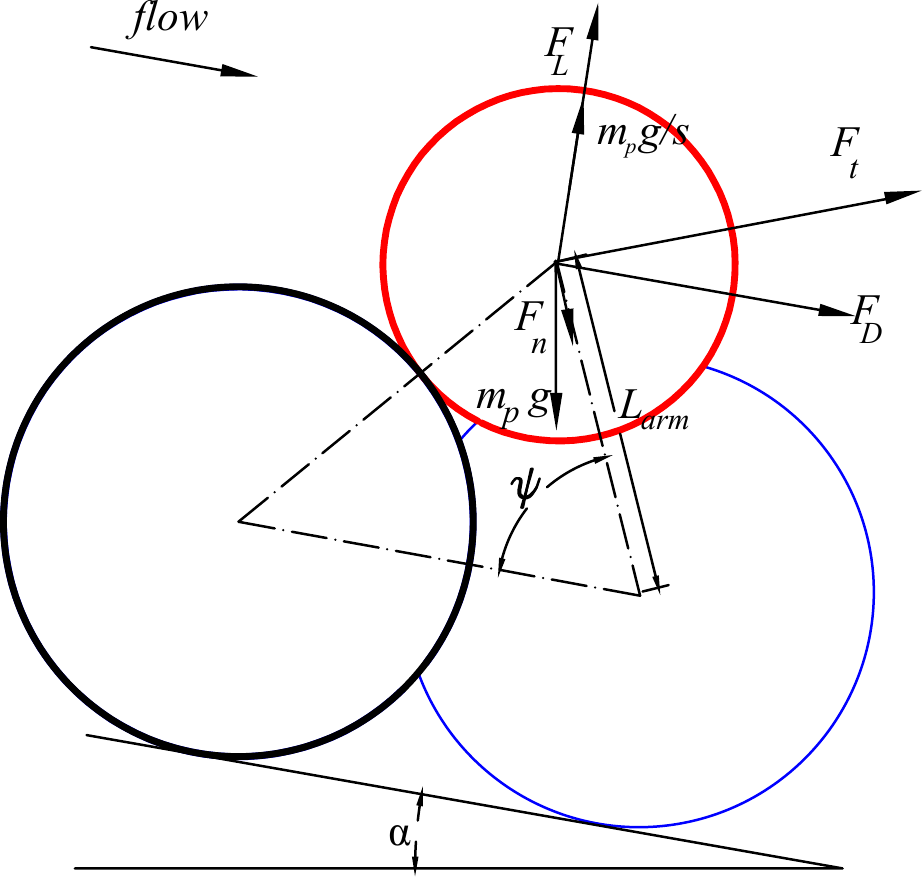}
 \end{center}
 \caption{Sketch of pocket geometry.}
\label{PocketSketch}
\end{figure}

In order to use equation~(\ref{ImpulseDef}) for predicting particle entrainment, one needs to know the impulse threshold $I_{fc}$. For entrainment into a rolling motion, \citet{Valyrakisetal10} derived an expression for the critical impulse $I_{fc}=F_tT_t$ ($F_t$ is defined below) assuming a constant pulse of a hydrodynamic force, separated into a horizontal drag and vertical lift component ($\mathbf{F}=(F_D,F_L)$), of short duration $T_t$ (so that the angular displacement $\Delta\psi$ of the particle remains small for $t\in(t_0,t_0+T_t)$):
\begin{linenomath*}
\begin{equation}
 I_{fc}=\frac{F_t}{g}\sqrt{2f(\psi,\alpha,s)L_{\rm arm}g\left(\frac{7}{5}+\frac{C_m}{s}\right)}\sqrt{\frac{-m_pg}{2\rho_\psi(F_n-F_{nc})}}\operatorname{arsinh}\left[\frac{\sqrt{-2\rho_\psi(F_n-F_{nc})(m_pg)}}{(F_t-F_{tc})}\right], \label{ImpulseCriticalRolling1}
\end{equation}
\end{linenomath*}
where $F_t=F_D\sin\psi+F_L\cos\psi$ and $F_n=-F_D\cos\psi+F_L\sin\psi$ are the tangential and normal components, respectively, of the driving flow force at the rest position, $m_p=\frac{1}{6}\rho_p\pi d^3$ is the particle mass, $F_{tc}=m_pg\cos(\psi+\alpha)/\sin\psi-(m_pg/s)\cot\psi$ the resisting force, $L_{\rm arm}$ the lever arm length, $C_m=1/2$ the added mass coefficient, and $f(\psi,\alpha,s)=\cos(\psi+\alpha)\sin\alpha+[1-\sin(\psi+\alpha)](\cos\alpha-1/s)$, with $\alpha$ the bed slope angle and $\psi$ the pivoting angle (Figure~\ref{PocketSketch}).
For many conditions, this expression can be well approximated by~\citep{Valyrakisetal10}
\begin{linenomath*}
\begin{equation}
 I_{fc}\approx m_p\left(\frac{F_t}{F_t-F_{tc}}\right)\sqrt{2f(\psi,\alpha,s)L_{\rm arm}g\left(\frac{7}{5}+\frac{C_m}{s}\right)}. \label{ImpulseCriticalRolling2}
\end{equation}
\end{linenomath*}
\citet{Leeetal12} derived an alternate expression for short turbulent fluctuation events. Instead of a pure rolling motion, they considered entrainment into a combined rolling and sliding motion (however, note that rolling is usually the preferred mode of entrainment) without bed slope ($\alpha=0$), assuming that the associated tangential motion is described by a Coulomb friction law with friction coefficient $\mu_C$. Furthermore, instead of the pivoting angle, they described the pocket geometry by the horizontal ($\Delta X$) and vertical ($\Delta Z$) particle displacement (in units of $d$) that is needed for the particle to escape (equivalent to $\psi+\alpha=\pi/2$ in Figure~\ref{PocketSketch}). The expression by \citet{Leeetal12} reads
\begin{linenomath*}
\begin{equation}
 I_{fc}\equiv(F_eT_e)_c=(\Delta Z+\mu_C\Delta X)m_p\sqrt{\frac{F_e}{F_e-F_{ec}}}\sqrt{2gd\left(1+\frac{C_m}{s}\right)\left(1+\frac{1}{s}\right)}, \label{ImpulseCriticalRolling3}
\end{equation}
\end{linenomath*}
where $F_e=F_D(\sin\psi-\mu\cos\psi)+F_L(\cos\psi+\mu_C\sin\psi)$ is an effective hydrodynamic force and $F_{ec}=m_pg(1-1/s)(\sin\psi+\mu_C\cos\psi)$ its critical value. For entrainment into a hopping motion, defined as a lift force-induced particle uplift by a vertical distance $\geq1d$, \citet{Valyrakisetal10} derived
\begin{linenomath*}
\begin{equation}
 I_{fc}\equiv(F_LT_L)_c=m_p\sqrt{\frac{F_L}{F_L-F_{Lc}}}\sqrt{2gd\cos\alpha\left(1+\frac{C_m}{s}\right)\left(1+\frac{1}{s}\right)}, \label{ImpulseCriticalHopping}
\end{equation}
\end{linenomath*}
where the resistance force is given by $F_{Lc}=m_pg(1-1/s)\cos\alpha$. Note that equation~(\ref{ImpulseCriticalHopping}) with $\alpha=0$ is equivalent to equation~(\ref{ImpulseCriticalRolling3}) if the critical dimensionless displacement $\Delta Z+\mu_C\Delta X=1$ and $F_{L(c)}$ replaced by $F_{e(c)}$.

Equations~(\ref{ImpulseCriticalRolling1})--(\ref{ImpulseCriticalHopping}) reveal that the impulse threshold $I_{fc}$ is constant only if the driving flow force is very strong ($F(t)\gg F_c$). However, for near-critical fluctuation events ($F(t)\rightarrow F_c$), $I_{fc}$ diverges. This motivates the introduction of an energy-based entrainment criterion.

\subsubsection{Energy Criterion} \label{EnergyCriterion}
\begin{figure}[!htb]
 \begin{center}
  \includegraphics[width=1.0\columnwidth]{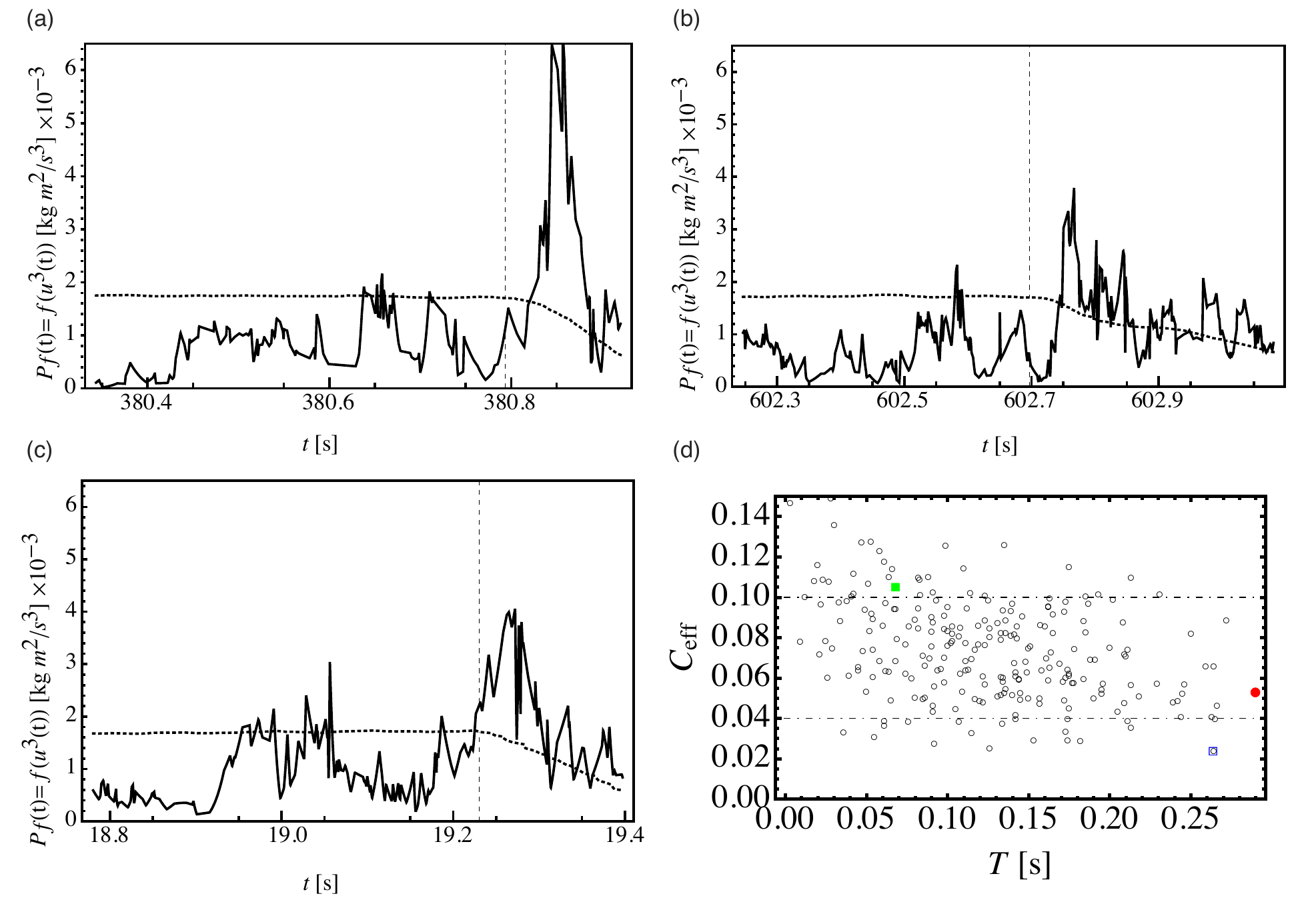}
 \end{center}
 \caption{(a--c) Flow power $P_f(t)$ versus time $t$ for three different turbulent fluctuation events that lead to entrainment of a target particle resting on a prearranged substrate. The solid lines corresponds to experimental data~\citep{Valyrakisetal13}. The dashed lines indicate the start of the respective fluctuation event. The dotted lines indicate the critical flow power that must be exceeded in order to overcome the resisting forces (i.e., $u>u_c(t)$), which depend on time because resisting forces weaken once the target particle starts to move. (d) Coefficient of energy transfer $C_{\rm eff}$ versus duration of turbulent fluctuation event ($T$) for various recorded entrainment events (symbols). The green closed square corresponds to the event shown in (a), the blue open square to the event shown in (b), and the red closed circle to the event shown in (c).}
\label{Ceff}
\end{figure}

The impulse criterion (equation~(\ref{ImpulseDef})) accounts for the available momentum of the turbulent fluctuation event in comparison to the momentum required for entrainment. However, close observation of near-bed turbulence reveals that fluctuation events are scarcely ever square pulses or even single-peaked~\citep{Valyrakis13}. Instead, turbulent flows in nature exhibit a wide range of flow patterns and structures, some of which may be more efficient for particle entrainment than others. For example, the transfer of energy from flow to particles in turbulent fluctuation events with large driving flow forces ($F(t)\gg F_c$) is expected to be much more efficient than in fluctuation events with near-critical flow forces ($F(t)\sim F_c$, see section~\ref{ImpulseCriterion}). This motivates the characterization of entrainment using the energy of the fluctuation event that is effectively transferred to the particle~\citep{Valyrakisetal13}:
\begin{linenomath*}
\begin{equation}
 C_{\rm eff}E_f=C_{\rm eff}\int\limits_{t_0}^{t_0+T}P_f(t)\mathrm{d}t\geq W_c, \label{EnergyDef}
\end{equation}
\end{linenomath*}
where $W_c$ is the minimal amount of work required for complete particle entrainment and $P_f(t)=f[u(t)^3]$ the instantaneous flow power, parameterized by the cube of the local flow velocity, and $C_{\rm eff}$ is the coefficient of energy transfer efficiency of the turbulent fluctuation event. The energy transfer coefficient $C_{\rm eff}$ is expected to increase with $\langle F\rangle/F_c$ (see section~\ref{ImpulseCriterion}), where $\langle\cdot\rangle$ denotes the time average over the event. Water flume experiments on the entrainment of a particle resting on an idealized substrate have confirmed that $C_{\rm eff}$ tends to increase with $\langle F\rangle/F_c$ (Figure~\ref{Ceff}).
However, one has to keep in mind that $C_{\rm eff}$ incorporates also other effects such as grain orientation and shape.

In order to use equation~(\ref{EnergyDef}) for predicting particle entrainment, one needs to know the energy threshold $W_c$. \citet{Valyrakisetal13} derived
\begin{linenomath*}
\begin{align}
 \text{Rolling:}&&W_c&=m_p\cos\alpha[1-\sin(\psi+\alpha)](1-1/s)gL_{\rm arm}, \label{EnergyCriticalRolling} \\
 \text{Hopping:}&&W_c&=m_p\cos\alpha(1-1/s)gd. \label{EnergyCriticalHopping}
\end{align}
\end{linenomath*}
For typical sediment beds, the ratio between both energy thresholds ($[1-\sin(\psi+\alpha)]L_{\rm arm}/d$) is of the order of $0.1$, demonstrating that a rolling motion is much more easily initiated upon entrainment than a hopping motion. Note that, in contrast to the expressions for the critical impulse for rolling (equations~(\ref{ImpulseCriticalRolling1}) and (\ref{ImpulseCriticalRolling2})), equation~(\ref{EnergyCriticalRolling}) does neither require the assumption of a small angular particle displacement $\Delta\psi$ during the acceleration phase of a turbulent fluctuation event nor the assumption of a short duration of this phase.

\subsection{Shear Stress Threshold of Incipient Motion and Initiation of Aeolian Saltation Transport} \label{ShearStressThreshold}
The entrainment criteria reviewed in section~\ref{EntrainmentCriteria} are able to predict whether a certain turbulent fluctuation event is capable of entraining a target particle, whereas a criterion based on a critical shear stress would not suffice for this purpose. However, one can still define a shear stress threshold $\tau^{\rm In}_t$ (the \textit{initiation threshold}) at which the fluid entrainment probability exceeds zero (i.e., below which entrainment never occurs). Such a threshold must exist because the size of turbulent flow eddies is limited by the system dimensions, such as the boundary layer thickness $\delta$. In fact, a limited size of turbulent flow eddies implies that also the magnitude of peaks of the flow force is limited. That is, one can always find a nonzero shear stress below which even the largest fluctuation peaks do not exceed the resisting forces acting on bed particles (however, note that the existence of sufficiently large flow force peaks does not guarantee a nonzero entrainment probability because their durations may always be too short). Like for $\Theta^{\rm max}_t$, transient behavior associated with the flow temporarily pushing particles from less stable to more stable pockets is excluded in the definition of $\tau^{\rm In}_t$, which implies $\Theta^{\rm In}_t\simeq\Theta^{\rm max}_t$ for laminar flows at sufficiently low shear Reynolds number $\mathrm{Re}_\ast$. Furthermore, surface inhomogeneities that can generate a lot of turbulence, such as vegetation~\citep{YagerSchmeeckle13,YangNepf18,YangNepf19}, are also not considered in the definition of $\tau^{\rm In}_t$. While $\tau^{\rm In}_t$ is usually not measured for turbulent fluvial bedload transport (it is much below the Shields curve~\citep{Paintal71}), it has often been measured in wind tunnel experiments (briefly reviewed in section~\ref{WindTunnel}), including those that sought to determine the initiation threshold of aeolian saltation transport. The reason is that as soon as the first particles of the initially quiescent bed surface are entrained (i.e., begin to roll as rolling requires the smallest flow forces), the flow is usually nearly sufficient to net accelerate them during their downstream motion, resulting in larger and larger particle hops (i.e., the initiation threshold of aeolian saltation transport is only slightly larger than $\tau^{\rm In}_t$)~\citep{Bagnold41,Iversenetal87,Burretal15}. This occurs because, for typical wind tunnels, $\tau^{\rm In}_t$ is significantly above the cessation threshold of saltation transport (see section~\ref{CessationThresholdModels}). However, it will become clear that this statement may not apply to aeolian field conditions. Section~\ref{SaltationInitiation} briefly reviews models of $\tau^{\rm In}_t$ derived from wind tunnel experiments, while section~\ref{BoundaryLayerThickness} reviews recent evidence that indicates that such models, in general, are unreliable, particularly when applied to field conditions.

\subsubsection{Wind Tunnel Experiments of the Initiation of Aeolian Rolling and Saltation Transport} \label{WindTunnel}
Two distinct experimental setups have been used to measure $\tau^{\rm In}_t$. In the first setup, small isolated patches of particles are placed at the bottom of a wind tunnel and then the fluid shear stress $\tau$ is increased until particles in such patches start to roll or detach~\citep{Williamsetal94,Merrisonetal07,DeVetetal14}. In the second setup, a complete bed of particles is prepared at the tunnel bottom and then the fluid shear stress $\tau$ is increased until saltation transport begins~\citep[e.g.,][]{Bagnold37,Chepil45,LylesKrauss71,Iversenetal76,Greeleyetal76,Greeleyetal80,Greeleyetal84,Gilletteetal80,GreeleyMarshall85,Nickling88,IversenRasmussen94,Dongetal03b,CornelisGabriels04,Burretal15,Carneiroetal15,Swannetal20} (see also \citet[][and references therein]{Raffaeleetal16}). It is worth noting that, according to the definition of $\tau^{\rm In}_t$, beginning saltation transport refers to the mere occurrence of saltation transport, even if very sporadic, which is also the definition used by \citet{Bagnold37}. However, many experimental studies defined beginning saltation transport through a critical loosely defined saltation transport activity (similar to the definition of the fluvial transport thresholds compiled in the Shields diagram), which yields slightly larger threshold values~\citep{Nickling88}. 

\subsubsection*{Open Problem: Qualitative Discrepancy Between Threshold Measurements}
For cohesionless particles ($d\gtrsim100~\mu$m), existing threshold measurements based on the second setup show that $\tau^{\rm In}_t$ increases relatively strongly with the particle diameter $d$~\citep{Raffaeleetal16}. In contrast, for the first setup, measurements indicate that $\tau^{\rm In}_t$ remains constant with $d$ for $d\gtrsim100~\mu$m~\citep{Merrisonetal07,DeVetetal14}. The reason for this qualitative inconsistency is not understood. \citet{Merrisonetal07} suggested that the initiation of rolling (measured in their experiments) may be different to that of saltation transport. However, this suggestion is inconsistent with the observation that saltation transport in wind tunnels is preceded by rolling further upwind~\citep{Bagnold41,Iversenetal87,Burretal15}. Furthermore, in contrast to standard wind tunnel experiments, for experiments in pressurized wind tunnels with Venusian air pressure, both an equilibrium rolling (lower initiation threshold) and an equilibrium saltation transport regime (higher initiation threshold) exist, and both initiation thresholds strongly increase with $d$~\citep{GreeleyMarshall85}.

\subsubsection{Models of the Initiation of Aeolian Rolling and Saltation Transport} \label{SaltationInitiation}
Nearly all existing models of the initiation of aeolian rolling and saltation transport (including sand transport~\citep{Bagnold41,Iversenetal76,Iversenetal87,IversenWhite82,ShaoLu00,CornelisGabriels04,Luetal05,ClaudinAndreotti06,KokRenno06,Merrisonetal07,Duranetal11,Duanetal13a,DeVetetal14,Burretal15,EdwardsNamikas15}, drifting snow~\citep{Schmidt80,Lehningetal00,HeOhara17}, and the transport of regolith dust by outgassed ice on the comet 67P/Churyumov-Gerasimenko~\citep{Jiaetal17}) predict $\tau^{\rm In}_t$ from the balance between aerodynamic forces and/or torques and resisting forces and/or torques acting on a bed particle. Even though many of these models do not consider peaks of the aerodynamic force, and some of them do not treat $\tau^{\rm In}_t$ as what it is (i.e., the threshold at which the fluid entrainment probability exceeds zero, see above), they are conceptually very similar and mainly differ in the empirical equations that they use for the aerodynamic and cohesive interparticle forces. For this reason, only one of the most popular and simple models, the model by \citet{ShaoLu00}, is discussed here. It reads
\begin{linenomath*}
\begin{equation}
 \Theta^{\rm In}_t=A_N\left(1+\frac{\gamma_C}{\rho_pgd^2}\right), \label{ShaoLuModel}
\end{equation}
\end{linenomath*}
where $A_N=0.0123$ is an empirical scaling factor and $\gamma_C=3\times10^{-4}~$kg/s$^2$ an empirical constant that accounts for cohesive interparticle forces. More complex models~\citep[e.g.,][]{IversenWhite82,ClaudinAndreotti06,Duranetal11} involve additional dependencies of $\Theta^{\rm In}_t$ on the shear Reynolds number $\mathrm{Re}_\ast$ or, equivalently, on the \textit{Galileo number} $\mathrm{Ga}\equiv\sqrt{(s-1)gd^3}/\nu_f\equiv\mathrm{Re}_\ast/\sqrt{\Theta}$ (also called \textit{Yalin parameter}~\citep{Yalin77}).

\subsubsection{Effects of the Boundary Layer Thickness on the Initiation of Aeolian Rolling and Saltation Transport} \label{BoundaryLayerThickness}
\begin{figure}[!htb]
\centering
\includegraphics[width=1.0\columnwidth]{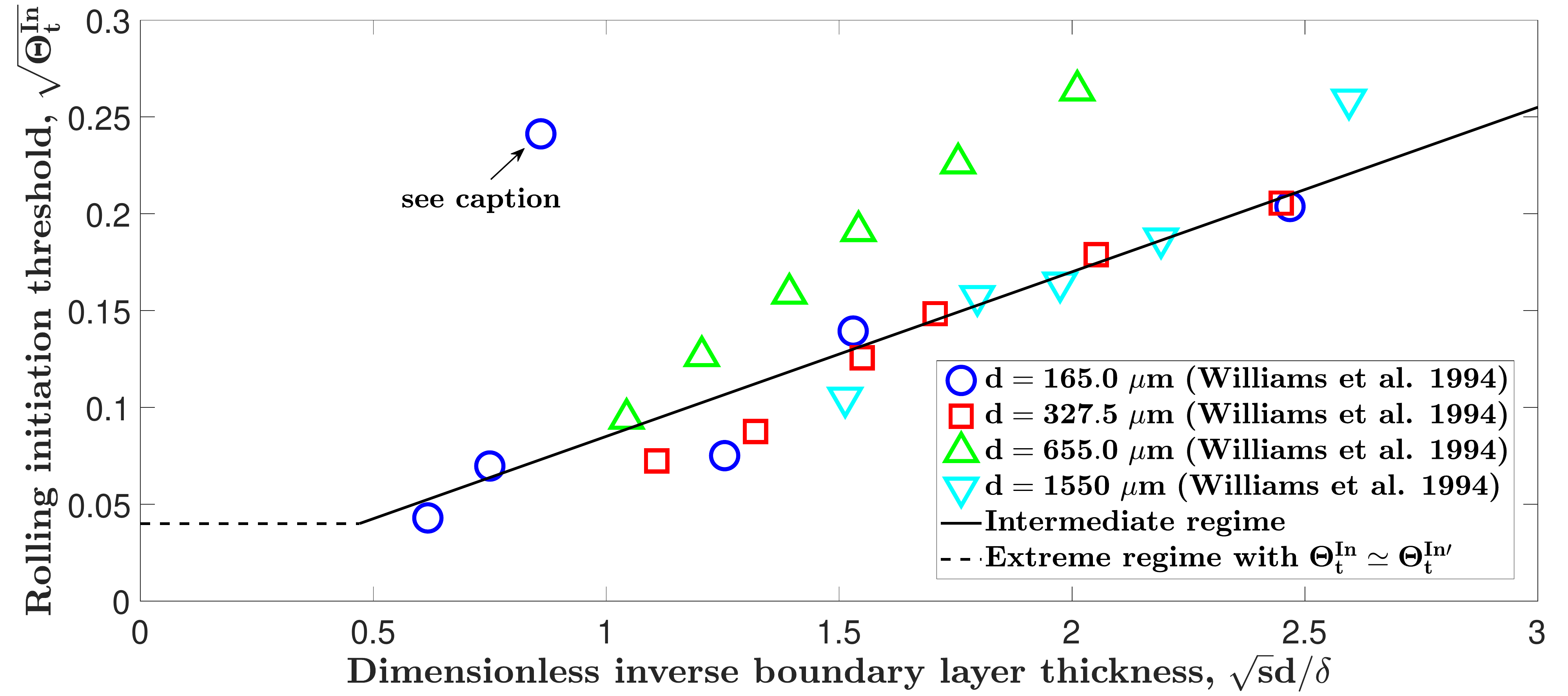}
\caption{From \citet{Pahtzetal18} (Creative Commons Attribution 4.0 International License), threshold parameter $\sqrt{\Theta^{\rm In}_t}$ versus dimensionless inverse boundary layer thickness $\sqrt{s}d/\delta$. Symbols correspond to the measurements of incipient rolling by \citet{Williamsetal94}, who set up their wind tunnel in a manner that produces a developing turbulent boundary layer, for four different sediments consisting of nearly uniform, cohesionless particles. The solid line corresponds to equation~(\ref{CriticalShearStress}) for the intermediate regime using $\sqrt{\Theta^{\rm In\prime}_t}\simeq\mathrm{const}$ (neglecting the weak logarithmic dependency of $\sqrt{\Theta^{\rm In\prime}_t}$ on $\delta/d$). This regime turns into the extreme regime in which $\sqrt{\Theta^{\rm In}_t}\simeq\sqrt{\Theta^{\rm In\prime}_t}$. This transition is shown by the dashed line assuming $\sqrt{\Theta^{\rm In\prime}_t}=0.04$ (only for illustration purposes as the actual values of $\sqrt{\Theta^{\rm In\prime}_t}$ in the experiments by \citet{Williamsetal94} are unknown). It is suspected that the one extreme outlier for $d=165~\mu$m may either have been a faulty measurement or be associated with the observation that the boundary layer for this particular sand sample was not always fully turbulent~\citep{Williamsetal94}.}
\label{WilliamsExperiments}
\end{figure}

\begin{figure}[!htb]
\centering
\includegraphics[width=1.0\columnwidth]{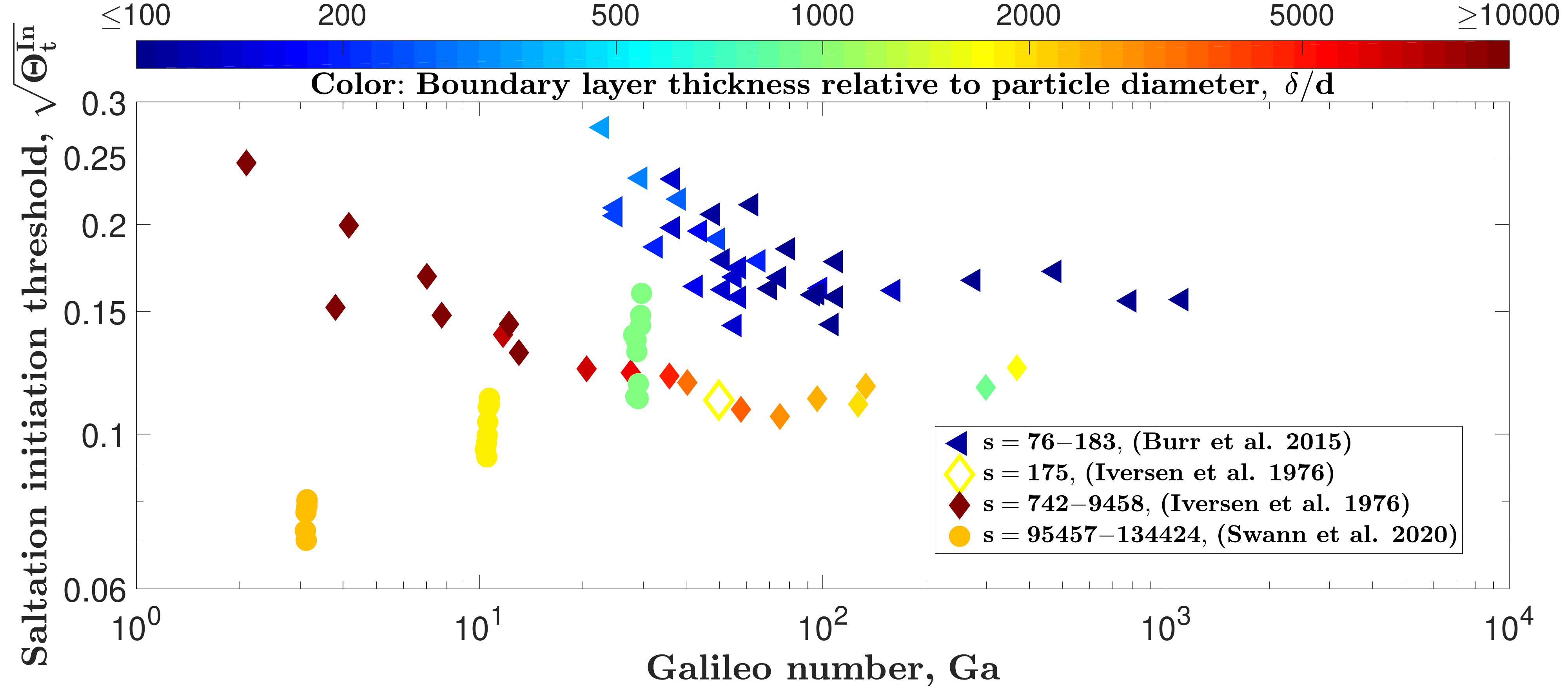}
\caption{Modified from \citet{Pahtzetal18} (Creative Commons Attribution 4.0 International License), compilation of measurements in wind tunnels with fully developed boundary layer of the initiation threshold parameter of saltation transport ($\approx\sqrt{\Theta^{\rm In}_t}$)~\citep{Iversenetal76,Burretal15,Swannetal20} versus the Galileo number $\mathrm{Ga}$. The color indicates the thickness of the boundary layer $\delta$ relative to the particle diameter $d$, which controls the relative amplitude of turbulent fluid velocity fluctuations for a constant $\mathrm{Ga}$. Circles correspond to threshold values obtained from the raw data by \citet{Swannetal20}. The threshold values for the experiments by \citet{Iversenetal76} are found in \citet{IversenWhite82}.}
\label{ThresholdMeasurements}
\end{figure}

The size of turbulent flow eddies, and thus the duration of turbulent fluctuation events, is limited by the system dimensions, more specifically, the boundary layer thickness $\delta$~\citep[see review by][and references therein]{Smitsetal11}. However, in most wind tunnel experiments and the field, the produced turbulent boundary layer should be so thick that any turbulent fluctuation has a nonzero probability to last sufficiently long for entrainment to occur~\citep{Pahtzetal18}. That is, the mere existence of aerodynamic force peaks that exceed resisting forces is sufficient for $\tau^{\rm In}_t$ to be exceeded. However, this is no longer true when $\delta$ becomes too small, at which point turbulent fluctuation events may cause particles to rock (i.e., vibrate or wobble or oscillate) within their bed pockets but no fluctuation lasts long enough for the particles to completely leave them. \citet{Pahtzetal18} physically modeled such situations and derived an expression for the ratio between $\tau^{\rm In}_t$ and the shear stress threshold $\tau^{\rm In\prime}_t$ of incipient rocking (equivalent to the Shields number ratio $\Theta^{\rm In}_t/\Theta^{\rm In\prime}_t$). These authors' derivation uses the impulse criterion of section~\ref{ImpulseCriterion} (even though \citet{Pahtzetal18} start with the energy criterion, their analysis is effectively equivalent to assuming a constant impulse threshold) and the fact that the maximal duration $T_{\rm max}$ of turbulent fluctuation events is controlled by $\delta$ and the local mean flow velocity $\overline{u}$ via $T_{\rm max}\propto\delta/\overline{u}$~\citep{AlhamdiBailey17}. The derived expression reads
\begin{linenomath*}
\begin{align}
 \sqrt{\frac{\Theta^{\rm In}_t}{\Theta^{\rm In\prime}_t}}&\simeq 
 \begin{matrix}
 1 & \text{if} & C<1  \\
 C & \text{if} & 1\leq C\leq\alpha_f  \\
 \alpha_f & \text{if} & C>\alpha_f \\
 \end{matrix}
 \label{CriticalShearStress} \\
 C&\equiv\alpha_f^{-1}f(G)\frac{\sqrt{s}d}{\delta}. \nonumber
\end{align}
\end{linenomath*}
where $\alpha_f\equiv u_m/\overline{u}\geq1$ is the ratio between the characteristic flow velocity $u_m$ associated with the largest positive fluctuations and $\overline{u}$, and $f(G)$ is a factor that encodes information about particle shape, orientation, and the pocket geometry. Equation~(\ref{CriticalShearStress}) encompasses three different regimes. In one extreme, if there is a nonzero probability that turbulent fluctuation events associated with the largest positive fluctuations last sufficiently long for particle entrainment, then there will be a nonzero probability that incipient rocking evolves into incipient rolling (i.e., $\Theta^{\rm In}_t/\Theta^{\rm In\prime}_t\simeq1$). In the other extreme, if all positive fluctuation events always last too short, the mean flow must exceed the torque balance for entrainment to occur (i.e., $\Theta^{\rm In}_t/\Theta^{\rm In\prime}_t\simeq\alpha_f^2$). In the intermediate regime between these two extremes, $\Theta^{\rm In}_t/\Theta^{\rm In\prime}_t$ is proportional to the square of the inverse dimensionless boundary layer thickness $(d/\delta)^2$. Although weak logarithmic dependencies on $\delta/d$ are also incorporated in $\alpha_f$ and $\Theta^{\rm In\prime}_t$~\citep{Luetal05}, they are dominated by this proportionality. In fact, Figure~\ref{WilliamsExperiments} shows that the prediction for the intermediate regime is roughly consistent with the experimental data by \citet{Williamsetal94} if one uses that the Shields number for incipient rocking ($\Theta^{\rm In\prime}_t$) is approximately constant, neglecting the logarithmic dependency of $\Theta^{\rm In\prime}_t$ on $\delta$ (and further minor dependencies on $\mathrm{Ga}$).
\citet{Williamsetal94} set up their wind tunnel in a manner that produces a relatively thin developing turbulent boundary layer (i.e., $\delta$ increases with downstream distance). However, once the intermediate regime is exceeded (i.e., $\Theta^{\rm In}_t\simeq\Theta^{\rm In\prime}_t$) because $\delta$ becomes too large, as for most wind tunnel experiments with fully developed boundary layers, the logarithmic dependency of $\Theta^{\rm In}_t$ on $\delta/d$ via $\Theta^{\rm In\prime}_t$ may become significant (Figure~\ref{ThresholdMeasurements}).
For example, for the same Galileo number $\mathrm{Ga}$, the threshold values measured by \citet{Burretal15} in Figure~\ref{ThresholdMeasurements}, which were carried out in a pressurized wind tunnel with $\delta\approx1.9~$cm, are significantly larger than those measured by \citet{Iversenetal76}, which were carried out in a wind tunnel with $\delta\approx1.2~$m.

\subsubsection*{Open Problem: Unexpected Behavior of Saltation Transport Initiation Threshold for Large Density Ratio}
The very recent measurements by \citet{Swannetal20}, who used a very-low pressure wind tunnel and three different beds of cohesionless particles ($d=[310,730,1310]~\mu$m) to mimic Martian conditions, indicate that $\sqrt{\Theta^{\rm In}_t}$ unexpectedly increases substantially with $\mathrm{Ga}$ and thus $d$ (Figure~\ref{ThresholdMeasurements}). A possible explanation could be that, because of the very large density ratio $s$, some of the experimental conditions may have been in the intermediate regime (i.e., $1\leq C\leq\alpha_f$ in equation~(\ref{CriticalShearStress})), in which $\sqrt{\Theta^{\rm In}_t}$ scales with $d$ (Figure~\ref{WilliamsExperiments}). In fact, $1/C\propto\delta/(\sqrt{s}d)\simeq[3.1,4.9,12.4]$ for the three conditions, where only the largest value (corresponding to $d=310~\mu$m) is larger than the critical value $\delta/(\sqrt{s}d)\approx6.6$ that \citet{Pahtzetal18} associated with the end of the intermediate regime. In other words, the measurements for $d=730~\mu$m and $d=1310~\mu$m may both have been in the intermediate regime, in which $\sqrt{\Theta^{\rm In}_t}$ roughly scales with $d$ (cf. Figure~\ref{WilliamsExperiments}).

\subsubsection*{Controversy: Dependency of Saltation Transport Initiation Threshold on Density Ratio}
Based on comparisons between experiments in pressurized wind tunnels with comparably very thin boundary layers but larger-than-normal air density~\citep{Greeleyetal84,Burretal15} and nonpressurized wind tunnels with comparably very large boundary layers~\citep{Iversenetal76} (and normal air density), \citet{Iversenetal87} and \citet{Burretal15} argued that there is an underlying decrease of the saltation transport initiation threshold (which is slightly larger than $\Theta^{\rm In}_t$ for aeolian transport in typical wind tunnels, see above) with the density ratio $s$ for a constant shear Reynolds number $\mathrm{Re}_\ast$ (equivalent to a constant $\mathrm{Ga}$). However, this dependency on $s$ may be an artifact of huge differences in the dimensionless boundary layer thickness $\delta/d$~\citep{Pahtzetal18}. In fact, even though the dependency of $\Theta^{\rm In}_t$ on $\delta/d$ is logarithmic once the intermediate regime is exceeded (like for the measurements in question), such weak dependencies can still have significant effects once differences in $\delta/d$ become very large. This point of view is supported by Figure~\ref{ThresholdMeasurements}, in which $\delta/d$ is color-coded. It can be seen that the yellow, open diamond (a measurement from a nonpressurized wind tunnel) exhibits a similar value of $s$ as the blue symbols (measurements from a pressurized wind tunnels), which was achieved by using a very light particle material ($\rho_p=210~$kg/m$^3$). Nonetheless, the threshold $\sqrt{\Theta^{\rm In}_t}$ of the former is significantly smaller than those of the latter. Also, the former measurement relatively smoothly connects to the other measurements carried out in the same nonpressurized wind tunnel, which exhibit much larger values of $s$. On the other hand, the measurements by \citet{Swannetal20}, for which $s$ is comparably very large and $\delta/d$ of a similar size as for the measurements by \citet{Iversenetal76}, support the density ratio hypothesis because of comparably small values of $\sqrt{\Theta^{\rm In}_t}$. Note that, for the discussion of threshold values, one has to keep in mind that threshold measurements are highly prone to measurement errors of various sources~\citep{Raffaeleetal16}. Such errors are likely much larger than often reported because measurements of $\sqrt{\Theta^{\rm In}_t}$ can vary by more than a factor of $2$ for a given condition, even for cohesionless particles~\citep{Raffaeleetal16}.

\subsubsection*{Open Problem: Aeolian Bedload Transport in the Field}
In wind tunnel experiments, rolling is being initiated at threshold values that are significantly above the cessation threshold of saltation transport (see section~\ref{CessationThresholdModels}). This is why rolling seems to always evolve into saltation transport (i.e., equilibrium rolling and thus aeolian bedload transport does not seem to exist)~\citep{Bagnold41,Iversenetal87,Burretal15}. However, atmospheric boundary layers are several orders of magnitude thicker than those of wind tunnels~\citep{Lorenzetal10,Petrosyanetal11,Koketal12,Lebonnoisetal18} and may therefore exhibit a significantly smaller rolling threshold. In contrast, the cessation threshold of saltation transport is predominantly a property of the mean turbulent flow (see section~\ref{CessationThresholdModels}) and therefore rather insensitive to the boundary layer thickness $\delta$. Hence, for atmospheric boundary layers, it is possible that equilibrium rolling transport exists. Note that equilibrium rolling transport has been observed in pressurized wind tunnels with Venusian air pressure for a narrow range of Shields numbers $\Theta$~\citep[e.g.,][]{GreeleyMarshall85}.

\subsubsection*{Open Problem: Reliable Models of the Initiation Threshold of Planetary Saltation Transport}
The most widely used models for the initiation of aeolian saltation transport (see section~\ref{SaltationInitiation}), which have been adjusted to wind tunnel measurements, do not take into account the dependency of the relative magnitude of turbulent fluctuations on the dimensionless boundary layer thickness $\delta/d$. This may be the reason why these models, when applied to Martian atmospheric conditions, predict threshold shear stresses for fine sand particles that are so large that transport should occur only during rare strong Mars storms~\citep{SullivanKok17}, in contradiction to modern observations indicating widespread and persistent sediment activity~\citep{Bridgesetal12a,Bridgesetal12b,Silvestroetal13,Chojnackietal15}, even of very coarse sand~\citep{Bakeretal18}. For example, for the Martian conditions reported by \citet{Bakeretal18} ($\rho_p=2900~$kg/m$^3$, $\rho_f=0.02~$kg/m$^3$, $g=3.71~$m/s$^2$, $d=1.5~$mm), equation~(\ref{ShaoLuModel}) predicts for the threshold shear velocity: $u^{\rm In}_{\ast t}\equiv\sqrt{\Theta^{\rm In}_t(\rho_p/\rho_f-1)gd}\simeq3.7~$m/s, which corresponds to winds that are more than twice as fast as the strongest Mars storms. Note that \citet{Luetal05} proposed a model for the initiation of rolling that includes the effect of $\delta/d$. The authors of this review therefore recommend to use the model by \citet{Luetal05} in combination with models of the cessation threshold of saltation transport (see section~\ref{CessationThresholdModels}) for the estimation of the occurrence of saltation transport in real atmospheric boundary layers. However, it remains to be demonstrated that this approach yields reliable predictions. In fact, in the field, atmospheric instability, topography gradients, and surface inhomogeneities, such as obstacles and vegetation, can dramatically enhance local turbulence and thus fluid entrainment. Likewise, sublimation of subsurface ice in cold environments (the so-called \textit{solid-state greenhouse effect}~\citep{Kaufmannetal06}) can generate airborne particles of carbon dioxide, methane, and nitrogen ice~\citep{Hansenetal90,Thomasetal15b,Jiaetal17,Telferetal18}. Given that even a few entrained particle can result in fully developed saltation transport provided that the fetch is sufficiently long~\citep{SullivanKok17}, it may well be that saltation transport in the field can almost always be initiated close to the cessation threshold~\citep{SullivanKok17,Pahtzetal18,Telferetal18}. Evidence for this hypothesis is seen on Pluto, where aeolian dunes and wind streaks have been observed even though saltation transport initiation had been thought to be virtually impossible because of Pluto's very thin atmosphere (pressure $P=1~$Pa) and relatively weak 10~m winds ($u^{\mathrm{max}}_{\mathrm{10m}}\approx10~$m/s)~\citep{Telferetal18}.

\subsubsection*{Open Problem: Lack of Direct Aeolian Sediment Transport Initiation Measurements in the Field}
The overarching problem associated with the rather poor current knowledge of aeolian sediment transport initiation in the field (see open problems above) is that, to the authors' knowledge, there are no \textit{direct} field measurements of the transport initiation threshold $\Theta^{\rm In}_t$. In fact, existing field experiments have focused on detecting aeolian saltation transport~\citep[][and references therein]{BarchynHugenholtz11} rather than on how the fluid entrainment of individual bed particles, which usually starts out as a rolling motion, leads to saltation transport. Hence, we currently do neither know the wind speeds that are required in the field to initiate rolling transport of individual bed particles nor whether such rolling transport, like in wind tunnels, always evolves into saltation transport (see open problems above). What adds to the problem is that existing field studies either obtain saltation transport threshold estimates using methods that do not seek to distinguish saltation transport initiation and cessation~\citep[][and references therein]{BarchynHugenholtz11} or assume that $\Theta^{\rm In}_t$ coincides with the continuous saltation transport threshold~\citep{MartinKok18} (which is a controversial assumption, see section~\ref{ImpactEntrainmentContinuousTransport}).

\section{The Role of Particle Inertia in Nonsuspended Sediment Transport} \label{ParticleInertia}
As discussed in section~\ref{Introduction}, old experimental studies~\citep[e.g.,][]{Ward69,GrafPazis77} strongly indicated that the fluvial transport threshold measurements that are compiled in the Shields diagram are to a nonnegligible degree affected by particle inertia. As the Shields diagram shows a rough data collapse of the threshold Shields number $\Theta_t$ as a function of the shear Reynolds number ${\rm Re}_\ast$, this raises the question of whether ${\rm Re}_\ast$ is in some way associated with particle inertia. Indeed, while ${\rm Re}_\ast$ has usually been interpreted as the ratio between the particle size and the size of the viscous sublayer of the turbulent boundary layer, \citet{Clarketal17} showed that it can also be interpreted as a number that compares the viscous damping time scale to the ballistic time scale between bed collisions. Importantly, these authors showed that the shape of the Shields curve can be partly explained by the fact that inertial particles at high ${\rm Re}_\ast$ are harder to stop.

In general, the role of particle inertia in nonsuspended sediment transport can be twofold. On the one hand, entrainment by or supported by particle-bed impacts may be able to supply the transport layer with bed particles and thus compensate captures of transported particles by the bed (section~\ref{ImpactEntrainment}). This mechanism gives rise to a shear stress threshold associated with impact entrainment. On the other hand, although the mean turbulent flow is usually too weak to initiate transport (which instead usually requires turbulent fluctuation events, see section~\ref{FluidEntrainment}), it may be able to sustain the motion of particles that are already in transport. This mechanism gives rise to a physical process-based definition of transport capacity and a shear stress threshold, which has often been misidentified as an entrainment threshold by \citet{Shields36} and others (section~\ref{ContinuousRebounds}). Various models for both shear stress thresholds that have been proposed in the literature are compared with one another in section~\ref{CessationThresholdModels}.

\subsection{Impact and Impact-Supported Entrainment} \label{ImpactEntrainment}
\citet{Bagnold41} was the first to recognize that impact entrainment is crucial for sustaining aeolian saltation transport. Based on his wind tunnel and field observations, he explained~\citep[][p.~102]{Bagnold41}, ``In air, the grains, when once set in motion along the surface, strike other stationary grains, and either themselves bounce high (a distance measured in hundreds if not thousands of grain diameters) into the relatively tenuous fluid, or eject other grains upwards to a similar height.'' Largely because of Bagnold's observations, the statistics of particle impacts onto a static granular packing have been subject of many experimental and theoretical investigations (section~\ref{StaticBedCollision}). \citet[][p.~102]{Bagnold41} also believed that impact entrainment is negligible for fluvial bedload transport: ``If the physics of this impact-ejection mechanism is applied to sand in water, it is found that the impact momentum of the descending grains is insufficient to raise surface grains to a height greater than a small fraction of one grain diameter.'' However, Bagnold, and numerous researchers after him, did not consider that even a marginal uplift of a bed particle can make it much easier for a turbulent fluctuation event to entrain it (section~\ref{ImpactEntrainmentBedload}) and that, once bedload transport becomes sufficiently strong, multiple particle-bed impacts occur in so short sequence that the bed can no longer be considered as static. In fact, for continuous transport, recent studies revealed that impact entrainment alone can sustain bedload transport (section~\ref{ImpactEntrainmentContinuousTransport}).

\subsubsection{Impact of an Incident Particle Onto a Static Granular Packing} \label{StaticBedCollision}
\begin{figure}[!htb]
 \begin{center}
  \includegraphics[width=1.0\columnwidth]{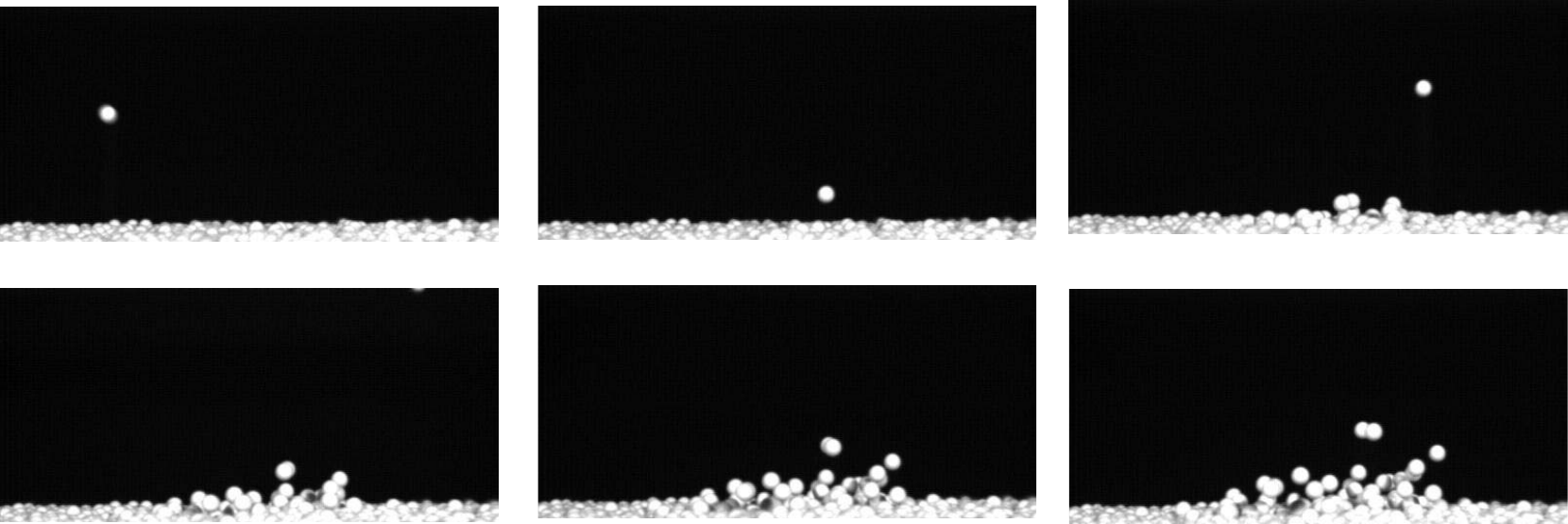}
 \end{center}
 \caption{From \citet{Beladjineetal07}, high-speed images of the impact of an incident particle on a static granular packing. The time step between two successive images is $4~$ms. Copyright 2007 American Physical Society.}
\label{Photo}
\end{figure}

\begin{figure}[!htb]
 \begin{center}
  \includegraphics[width=0.5\columnwidth]{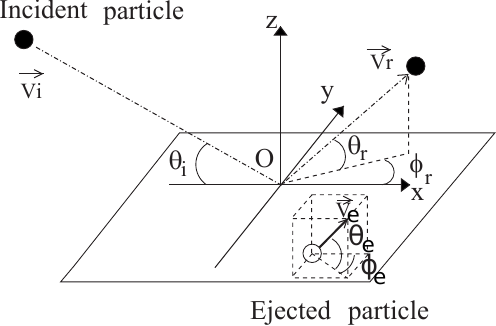}
 \end{center}
 \caption{Sketch of collision process.}
\label{CoordinateSystem}
\end{figure}

The collision process between an incident particle and a static granular packing has been investigated in many experimental~\citep{Mithaetal86,Werner90,Rioualetal00,Rioualetal03,Tanakaetal02,Nishidaetal04,Beladjineetal07,Ogeretal08,Ammietal09,Clarketal12,Clarketal15b,Clarketal16,Bacheletetal18,Chenetal19} and theoretical~\citep{WernerHaff88,AndersonHaff88,AndersonHaff91,HaffAnderson93,McElwaineetal04,Ogeretal05,Ogeretal08,Zhengetal05,Zhengetal08,Namikas06,Crassousetal07,Bourrieretal08,KokRenno09,ValanceCrassous09,Hoetal12,Duanetal13b,XingHe13,ComolaLehning17,Huangetal17,Tanabeetal17,Lammeletal17} studies in order to better understand aeolian saltation transport and other geophysical phenomena (e.g., rockfall~\citep{Bourrieretal08,Bacheletetal18}); see also \citep{WhiteSchulz77,WilletsRice86,WilletsRice89,McEwanetal92,Nalpanisetal93,Riceetal95,Riceetal96,Dongetal02,McElwaineetal04,GordonMcKennaNeuman09,GordonMcKennaNeuman11} for collision statistics during ongoing aeolian saltation transport. In typical experiments, a spherical incident particle of diameter $d$ and mass $m$ is shot (e.g., by an airgun) at a given speed $\mathbf{v_i}$ and angle $\theta_i$ onto a static packing of spheres of the same size. As shown in Figure~\ref{Photo} and sketched in Figure~\ref{CoordinateSystem}, as a result of its impact on the packing, the incident particle may rebound (velocity $\mathbf{v_r}$, angles $\theta_r$, $\phi_r$) and/or eject bed particles into motion (number $N_e$, velocity $\mathbf{v_e}$, angles $\theta_e$, $\phi_e$), where a particle is typically counted as ejected if its center is lifted by more than $d$ above the top of the bed surface.
The statistics of this process has been the subject of several recent experimental and numerical studies~\citep[e.g.,][]{Beladjineetal07,Ammietal09,Tanabeetal17} (note that experimental studies that used only one camera measured quantities projected into the incident plane: $v^{2D}_{r(e)}\equiv\sqrt{v_{r(e)x}^2+v_{r(e)z}^2}$ and $\tan\theta^{2D}_{r(e)}\equiv\tan\theta_{r(e)}/\cos\phi_{r(e)}$). These studies have yielded the following insights:

(i) The incident particle loses much more energy in head-on than in grazing collisions. In fact, the average restitution coefficient and its two-dimensional projection obey the following empirical relationships for $10^\circ\leq\theta_i\leq90^\circ$:
\begin{linenomath*}
\begin{subequations}
\begin{align}
 \overline{e}&\equiv\overline{|\mathbf{v_r}|}/|\mathbf{v_i}|=A-B\sin\theta_i, \\
 \overline{e^{2D}}&\equiv\overline{v^{2D}_r}/|\mathbf{v_i}|=A^{2D}-B^{2D}\sin\theta_i, \label{e2Dempirical}
\end{align}
\end{subequations}
\end{linenomath*}
where the overbar denotes an ensemble average over collision experiments, and the $A$ and $B$ coefficients are empirical constants that vary slightly between the studies (e.g., $A\approx A^{2D}\approx0.87$, $B\approx0.62$~\citep{Ammietal09}, and $B^{2D}\approx0.72$~\citep{Beladjineetal07}).

(ii) The average vertical restitution coefficient exceeds unity at small impact angles and obeys the following empirical relationship for $10^\circ\leq\theta_i\leq90^\circ$:
\begin{linenomath*}
\begin{equation}
 \overline{e_z}\equiv\overline{v_{rz}}/v_{iz}=A_z/\sin\theta_i-B_z, \label{ezempirical}
\end{equation}
\end{linenomath*}
where $A_z\approx0.3$ and $B_z\approx0.15$ for the experiments by \citet{Beladjineetal07}. \citet{Pahtzetal20b} suggested the following modification of equation~(\ref{ezempirical}):
\begin{linenomath*}
\begin{equation}
 \overline{e_z}=A^{2D}/\sqrt{\sin\theta_i}-B^{2D}. \label{ezempirical2}
\end{equation}
\end{linenomath*}
This modification, which is also consistent with the experimental data, ensures the correct asymptotic behavior of the average rebound angle, $\overline{\theta_r}\sim\sqrt{\theta_i}$~\citep{Lammeletal17}, in the limit $\theta_i\rightarrow0$.

(iii) The average rebound angle and its two-dimensional projection are independent of the incident speed, increase with the impact angle, and obey the following empirical relationships for $10^\circ\leq\theta_i\leq90^\circ$:
\begin{linenomath*}
\begin{subequations}
\begin{align}
 \overline{\theta_r}&=\theta_0+\chi\theta_i, \\
 \sin\overline{\theta^{2D}_r}&=\overline{e_z}\sin\theta_i/\overline{e^{2D}},
\end{align}
\end{subequations}
\end{linenomath*}
where $\theta_0\approx20^\circ$ and $\chi\approx0.19$ for the experiments by \citet{Ammietal09}.

(iv) The average energy that the incident particle transfers to the bed is spent for the ejection of bed particles. That is, it is proportional to the average of the sum of the kinetic energy of ejected particles ($E_e=\frac{1}{2}m\mathbf{v_e}^2$ and $E^{2D}_e=\frac{1}{2}mv^{2D2}_e$). In fact, the following empirical relationships are obeyed for $10^\circ\leq\theta_i\leq90^\circ$:
\begin{linenomath*}
\begin{subequations}
\begin{align}
 \overline{N_e}\,\overline{E_e}&=r(1-\overline{e}^2)E_i,  \\
 \overline{N_e}\overline{E^{2D}_e}&=r^{2D}(1-\overline{e^{2D}}^2)E_i,
\end{align}
\end{subequations}
\end{linenomath*}
where $r\approx0.04$ and $r^{2D}\approx0.038$ for the experiments by \citet{Ammietal09}. Note that $r$ and $r^{2D}$ decrease with the coordination number of the particle packing~\citep{Rioualetal03}.

(v) The average number of ejected particles is a linear function of the incident speed for $10^\circ\leq\theta_i\leq90^\circ$:
\begin{linenomath*}
\begin{equation}
 \overline{N_e}=n_0(1-\overline{e}^2)[|\mathbf{v_i}|/(\zeta\sqrt{gd})-1]\simeq n_0(1-\overline{e^{2D}}^2)[|\mathbf{v_i}|/(\zeta\sqrt{gd})-1], \label{Neemp}
\end{equation}
\end{linenomath*}
where $n_0\approx13$ and $\zeta\approx40$ for the experiments by \citet{Ammietal09}. Note that $n_0$ decreases with the coordination number of the particle packing~\citep{Rioualetal03}.

(vi)  The average horizontal and lateral velocities of ejected particles are nearly independent of the incident velocity, but the average vertical velocity increases slightly with the incident velocity and is independent of the impact angle for $10^\circ\leq\theta_i\leq90^\circ$~\citep{Ammietal09}:
\begin{linenomath*}
\begin{subequations}
\begin{align}
 \overline{v_{rx}^2}\approx\overline{v_{ry}^2}&\approx4gd,  \\
 \overline{v_{rz}}/\sqrt{gd}&\approx1.06(|\mathbf{v_i}|/\sqrt{gd})^{1/4}, \\
 \overline{v_{rz}^2}/gd&\approx1.46(|\mathbf{v_i}|/\sqrt{gd})^{1/2}.
\end{align}
\end{subequations}
\end{linenomath*}

(vii) The average ejection angle $\overline{\theta_e}$ is constant for $10^\circ\leq\theta_i\leq90^\circ$~\citep{Ammietal09}. However, its projection into the incident plane increases with the impact angle~\citep{Beladjineetal07}:
\begin{linenomath*}
\begin{equation}
 \overline{\theta^{2D}_e}\approx\frac{\pi}{2}+0.1\left(\theta_i-\frac{\pi}{2}\right).
\end{equation}
\end{linenomath*}

\subsubsection*{Open Problem: Behavior of the Rebound Probability}
\citet{Mithaetal86} measured that about $94\%$ of all impacting particles are not captured by the bed (i.e., they successfully rebound). However, the range of impact velocities in their experiments was very narrow ($|\mathbf{v_i}|\in(106,125)\sqrt{gd}$). More systematic measurements of the rebound probability $P_r$ are needed.

\begin{figure}[!htb]
 \begin{center}
  \includegraphics[width=1.0\columnwidth]{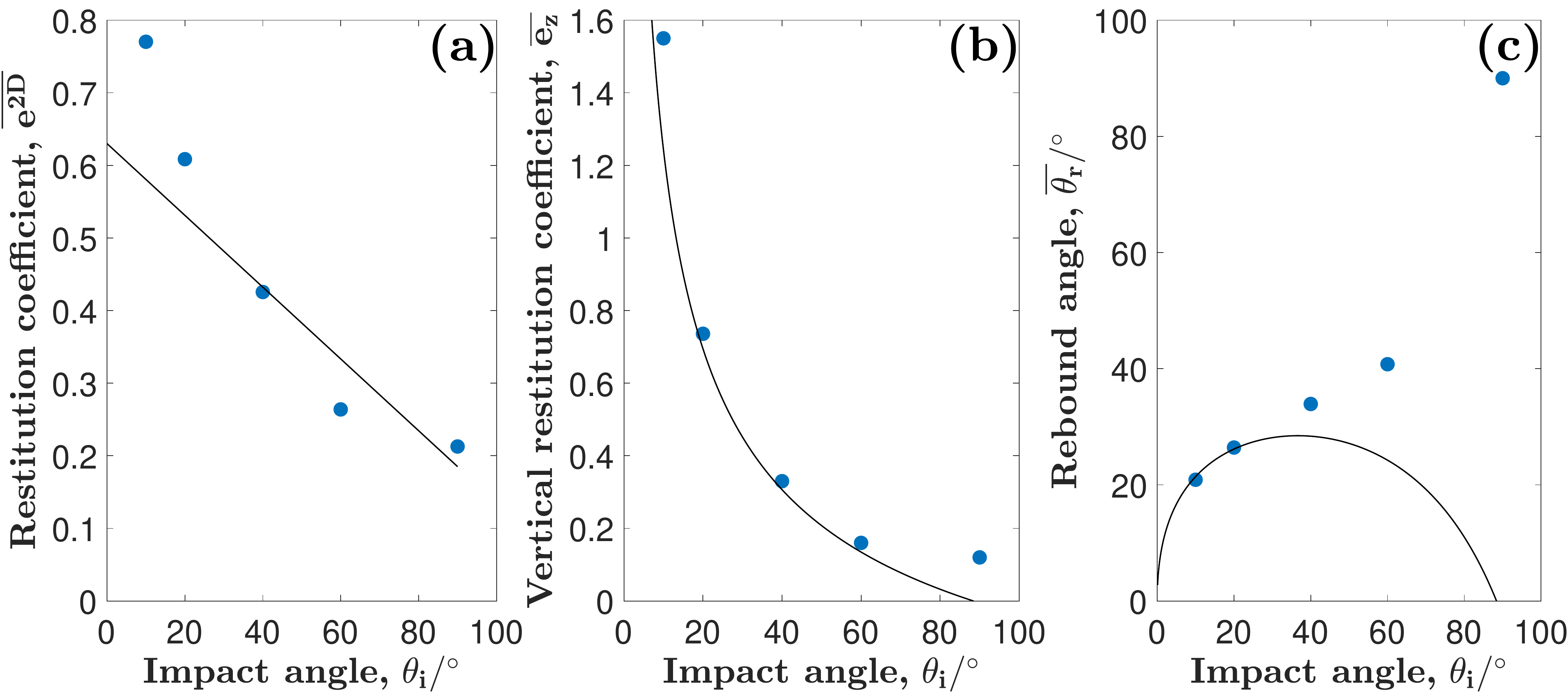}
 \end{center}
 \caption{Test of the analytical expressions by \citet{Lammeletal17} that describe the particle rebound of an impacting particle in the limit of shallow impact angles ($\theta_i\lesssim20^\circ$). (a) Average rebound restitution coefficient $\overline{e^{2D}}$, (b) average vertical rebound restitution coefficient $\overline{e_z}$, and (c) average rebound angle $\overline{\theta_r}$ versus impact angle $\theta_i$. Symbols correspond to experimental data by \citet{Beladjineetal07}. Solid lines correspond to equations~(\ref{e2D})-(\ref{thetar2D}).}
\label{LammelRebound}
\end{figure}

Studies have attempted to physically describe both the rebound~\citep{Zhengetal05,Zhengetal08,Namikas06,Lammeletal17} and ejection dynamics~\citep{McElwaineetal04,Crassousetal07,KokRenno09,ValanceCrassous09,Hoetal12,ComolaLehning17,Lammeletal17}. For example, the rebound dynamics can be analytically calculated for an idealized packing geometry and a given rebound location assuming a binary collision between the incident particle and hit bed particle. From averaging over all possible rebound locations, one can then determine the rebound angle and restitution coefficient distributions. Using this procedure, \citet{Lammeletal17} derived the following expressions for $\overline{e^{2D}}$, $\overline{e_z}$, $\overline{\theta^{2D}_r}$, and $P_r$ in the limit of shallow impact angles ($\theta_i\lesssim20^\circ$):
\begin{linenomath*}
\begin{align}
 \overline{e^{2D}}&=\beta_r-(\beta_r^2-\alpha_r^2)\theta_i/(2\beta_r), \label{e2D} \\
 \overline{e_z}&=-\beta_r+(2/3)(\alpha_r+\beta_r)\sqrt{2/\theta_i}, \label{ez} \\
 \overline{\theta^{2D}_r}&=\overline{e_z}\theta_i/\overline{e^{2D}}\approx(2/3)(1+\alpha_r/\beta_r)\sqrt{2\theta_i}-\theta_i, \label{thetar2D} \\
 P_r&=1-\frac{1+\ln\xi}{\xi},\quad\text{with}\quad\xi\equiv\max\left[1,\frac{9\sqrt{2}(1+\alpha_r/\beta_r)^2\theta_i\mathbf{v_i}^2}{4\sqrt{3}gd}\right], \label{Pr}
\end{align}
\end{linenomath*}
where $\alpha_r$ and $\beta_r$ are the normal and tangential rebound restitution coefficients, respectively, in the impact plane, which depend on the binary normal and tangential restitution coefficient (i.e., the ratio between the postcollisional and precollisional relative particle velocity component normal and tangential, respectively, to the contact plane). Figure~\ref{LammelRebound} compares equations~(\ref{e2D})--(\ref{thetar2D}) with the experimental data by \citet{Beladjineetal07} using the values $\alpha_r=0.2$ and $\beta_r=0.63$, which \citet{Lammeletal17} obtained from fitting the numerical solution of the full problem (i.e., not limited to $\theta_i\lesssim20^\circ$) to the experimental data.
The agreement with the data with $\theta_i\lesssim20^\circ$ is acceptable considering that the theory has been derived mostly from first physical principles. Equation~(\ref{Pr}), which is the modified version of equations~(41) and (42) of \citet{Lammeletal17} that these authors describe in their text, cannot be tested because of the lack of systematic measurements of the rebound probability $P_r$. A widely used alternative expression for $P_r$ was given by \citet{AndersonHaff91}: $P_r\approx0.95[1-\exp(-\gamma_r|\mathbf{v_i}|)]$. However, this expression is empirical and contains the dimensional parameter $\gamma_r$ (note that \citet{Andreotti04} assumed $\gamma_r\propto1/\sqrt{gd}$). Because $\overline{e_z}\gtrsim1$, which is a precondition for sustained aeolian saltation transport (from energy conservation), requires shallow impact angles, equations~(\ref{e2D})--(\ref{Pr}) can be used for the theoretical modeling of aeolian saltation transport.

\begin{figure}[!htb]
 \begin{center}
  \includegraphics[width=1.0\columnwidth]{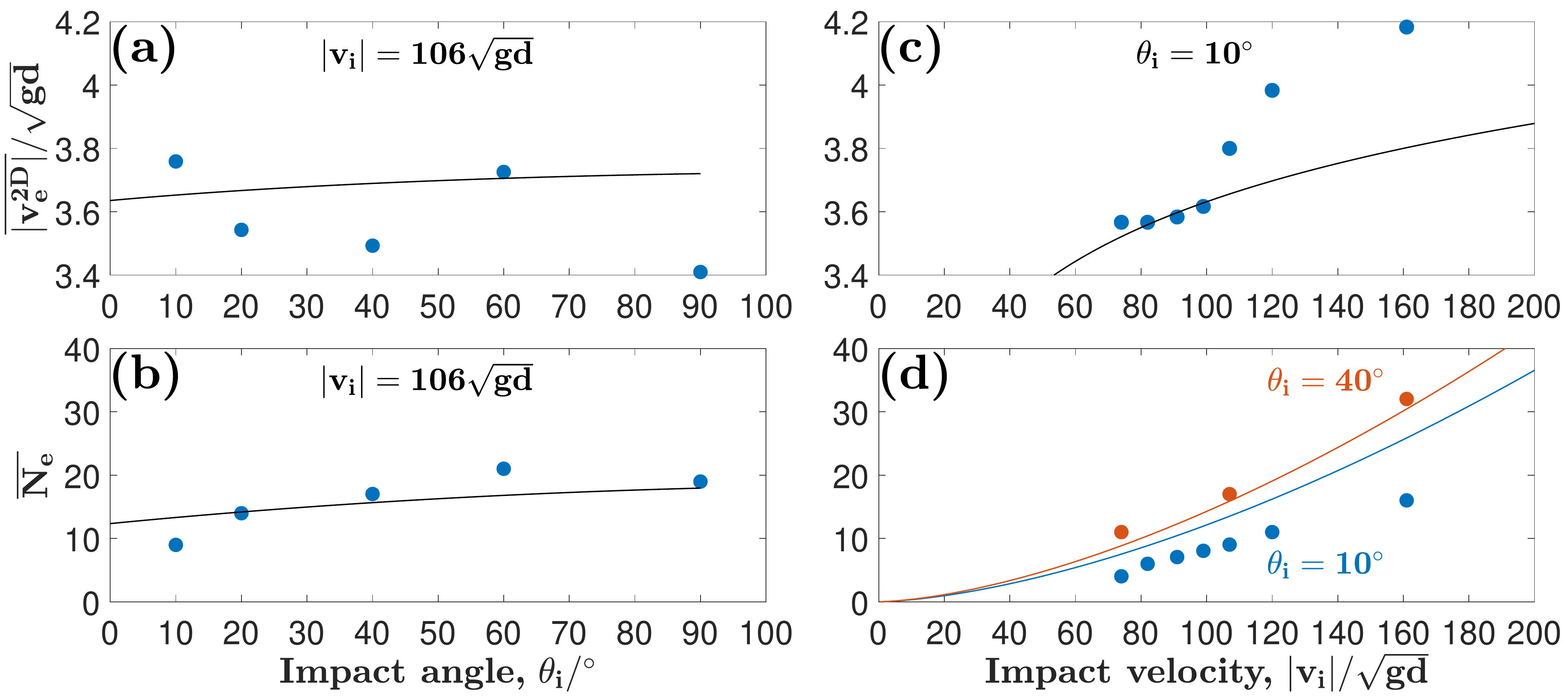}
 \end{center}
 \caption{Test of the analytical expressions by \citet{Lammeletal17} that describe the ejection of bed surface particles caused by the splash of an impacting particle. (a, c) Nondimensionalized average ejection velocity $\overline{|\mathbf{v^{2D}_e}|}$ and (b, d) average number of ejected particles $\overline{N_e}$ versus (a, b) impact angle $\theta_i$ and (c, d) nondimensionalized impact velocity $|v_i|/\sqrt{gd}$. Symbols correspond to experimental data by \citet{Beladjineetal07}. Solid lines correspond to equations~(\ref{Ne}) and (\ref{ve}) combined with the approximation $\overline{|\mathbf{v_e}|}\simeq\overline{|\mathbf{v^{2D}_e}|}$.}
\label{LammelSplash}
\end{figure}

For the description of the ejection dynamics, there have been two distinct approaches: solving an underdetermined momentum and/or energy balance of the particles involved in the collision process~\citep{KokRenno09,ComolaLehning17} and treating the collision process as a sequence of binary collisions, in which the energy is split between the collisional partners (i.e., incident and bed particle or two bed particles)~\citep{McElwaineetal04,Crassousetal07,ValanceCrassous09,Hoetal12,Lammeletal17}. A minimal numerical model that is based on the latter approach has been able to reproduce experimental data of both the rebound and ejection dynamics, including the measured log-normal distribution of the vertical ejection velocity~\citep{Crassousetal07}. Furthermore, based on this approach and the derivation by \citet{Hoetal12}, \citet{Lammeletal17} derived the following analytical expression for the distribution of the ejection energy $E_e$:
\begin{linenomath*}
\begin{align}
 P(E_e)&=\frac{1}{\sqrt{2\pi}\sigma E_e}\exp\left[-\frac{(\ln E_e-\mu)^2}{2\sigma^2}\right],\quad\text{with} \\
 \sigma&=\sqrt{\lambda}\ln 2, \nonumber \\
 \mu&=\ln[(1-\overline{e}^2)E_i]-\lambda\ln 2, \nonumber \\
 \lambda&=2\ln[(1-\overline{e}^2)E_i/(mgd)], \nonumber
\end{align}
\end{linenomath*}
from which they further obtained expressions for $\overline{N_e}$, $\overline{E_e}$, and $\overline{|\mathbf{v_e}|}$:
\begin{linenomath*}
\begin{align}
 \overline{N_e}&=r\frac{(1-\overline{e}^2)E_i}{2\overline{E_e}}\mathrm{erfc}\left[\frac{\ln(mgd)-\mu}{\sqrt{2}\sigma}\right], \label{Ne} \\
 \overline{E_e}&=mgd[(1-\overline{e}^2)E_i/(mgd)]^{1-(2-\ln2)\ln2}, \\
 \overline{|\mathbf{v_e}|}&=\frac{\mathrm{erfc}\{[\ln(mgd)-\mu-\sigma^2/2]/(\sqrt{2}\sigma)\}}{\mathrm{erfc}\{[\ln(mgd)-\mu]/(\sqrt{2}\sigma)\}}\sqrt{2}\exp(\mu/2+\sigma^2/8), \label{ve}
\end{align}
\end{linenomath*}
where $r=0.06$. Figure~\ref{LammelSplash} shows that these expressions are roughly consistent with experimental data considering that they have been derived mostly from first physical principles.
Note that equations~(\ref{e2D})--(\ref{ve}), after some minor modifications, can also be applied to situations in which the size of the impacting particle differs from the size of the particles of the granular packing~\citep{Lammeletal17}. Further note that equation~(38) of \citet{Lammeletal17}, which is the equivalent of equation~(\ref{ve}), contains a typo (a $\sigma$ is missing in the denominator).

\subsubsection*{Open Problem: Impacts Onto Mobile Beds}
The findings from collision experiments with static beds are often applied to model fluvial bedload~\citep{Berzietal16,Pahtzetal20b} and aeolian saltation transport~\citep{Andreotti04,ClaudinAndreotti06,Creysselsetal09,KokRenno09,Kok10b,Jenkinsetal10,Lammeletal12,Pahtzetal12,Huangetal14,JenkinsValance14,JenkinsValance18,WangZheng14,WangZheng15,Berzietal16,Berzietal17,Boetal17,LammelKroy17,Pahtzetal20b}. However, if the time between successive particle-bed impacts is too short for a bed particle to fully recover from each impact, it can accumulate more and more kinetic energy with each impact. Hence, for a sufficiently large impact frequency and impact energy (both increase with the sediment transport rate $Q$), the bed can no longer be treated as static and the findings from such collision experiments may no longer apply. For example, the simultaneous impact of two particles onto the bed leads to a significantly different outcome compared with the situation in which each particle impacts separately~\citep{Duanetal13b}. For these reasons, future studies should try to systematically investigate the effects of disturbances from the static bed on the outcome of a particle-bed impact.

\subsubsection*{Open Problem: Effects of Particle Shape and Size Distribution}
\citet{Chenetal19} investigated the particle-bed collision process for natural sand particles, which exhibit nonspherical shapes and nonuniform particle size distributions. They found significant quantitative and qualitative deviations from the laws describing spherical, uniform particles. More systematic experimental studies are needed to pinpoint the exact manner in which particle shape and size distribution affect the collision process.

\subsubsection*{Controversy: Effects of Viscous Damping}
Binary collisions that occur within an ambient fluid can be significantly damped depending on the Stokes number $\mathrm{St}\equiv s|\mathbf{v_r}|d/(9\nu_f)$~\citep{Gondretetal02,YangHunt06,Schmeeckle14,Maurinetal15}, where $\mathbf{v_r}$ is the relative particle velocity just before a collision. For example, experiments suggest that the effective normal restitution coefficient $\epsilon$ of a damped binary collision vanishes for $\mathrm{St}\lesssim10$~\citep{Gondretetal02}. The question that then arises is how does viscous damping affect the rebound and ejection dynamics of a particle-bed impact. \citet{Berzietal16,Berzietal17} assumed that the rebound restitution coefficients $\overline{e^{2D}}$ and $\overline{e_z}$, like $\epsilon$, also vanish when $\mathrm{St}$ falls below a critical value. In contrast, DEM-based simulations indicate that the dynamics of saltation~\citep{PahtzDuran18a} and particularly bedload transport~\citep{DrakeCalantoni01,Maurinetal15,ElghannayTafti17b,PahtzDuran17,PahtzDuran18a,PahtzDuran18b} are not much affected by the value of $\epsilon$, which suggests that the rebound and ejection dynamics of a particle-bed impact may not be much affected by viscous damping. A possible explanation for this unexpected behavior could be that a nearly elastic particle-bed impact may be roughly equivalent to a sequence of binary collisions between particles in contact at the instant of impact. In fact, a theoretical model based on this hypothesis reproduced experiments of the collision process~\citep{Crassousetal07,ValanceCrassous09}. For the perfectly elastic case ($\epsilon=1$), the impactor would then transfer all of its momentum in the direction normal to the contact plane to the particle it hit (which is the expected result of an elastic binary collision between a mobile and a resting particle) and, therefore, rebound with zero normal momentum. A complete loss of normal momentum is also expected for the completely inelastic case ($\epsilon=0$). This suggests that the rebound process is not much affected by $\epsilon$, which would imply that the momentum in the direction tangential to the contact plane is what mainly matters. Collision experiments in an ambient viscous liquid could resolve this controversy.

\subsubsection*{Open Problem: Effects of Cohesion}
Cohesive interparticle forces, including van der Waals~\citep{Castellanos05}, water adsorption~\citep{Herminghaus05}, and electrostatic forces~\citep{LacksSankaran11}, become significant in the collision process for sufficiently small particles (on Earth, for $d\lesssim100~\mu$m) because they scale with a lower power $p$ in the particle diameter ($F_{\mathrm{coh}}\sim d^p$) than the gravity force ($F_g\sim d^3$). However, collision experiments with so small particles have not been carried out because it is very difficult to detected their dynamics with cameras. Numerical studies are also very scarce. To the authors' knowledge, only the very recent study by \citet{Comolaetal19a} studied cohesive forces, by implementing them in a numerical DEM-based model of aeolian saltation transport. These authors investigated the impact of a particle onto the bed for a large range of the strength of cohesive forces and found that cohesion decreases $\overline{N_e}$ via solidifying the bed, while $\overline{e}$ slightly and $|\mathbf{v_e}|/|\mathbf{v_i}|$ considerably increase. However, more systematic studies are needed to confirm these results and determine scaling laws describing the effects of cohesion on the outcome of a particle-bed impact.

\subsubsection{Collision-Enhanced Turbulent Entrainment in Fluvial Bedload Transport} \label{ImpactEntrainmentBedload}
\begin{figure}[!htb]
 \begin{center}
  \includegraphics[width=1.0\columnwidth]{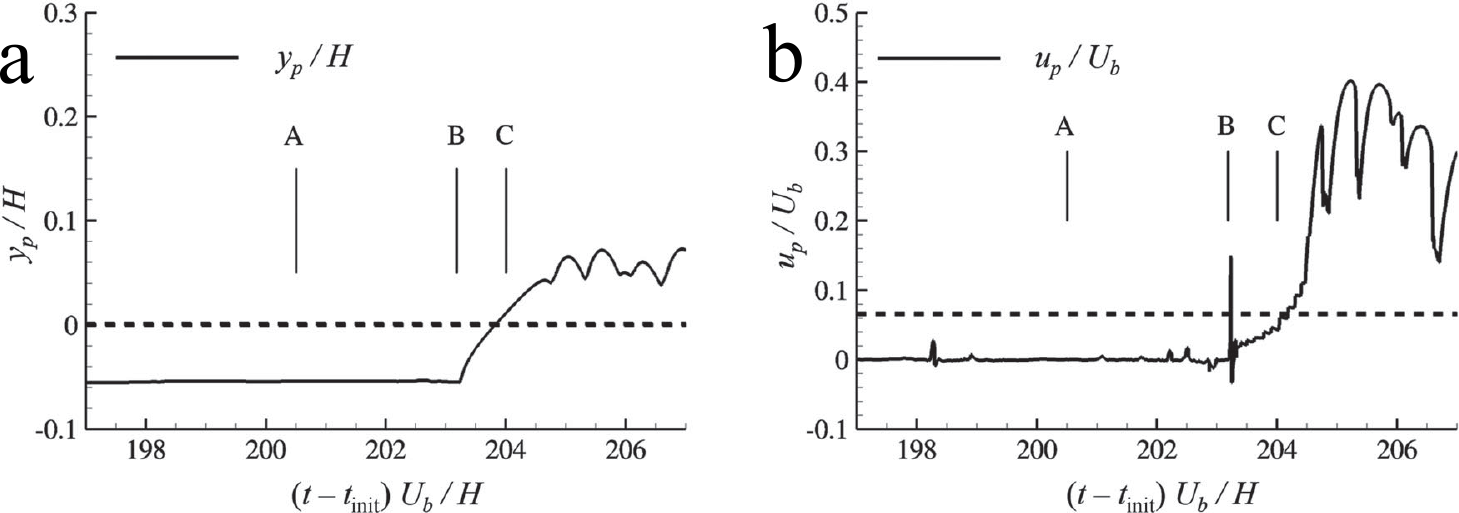}
 \end{center}
 \caption{From \citet{Vowinckeletal16}, time evolution of a typical erosion event. At time instant $A$, a bed surface particle is at rest. At time instant $B$, it is hit by an impacting transported particle. The impact causes a slight dislocation off its initial position. Once slightly lifted, the particle protrudes into the flow, enhancing the flow forces acting on it. This enhancement in combination with much-larger-than-average flow velocities during a turbulent fluctuation event (Figure~\ref{ErosionEvent2}) leads to entrainment (time instant $C$), as indicated by the nondimensionalized (a) vertical displacement ($y_p/H$) and (b) particle velocity ($u_p/U_b$) exceeding critical values (dashed lines). Copyright 2016 Taylor \& Francis Group.}
\label{ErosionEvent1}
\end{figure}

\begin{figure}[!htb]
 \begin{center}
  \includegraphics[width=1.0\columnwidth]{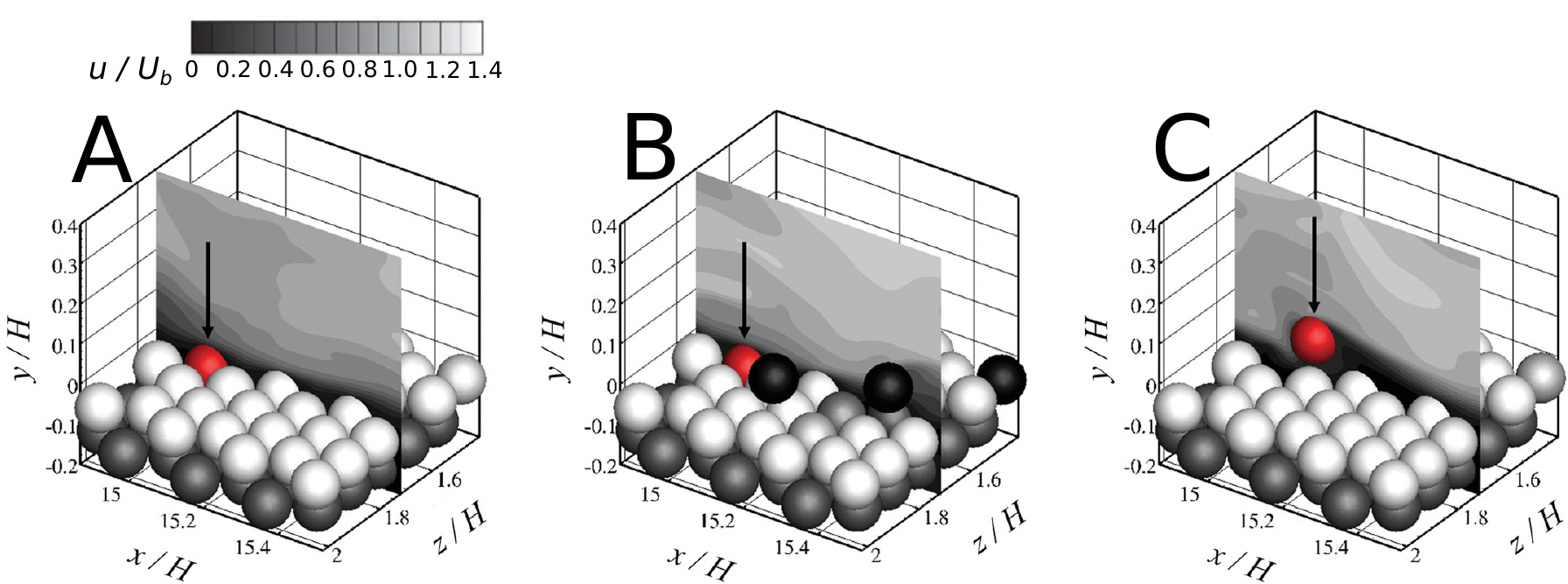}
 \end{center}
 \caption{From \citet{Vowinckeletal16}, zoom into the simulation domain and contour plots of the instantaneous streamwise flow field ($u/U_b$) during a typical erosion event of a target particle (red). The color (gray, white, and black) corresponds to (nonerodible, resting, and transported) particles. Time instants $A$, $B$, and $C$ are as in Figure~\ref{ErosionEvent1}. Copyright 2016 Taylor \& Francis Group.}
\label{ErosionEvent2}
\end{figure}

To the authors' knowledge, only a single study has resolved the effects of particle-bed impacts on entrainment by turbulent fluctuation events in bedload transport~\citep{Vowinckeletal16}. However, this study provided one of the largest, if not the largest, data sets of entrainment events associated with fluvial bedload transport with a very high resolution in space and time. \citet{Vowinckeletal16} coupled direct numerical simulations (DNS) for the fluid phase (i.e., the Navier-Stokes equations are directly solved without using turbulent closure assumptions) with DEM simulations for the particle phase (i.e., particles interact with each other according to a contact model) using the immersed boundary method, which fully resolves the geometry of particles (and thus the hydrodynamic forces acting on them) without remeshing the grid during their motion~\citep{Vowinckeletal14}. Because of the sophistication of this numerical method (i.e., resolving all relevant physical processes at very small scale), the produced data can be considered to be very reliable. The simulated setup consisted of two layers of grains resting on the simulation bottom wall, the lower of which was fixed, arranged in a hexagonal packing, and exposed to a unidirectional open channel flow of thickness $H=9d$ (Reynolds number $\mathrm{Re}\equiv U_bH/\nu_f=2941$, where $U_b$ is the bulk flow velocity). The Shields number was at $\Theta=0.0255$, which is about $25\%$ below the Shields curve for the simulated condition. That is, the nondimensionalized transport rate $Q_\ast$ was likely below the value associated with critical transport conditions (see section~\ref{Introduction}), which is consistent with \citet{Vowinckeletal16} reporting that only $3\%$ of all particles were in motion on average. For these conditions, it was found that, in the vast majority of cases (overall $96.5\%$), a particle-bed impact and a subsequent turbulent fluctuation event are responsible for entrainment, even when one or more of the six pockets surrounding the target particle were not occupied by other particles (in which case the target particle experiences a larger exposure to the flow). For an entrainment event following this pattern, Figure~\ref{ErosionEvent1} shows the time evolution of (a) the vertical displacement ($y_p$) and (b) velocity of a bed surface particle ($u_p$), while Figure~\ref{ErosionEvent2} shows the simulation domain and contour plots of the instantaneous flow field.
It can be seen that, at the instant of entrainment, the instantaneous streamwise flow velocity ($u$) exhibits larger-than-average values (Figure~\ref{ErosionEvent2}c). In fact, \citet{Vowinckeletal16} reported that $82\%$ of the entrainment events were caused by \textit{sweep}, characterized by positive fluctuations of $u$ and negative fluctuations of the flow velocity component in the direction normal to the bed.

The results by \citet{Vowinckeletal16} were obtained for an idealized hexagonal packing and may not necessarily apply in their full extent to realistic sediment beds found in nature. While for a hexagonal packing, the vast majority of entrainment events are initiated by particle-bed impacts, it remains unclear whether this holds true also for natural sediment beds, in which bed surface particles tend to protrude much more strongly into the flow. On the one hand, a larger protrusion makes it easier for a turbulent fluctuation event to entrain a bed surface particle without a preceding particle-bed impact. On the other hand, particle-bed impacts can result in entrainment without the need of a turbulent fluctuation event (see section~\ref{ImpactEntrainmentContinuousTransport}).

\subsubsection{The Role of Particle-Bed Impacts in Sustaining Continuous Sediment Transport} \label{ImpactEntrainmentContinuousTransport}
\citet{PahtzDuran17} numerically studied the role of particle-bed impacts in sustaining continuous nonsuspended sediment transport for transport conditions characterized by a large range of the Shields number $\Theta$, density ratio $s$, and Galileo number $\mathrm{Ga}$. These authors coupled quasi-two-dimensional DEM simulations for the particle phase with a Reynolds-averaged description of the fluid hydrodynamics that neglects turbulent fluctuations around the mean turbulent flow. While such simulations cannot resolve entrainment by turbulent fluctuation events, they are able to elucidate the importance of entrainment by particle-bed impacts relative to entrainment by the mean turbulent flow. Also, the absence of turbulent fluctuations eliminates transport intermittency in the sense that transport in the simulation domain is either continuous (i.e., periods of rest are absent) or it completely stops after a finite time (except for potential creeping, see section~\ref{Creeping}). From their simulations, \citet{PahtzDuran17} determined an effective value of the local particle velocity averaged over elevations near the bed surface ($V_b$) relative to the critical velocity that is needed to escape the potential wells set by the pockets of the bed surface ($\propto\sqrt{\hat gd}$, where $\hat g=[1+1/(s+C_m)]$ is the value of the gravity constant reduced by the buoyancy and added mass force, with $C_m=1/2$ the added mass coefficient). They found that $V_b/\sqrt{\hat gd}$ exhibits a universal approximately constant value of order unity for continuous nonsuspended sediment transport if the following constraint is obeyed:
\begin{linenomath*}
\begin{equation}
 \mathrm{Im}\equiv\mathrm{Ga}\sqrt{s+C_m}\gtrsim20\quad\text{or}\quad\Theta\gtrsim5/\mathrm{Im}. \label{Constraint}
\end{equation}
\end{linenomath*}
The interpretation of $V_b/\sqrt{\hat gd}\approx\mathrm{const}$ is that particles located near the bed surface (which includes both particles of the bed and transported particles) are on average at the verge of leaving it or being captured by its potential wells, consistent with a dynamic equilibrium that is solely controlled by particle inertia. This implies that entrainment occurs solely due to the action of particle-bed impacts. Consistently, \citet{PahtzDuran17} observed from visually inspecting simulations that obey equation~(\ref{Constraint}) that every entrainment event is initiated by a particle-bed impact, usually with a small time delay between the instant of impact and beginning visible motion. In contrast, for transport conditions that do not obey equation~(\ref{Constraint}), $V_b/\sqrt{\hat gd}$ exhibits a smaller value, which means that the mean turbulent flow must assist particles located near the bed surface in escaping the potential wells. For bedload transport, the findings by \citet{PahtzDuran17} were independent of the effective normal restitution coefficient $\epsilon$ for a damped binary collision, which indicates that viscous damping does not suppress impact entrainment (see also the discussion of viscous damping in section~\ref{StaticBedCollision}).

The constraint set by equation~(\ref{Constraint}) is obeyed by the vast majority of sediment transport regimes, including turbulent fluvial bedload transport. That is, for the absence of turbulent fluctuation events, only viscous fluvial bedload transport is significantly affected by the entrainment of bed sediment by the mean turbulent flow. The numerical prediction that impact entrainment dominates entrainment by the mean turbulent flow in turbulent fluvial bedload transport is supported by experiments~\citep{Heymanetal16,LeeJerolmack18}. \citet{LeeJerolmack18} studied bedload transport driven by a water flow in a quasi-two-dimensional flume (i.e., its lateral dimension was only slightly larger than the particle diameter $d$). Because the size of turbulent structures, and thus turbulent fluctuation events, is strongly suppressed when the system dimensions are so strongly narrowed down, their experiments are somewhat comparable to the numerical simulations by \citet{PahtzDuran17} described above. \citet{LeeJerolmack18} fixed the water discharge and fed particles at the flume entrance with varying frequency $f_{\mathrm{in}}$ (the tested range of $f_{\mathrm{in}}$ was likely associated with a transport rate below capacity). In contrast to similar older experiments~\citep{Bohmetal04,Anceyetal08,Heymanetal13}, the bed was relatively deep, which ensured the complete dissipation of shock waves associated with particle-bed impacts~\citep{Rioualetal03}. \citet{LeeJerolmack18} reported that, for all tested conditions, every entrainment event is initiated by a particle-bed impact, exactly as numerically predicted, and that the number of transported particles roughly scales with the energy transferred to the bed by rebounding particles. The latter finding is remarkably similar to the scaling of the average ejected particle number $\overline{N_e}$ in static bed experiments (e.g., see equation~(\ref{Ne})). \citet{LeeJerolmack18} also measured the frequency of particles passing an illuminated window near the flume exit ($f_{\mathrm{out}}$). They found that $f_{\mathrm{out}}<f_{\mathrm{in}}$ for sufficiently small $f_{\mathrm{in}}$ and that $f_{\mathrm{out}}\approx f_{\mathrm{in}}$ once $f_{\mathrm{in}}$ exceeds a critical value.

Similar observations were made by \citet{Heymanetal16}, who used a water flume with a narrow but larger width ($W=5d$) than \citet{LeeJerolmack18} and who also used a relatively deep bed. \citet{Heymanetal16} measured that the entrainment rate was proportional to the number of transported particles per unit bed area, which is indirect evidence supporting that the majority of entrainment events is caused by particle-bed impacts. These authors also reported for all their tested feeding frequencies $f_{\mathrm{in}}$ that the entrainment and deposition rate are equal to one another, in resemblance of the measurement $f_{\mathrm{out}}\approx f_{\mathrm{in}}$ for sufficiently large $f_{\mathrm{in}}$ by \citet{LeeJerolmack18}. Note that one expects the approximate equality $f_{\mathrm{out}}\approx f_{\mathrm{in}}$ to break down for large $f_{\mathrm{in}}$ (when the influx exceeds transport capacity) because increasing momentum transfer from fluid to particles slows down the flow, which at some point can no longer sustain the particle motion.

The results by \citet{Heymanetal16} and \citet{LeeJerolmack18} suggest that mainly (but not solely) particles that were previously in motion are being entrained by particle-bed impacts. Otherwise, there would be no reason to expect that the entrainment and deposition rate are relatively equal to one another for a large range of $f_{\mathrm{in}}$ (instead, one would expect that only for transport capacity). This can be explained when assuming that particle-bed impacts are effective in mobilizing a bed particle almost only when the bed particle exceeds a critical energy level just before the impact. On the one hand, this assumption would explain why bed particles that have never been transported only rarely become mobilized by particle-bed impacts. On the other hand, this assumption is consistent with the fact that a transported particle that has just been captured by a bed pocket exhibits a residual kinetic energy that takes some time to be completely dissipated, during which it can be remobilized by an impact from a particle coming from behind. It seems that, once $f_{\mathrm{in}}$ exceeds a critical value, there is usually a particle coming from behind in time and transported particles can only rarely settle completely even though they may temporarily stop. Temporary particle stops and reentrainment make transported particles tend to move in clusters near the flume exit even though they are apart from one another at the flume entrance, which is exactly what \citet{LeeJerolmack18} reported and what can be observed in the numerical simulations by \citet[][Movie~S2]{PahtzDuran18a}.

There is evidence that the presumed impact entrainment mechanism described above may play an important role in nonsuspended sediment transport in general. In fact, in simulations of steady, homogenous sediment transport using DEM-based numerical models that \textit{neglect turbulent fluctuations around the mean turbulent flow}, the steady state transport rate $Q$ exhibits a discontinuous jump at a fluid shear stress $\tau^{\rm ImE}_t$~\citep{Carneiroetal11,Carneiroetal13,Clarketal15a,Clarketal17,PahtzDuran18a}. That is, for $\tau\geq\tau^{\rm ImE}_t$, transport is significantly larger than zero ($Q>0$) and continuous, whereas $Q\simeq0$ when $\tau<\tau^{\rm ImE}_t$. Assuming that only impact entrainment took place in all these simulations (as the mean turbulent flow is too weak for entrainment, see above), $\tau^{\rm ImE}_t$ can be identified as the impact entrainment threshold. The discontinuous jump of $Q$ thus means that, in order for impact entrainment to sustain transport, a critical transport rate must be exceeded. Like the critical feeding frequency in the experiments by \citet{LeeJerolmack18}, this critical transport rate may be interpreted as the value above which most transported particles can be captured only temporarily by bed pockets as they are usually hit in time and thus reentrained by an impact from a particle coming from behind before dissipating too much of their kinetic energy. However, it is crucial to point out that impact entrainment of bed particles that have never been transported occasionally occurs in DEM-based sediment transport simulations as well, which is why a further interpretation of the physical origin of the discontinuous jump of $Q$ has been proposed~\citep{PahtzDuran18a}. It states that, at a critical transport rate, bed surface particles do no longer sufficiently recover between successive particle-bed impacts. They thus accumulate energy between successive impacts until they are eventually entrained. In contrast, for subcritical transport rates, particles sufficiently recover between impacts so that impact entrainment is inefficient, causing transport to eventually stop. The two interpretations above are based only on the energy of bed particles or temporarily captured transported particles. In contrast, in the context of an idealized continuous rebound modeling framework (see section~\ref{ContinuousRebounds}), an alternative mechanism based on the critical amount of energy $E_c$ that bed particles need to acquire for entrainment (more precisely, for entering a quasi-continuous motion) can explain the discontinuous jump of $Q$ without further assumptions (see section~\ref{ThoughtExperiment}).

\subsubsection*{Open Problem: Precise Mechanism of Impact Entrainment in Continuous Transport}
The proposed impact entrainment mechanisms described above and in section~\ref{ThoughtExperiment} are mostly speculative and based on indirect experimental or theoretical evidence, or idealized models. More direct investigations are therefore needed to uncover the precise nature of impact entrainment and the degree to which each of these mechanisms contributes. Such investigations may also help to better understand fluctuations of nonsuspended sediment transport. For example, the longer the average time $t_{\rm conv}$ it takes for transport to stop (in the absence of turbulent fluctuations around the mean turbulent flow) when $\tau<\tau^{\rm ImE}_t$ ($t_{\rm conv}$ obeys a critical scaling behavior at $\tau^{\rm ImE}_t$, see section~\ref{CriticalBehavior}), the larger are the transport autocorrelations, which can be quite substantial in fluvial bedload transport~\citep{HeathershawThorne85,Drakeetal88,Dinehart99,Anceyetal06,Anceyetal08,Anceyetal15,Martinetal12}.

\subsubsection*{Open Problem: Precise Definitions of Intermittent and Continuous Transport}
As explained above, in simulations of steady, homogenous sediment transport using DEM-based numerical models that neglect turbulent fluctuations, the steady transport rate $Q$ (in a time-averaged sense) exhibits a discontinuous jump at the impact entrainment threshold $\tau^{\rm ImE}_t$. In contrast, for most natural conditions, fluid entrainment by turbulent events can reinitiate transport whenever it temporarily stops, meaning that $Q$ remains significant below $\tau^{\rm ImE}_t$~\citep{Carneiroetal11}. Hence, since turbulent events capable of fluid entrainment occur only at an intermittent basis (see section~\ref{FluidEntrainment}), \citet{PahtzDuran18a} suggested that $\tau^{\rm ImE}_t$ is equivalent to the continuous transport threshold for most natural conditions and that transport becomes intermittent below $\tau^{\rm ImE}_t$. However, provided that fluid entrainment does occur, it is certain to find particles being in transport below $\tau^{\rm ImE}_t$ at any given instant in time in the large-system limit, which renders the distinction between intermittent and continuous transport somewhat ambiguous. For this reason, \citet{PahtzDuran18a} referred to intermittent conditions as those that deviate significantly from transport capacity (defined as in section~\ref{TransportCapacity}). Consistently, \citet{MartinKok18} and \citet{Comolaetal19b} found from aeolian field experiments that the long-term-averaged transport remains at capacity when the fraction $f_Q$ of active saltation transport is close to unity, that is, when transport quantified over a short but somewhat arbitrary time interval ($2~$s~\citep{MartinKok18} or $0.04~$s~\citep{Comolaetal19b}) almost never stops. Interestingly, \citet{Comolaetal19b} showed that the value of $f_Q$ can be indirectly estimated from the lowpass-filtered wind speed associated with large and very large scale turbulent structures (cutoff frequency $\Omega\approx0.04~$Hz). Alternatively, for their coupled DNS/DEM simulations of fluvial bedload transport, \citet{Gonzalezetal17} fitted continuous functions to the distributions of the discrete transported particle number (defined as the number of particles faster than a somewhat arbitrary velocity threshold) at different $\tau$ and identified the onset of continuous transport as the value of $\tau$ at which these fitting functions predict a zero probability for a vanishing particle number. Future studies should investigate the compatibility of these and other definitions of continuous transport.

\subsubsection*{Controversy: Threshold of Continuous Aeolian Saltation Transport}
In the opinion of the authors, the evidence reviewed above for the hypothesis that continuous transport occurs once impact entrainment alone is sufficient in compensating random captures of transported particles is quite strong. (In other words, significant fluid entrainment may occur in continuous transport---and does so quite likely in aeolian saltation transport given that the turbulent intensity within the saltation transport layer increases with the sediment transport rate~\citep{LiMcKennaNeuman12}---but it is not needed to sustain continuous transport.) However, it is worth pointing out that most aeolian researchers prefer a different narrative for aeolian saltation transport. For example, \citet{MartinKok18} assumed that continuous aeolian saltation transport in the field occurs once the saltation transport initiation threshold ($\approx\tau^{\rm In}_t$) is exceeded, whereas the impact entrainment threshold describes the cessation of intermittent saltation transport. This assumption is based on the idea that fluid entrainment continuously provides the transport layer with bed particles. However, this idea is problematic because turbulent events capable of fluid entrainment occur only at an intermittent basis (see section~\ref{FluidEntrainment}). The interested reader is also referred to the commentary by \citet{Pahtz18}, in which this controversy is extensively discussed.

\subsection{Continuous Particle Rebounds and Transport Capacity} \label{ContinuousRebounds}
In order for the mean turbulent flow to sustain the motion of particles that are already in transport, it needs to compensate, on average, the energy dissipated in particle-bed rebounds via drag acceleration during the particle trajectories. This mechanism, which is illustrated in detail by means of a thought experiment in section~\ref{ThoughtExperiment}, gives rise to a shear stress threshold of sediment transport (henceforth termed \textit{rebound threshold}), as was already noted by \citet[][p.~94]{Bagnold41} for aeolian saltation transport: ``Physically [the rebound threshold] marks the critical stage at which the energy supplied to the saltating grains by the wind begins to balance the energy losses due to friction when the grains strike the ground [and rebound].'' It also suggests a clear-cut definition of transport capacity, which is otherwise difficult to define~\citep[see review by][and references therein]{Wainwrightetal15}, that leads to an experimentally and numerically validated universal scaling of the transport load $M$ (i.e., the mass of transported sediment per unit bed area) with the fluid shear stress $\tau$ (section~\ref{TransportCapacity}). From the appearance of the rebound threshold in this scaling of $M$, one can conclude that at a significant if not predominant portion of the threshold measurements by \citet{Shields36} and others have been misidentified as measurements of the entrainment threshold (section~\ref{ShieldsDiagram}).

\subsubsection{Particle Rebounds Along a Flat Wall} \label{ThoughtExperiment}
\begin{figure}[!htb]
 \begin{center}
  \includegraphics[width=0.5\columnwidth]{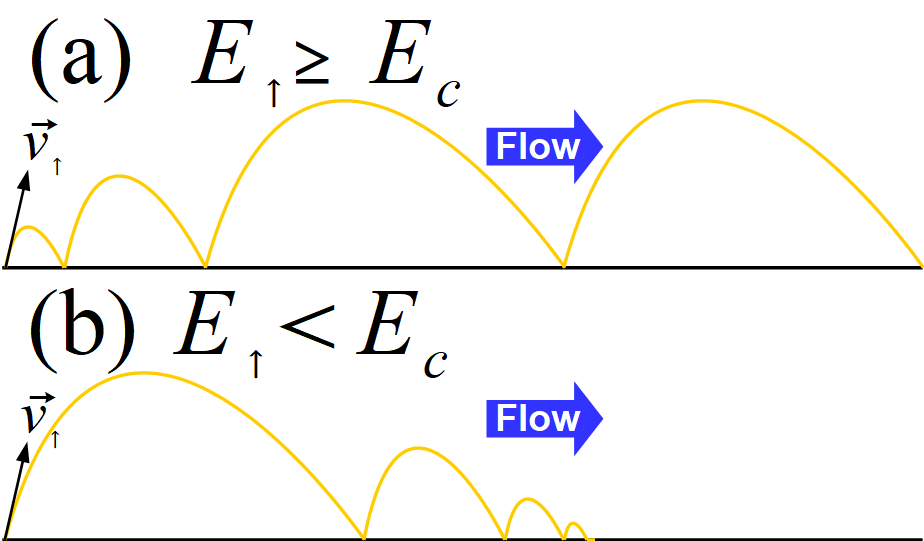}
 \end{center}
 \caption{Sketch of continuous rebound mechanism. Depending on its initial kinetic energy $E_\uparrow$ relative to a critical energy level $E_c$ that depends on the properties of the flow, a particle (yellow lines) either (a) gains sufficient energy in its hops along a flat wall (black lines) to approach a steady, periodic hopping motion or (b) net loses energy until it stops.}
\label{CriticalTrajectory}
\end{figure}

To illustrate the concept of continuous particle rebounds, the motion of a particle along a flat wall driven by a constant flow (e.g., the mean turbulent flow) is considered. This particle shall never be captured and instead, for illustration purposes, always rebound with a constant angle and lose a constant fraction of its impact energy (the core of the argument will not significantly change if more sophisticated rebound laws, such as equations~(\ref{e2D})--(\ref{thetar2D}), are considered). For this idealized scenario, there are two extremes of possible particle trajectories depending on the initial particle velocity $\mathbf{v_\uparrow}$, which are sketched in Figure~\ref{CriticalTrajectory}.
First, if the corresponding initial kinetic energy $E_\uparrow$ exceeds a critical value $E_c$, the particle will spend sufficiently long within the flow so that it gains sufficient energy via fluid drag during its hops to approach a steady, periodic hopping motion, in which its energy gain via fluid drag is exactly balanced by its energy loss during its rebounds (Figure~\ref{CriticalTrajectory}a). Henceforth, such particles are termed \textit{continuous rebounders}. Second, if $E_\uparrow<E_c$, the particle loses net energy in its initial and all subsequent hops until it stops (Figure~\ref{CriticalTrajectory}b). The critical energy $E_c$ depends on properties of the flow. Crucially, if the flow is too weak, all possible trajectories fall into the second category (i.e., $E_c=\infty$).

There are a few takeaways from the this simple thought experiment for realistic systems. First, as the mean turbulent flow is controlled by the fluid shear stress $\tau$, it suggests the existence of a rebound threshold $\tau^{\rm Rb}_t$ below which the energy losses in particle-bed rebounds cannot be compensated by the flow on average~\citep{JenkinsValance14,Berzietal16,Berzietal17,PahtzDuran18a,Pahtzetal20b}. Second, the randomness introduced by inhomogeneities of the bed and turbulent fluctuations of the flow introduce trajectory fluctuations that can lead to random losses of continuous rebounders, particularly when the lift-off energy accidentally falls below $E_c$~\citep{PahtzDuran18a}. Such losses must be compensated by the entrainment of bed particles into the continuous rebound layer. Hence, the mere mobilization of bed particles is not sufficient because the lift-off energy of mobilized particles must also exceed $E_c$. In particular, for rebound threshold models (see section~\ref{CessationThresholdModels}), it has been shown that $E_c$ becomes equal to the average rebound energy of continuous rebounders in the limit $\tau\rightarrow\tau^{\rm Rb}_t$~\citep{Pahtzetal20b}. This implies that the impact entrainment threshold $\tau^{\rm ImE}_t$ must be strictly larger than $\tau^{\rm Rb}_t$, since the energy of an entrained particle is much smaller than the energy of the particle that caused its entrainment (i.e., a continuous rebounder) because of energy conservation. In particular, $\tau^{\rm ImE}_t>\tau^{\rm Rb}_t$ automatically explains the discontinuous jump of the sediment transport rate $Q$ at $\tau^{\rm ImE}_t$ that has been observed in the absence of fluid entrainment by turbulent fluctuation events (see section~\ref{ImpactEntrainmentContinuousTransport}) because $Q(\tau^{\rm ImE}_t)$ is controlled by the excess shear stress $\tau^{\rm ImE}_t-\tau^{\rm Rb}_t>0$ in the absence of such events (see section~\ref{TransportCapacity}).

\subsubsection{Transport Capacity Interpretation Based on Continuous Rebounds} \label{TransportCapacity}
A third takeaway for realistic systems of the thought experiment described in section~\ref{ThoughtExperiment} involves the fact that, because of momentum transfer from flow to particles, the flow slows down with increasing transport load $M$. Hence, for a given $\tau>\tau^{\rm Rb}_t$, provided that there is an abundance of impact and/or fluid entrainment, the system tends to entrain bed material until the mean turbulent flow becomes so weak that it can barely sustain the average motion of continuous rebounders~\citep{PahtzDuran18b}. Any further slowdown of the flow would then spike the deposition rate, leading to a decrease of $M$ and subsequent increase of the flow speed. That is, the system is at a dynamic equilibrium that may be interpreted as transport capacity.

\citet{PahtzDuran18b} analytically showed that this interpretation of transport capacity leads to the capacity scaling
\begin{linenomath*}
\begin{equation}
 M\simeq\mu_b^{-1}\tilde g^{-1}(\tau-\tau^{\rm Rb}_t), \label{CapacityM}
\end{equation}
\end{linenomath*}
where $\tilde g=(1-1/s)g$ is the buoyancy-reduced value of the gravitational constant $g$ and $\mu_b=\tau_{pb}/P_b$ an approximately constant bed friction coefficient (i.e., the ratio between the particle shear stress $\tau_{pb}$ and normal-bed particle pressure $P_b\simeq M\tilde g$ evaluated at the bed surface). Note that the definitions of $\tau_{pb}$ and $P_b$ (and thus $\mu_b$), in contrast to the definitions of $\tau_p$ and $P$ (and thus the yield stress ratio $\mu_s$, see section~\ref{Yielding}), include contributions from stresses associated with the particle fluctuation motion in addition to contributions from intergranular contacts. The derivation of equation~(\ref{CapacityM}) by \citet{PahtzDuran18b} is based on two main steps: showing the approximate constancy of $\mu_b$ starting from a geometric constraint on particle-bed rebounds in the steady state and assuming $\tau_{gb}\simeq\tau-\tau^{\rm Rb}_t$, which expresses the aforementioned dynamic equilibrium condition associated with the continuous rebound motion. Interestingly, $\tau_{gb}$ describes the momentum that is transferred from flow to transported particles per unit bed area per unit time, which implies that high-buoyant fluids (small $\tilde g$), such as water, require a larger transport load $M$ for a given rate of momentum transfer (i.e., for a given $M\tilde g\propto\tau_{gb}$) than low-buoyant fluids (large $\tilde g$), such as air. \citet{PahtzDuran18b} tested these derivation steps with numerical data from DEM-based simulations of nonsuspended sediment transport (the same as those by \citet{PahtzDuran17}, see section~\ref{ImpactEntrainmentContinuousTransport}). It turned out that these steps, and thus equation~(\ref{CapacityM}), are obeyed across nonsuspended sediment transport conditions with $\mathrm{Ga}\sqrt{s}\gtrsim10$ (all but relatively viscous bedload transport) provided that the bed surface is defined as the effective elevation of energetic particle-bed rebounds.

The functional form of equation~(\ref{CapacityM}) is the foundation of the majority of theoretical and experimental shear stress threshold-based expressions for the capacity transport rate, $Q\simeq M\overline{v_x}$ (where $\overline{v_x}$ is the average streamwise velocity of particles moving above the bed surface), and goes back to the pioneering theoretical descriptions of nonsuspended sediment transport by \citet{Bagnold56,Bagnold66,Bagnold73}. However, Bagnold's physical interpretation of the assumptions leading to this scaling was inaccurate: $\mu_b$ is not equal to $\mu_s$ and $\tau^{\rm Rb}_t$ is not an entrainment threshold, as Bagnold assumed~\citep{PahtzDuran18b}. In fact, equation~(\ref{CapacityM}) has no association with sediment entrainment whatsoever, except for the fact that sediment entrainment is a necessary requirement to keep transport at capacity~\citep{PahtzDuran18b}.

\begin{figure}[!htb]
 \begin{center}
  \includegraphics[width=1.0\columnwidth]{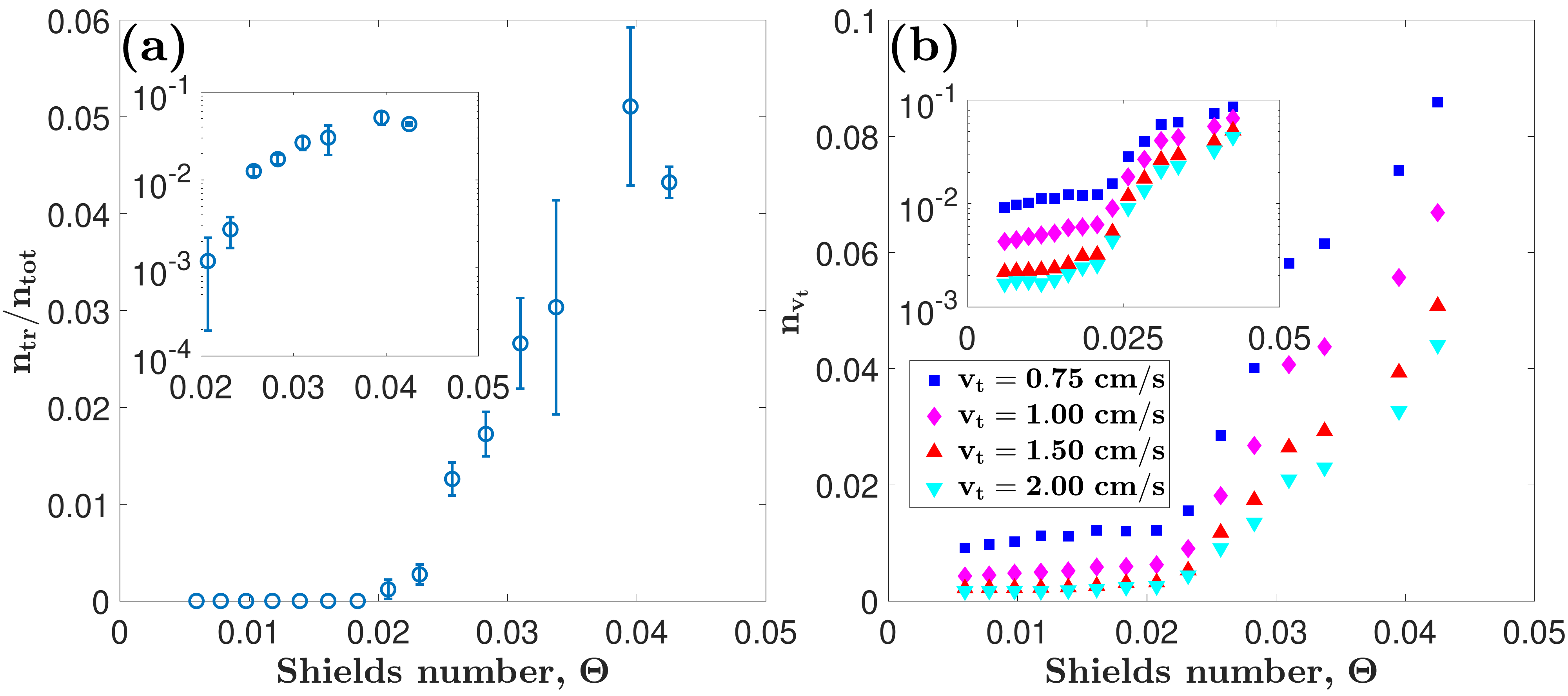}
 \end{center}
 \caption{Measurements of particle activity by \citet{Salevanetal17}. (a) Number of transported particles relative to the total number of bed surface particles ($n_{\mathrm{tr}}/n_{\mathrm{tot}}$) and (b) number of particles $n_{v_t}$ that are faster than a certain velocity threshold $v_t$ versus Shields number $\Theta$. Error bars in (a) correspond to the standard error computed from six experimental runs.}
\label{SalevanParticleActivity}
\end{figure}

As explained in section~\ref{ImpactEntrainmentContinuousTransport}, $Q$, and thus $M$, is significantly larger than zero at the impact entrainment threshold $\tau^{\rm ImE}_t$. In particular, transport becomes intermittent for $\tau<\tau^{\rm ImE}_t$ or even stops in the absence of entrainment by turbulent fluctuation events (i.e., transport capacity cannot be sustained). Hence, equation~(\ref{CapacityM}) is, in general, valid only for $\tau\geq\tau^{\rm ImE}_t$ and also consistent with the rebound threshold model prediction $\tau^{\rm ImE}_t>\tau^{\rm Rb}_t$ (see section~\ref{ThoughtExperiment}). Note that aeolian saltation transport experiments~\citep{Carneiroetal15,MartinKok18} and coupled DNS/DEM fluvial bedload transport simulation~\citep{Gonzalezetal17}, indeed, very roughly suggest $\tau^{\rm ImE}_t\approx1.5\tau^{\rm Rb}_t$ and $\tau^{\rm ImE}_t\approx2\tau^{\rm Rb}_t$, respectively. In order to extend the validity of equation~(\ref{CapacityM}), and thus of standard sediment transport rate relationships, to shear stresses $\tau$ with $\tau^{\rm Rb}_t<\tau<\tau^{\rm ImE}_t$, one must abandon long-term averaging sediment transport data. Instead, it is necessary to conditionally average $M$ (or $Q$) only over periods of near-capacity transport (on short-term average), but ignore periods with transport significantly below capacity or even at rest~\citep{BunteAbt05,Singhetal09,ShihDiplas18,Comolaetal19b}. Likewise, for realistic fluvial bedload transport, it is necessary to exclude the turbulence-driven fluctuation motion (including turbulent entrainment events) when measuring $M$ for equation~(\ref{CapacityM}) to remain valid; otherwise, transport does not vanish for $\tau\rightarrow\tau^{\rm Rb}_t$. \citet{Salevanetal17} demonstrated that implementing such constraints in the analysis of experimental data is, in principle, possible. By separating the velocity distribution of all measurable particles (including those that are visually perceived as resting) into a Student's $t$-distribution associated with the turbulence-driven fluctuation motion and an exponential distribution associated with the bulk transport of particles (which automatically implies conditional averaging as periods of rest do not affect this distribution), they obtained a measure for the number of transported particles relative to the total number of bed surface particles ($n_{\mathrm{tr}}/n_{\mathrm{tot}}$). This measure, indeed, vanishes within experimental precision below a Shields number threshold (Figure~\ref{SalevanParticleActivity}a), which can be interpreted as $\Theta^{\rm Rb}_t$, whereas the number of particles $n_{v_t}$ that are faster than a certain velocity threshold $v_t$ remains nonzero for the entire range of $\Theta$ because of the turbulence-driven fluctuation motion (Figure~\ref{SalevanParticleActivity}b).

\subsubsection{Does the Shields Diagram Truly Show Incipient Motion Thresholds?} \label{ShieldsDiagram}
The Shields diagram is a compilation of measurements of the threshold Shields number $\Theta_t$ as a function of the shear Reynolds number $\mathrm{Re}_\ast$, which have been labeled as measurements of incipient sediment motion by numerous studies and reviews~\citep[e.g.,][and references therein]{Shields36,Milleretal77,YalinKarahan79,ParkerKlingeman82,VanRijn84a,WibergSmith87,Ling95,BuffingtonMontgomery97,Dey99,Paphitis01,Caoetal06,DeyPapanicolaou08,AliDey16,DeyAli18,DeyAli19,Yangetal19}. However, incipient motion of turbulent fluvial bedload transport is much better characterized by impulse and energy-based criteria (section~\ref{EntrainmentCriteria}), unless one refers to the Shields number $\Theta^{\rm In}_t$ at which the fluid entrainment probability exceeds zero (section~\ref{ShearStressThreshold}), which is much below the Shields curve~\citep{Paintal71}. Furthermore, in steady, homogenous turbulent fluvial bedload transport in which turbulence is suppressed (e.g., in narrow water flumes), the vast majority of entrainment events is caused by particle-bed impacts (see section~\ref{ImpactEntrainmentContinuousTransport}). It is therefore here argued, based on the results reviewed in section~\ref{TransportCapacity}, that many of the threshold data compiled in the Shields diagram are actually measurements of the rebound threshold $\Theta^{\rm Rb}_t$.

The Shields diagram shows two kinds of threshold measurements obtained using two different methods. The first method is the \textit{reference method}, where one takes paired measurements of $\Theta$ and the nondimensionalized transport rate $Q_\ast$ (or transport load $M_\ast\equiv M/(\rho_pd)$) and extrapolates them to the Shields number at which $Q_\ast$ (or $M_\ast$) either vanishes~\citep[e.g.,][]{Shields36} (it is slightly controversial whether Shields really used this method~\citep{Buffington99}) or equals a small reference value~\citep[e.g.,][]{ParkerKlingeman82}. This method yields approximately the rebound threshold $\Theta^{\rm Rb}_t$ if an expression for $Q_\ast$ (or $M_\ast$) based on equation~(\ref{CapacityM}) is used for the extrapolation and provided that the data used for the extrapolation are at capacity (i.e., $\Theta\geq\Theta^{\rm ImE}_t$). For example, \citet{Lajeunesseetal10} extrapolated their measurements (many data points obeyed $\Theta\geq2\Theta^{\rm Rb}_t\approx\Theta^{\rm ImE}_t$) to $M_\ast=0$ using exactly equation~(\ref{CapacityM}), yielding exactly $\Theta^{\rm Rb}_t$. That the reference method yields the rebound threshold $\Theta^{\rm Rb}_t$ is further supported by the fact that the values of $\Theta^{\rm Rb}_t$ obtained from the DEM-based fluvial bedload transport simulations by \citet{PahtzDuran18a} are consistent with the compilation of reference method-based threshold measurements by \citet{BuffingtonMontgomery97}.

The second method is the \textit{visual method}, where one increases $\Theta$ until criteria defining what is considered critical transport are obeyed~\citep[e.g.,][]{Kramer35} (see section~\ref{Introduction}). The threshold values obtained from this method depend significantly on the chosen criterion and are, on average, close to those obtained from the reference method~\citep{BuffingtonMontgomery97}. For example, the transition point $(\Theta,Q_\ast)\approx(0.05,0.007)$ at which the function $Q_\ast(\Theta)$ measured in the gravel-bed experiments by \citet{Paintal71} changed from $Q_\ast\propto\Theta^{16}$ to $Q_\ast\propto\Theta^{2.5}$ (see section~\ref{Introduction}) is indistinguishable from the reference threshold for the same conditions within measurement uncertainty. In particular, a close examination of Paintal's and other gravel bed data has revealed that Paintal's power-16 region can actually be subdivided into two regions~\citep[][Figure~5]{DeyAli19} (see also \citep[][Figure~8b]{ShihDiplas19}): one region ($\Theta\lesssim0.04$) with a milder power law and one with a stronger power law ($0.04\lesssim\Theta\lesssim0.05$), which includes a jump of $Q_\ast$ by an order of magnitude at $\Theta\simeq0.04$. Such a jump is consistent with exceeding the rebound threshold $\Theta^{\rm Rb}_t$ because transported particles suddenly become able to move along the surface for comparably long times before being captured by the bed. Hence, it seems that also the visual method, at least for typical critical transport criteria, approximately yields the rebound threshold $\Theta^{\rm Rb}_t$ rather than an entrainment threshold.

The hypothesis that the Shields diagram shows measurements of the rebound threshold is further supported by the fact that certain rebound threshold models~\citep{PahtzDuran18a,Pahtzetal20b} reproduce the Shields curve without fitting to the experimental data compiled in the Shields diagram (see section~\ref{CessationThresholdModels}), even when limited to only visually measured data~\citep{Pahtzetal20b}.

\subsection{Sediment Transport Cessation Models} \label{CessationThresholdModels}
This section reviews theoretical models for both the rebound threshold $\Theta^{\rm Rb}_t$ and impact entrainment threshold $\Theta^{\rm ImE}_t$. One of the early motivations for developing such models was to better understand the hysteresis between the initiation and cessation of aeolian saltation transport observed in wind tunnel experiments~\citep[e.g.,][]{Bagnold41,Chepil45,IversenRasmussen94,Carneiroetal15}. While the difference between transport initiation and cessation is relatively small on Earth, wind tunnel experiments and observations suggested a substantial difference on Mars, which needed to be explained~\citep{Almeidaetal08,Kok10a}. (However, note that extrapolating wind tunnel measurements of the initiation threshold $\Theta^{\rm In}_t$ to field conditions using standard initiation threshold models is actually inappropriate because $\Theta^{\rm In}_t$ depends on the boundary layer thickness $\delta$, as discussed in section~\ref{ShearStressThreshold}.) Later on, cessation threshold models were developed with the purpose to unify fluvial bedload and aeolian saltation transport in a single theoretical framework~\citep{Berzietal16,PahtzDuran18a,Pahtzetal20b}.

As cessation threshold models are associated with a sustained motion of transported particles, they require a physical description of the particle motion within the transport layer that is coupled with boundary conditions that describe the interaction between transported particles and the bed surface. In general, there have been two approaches to describe the transport layer and bed interactions. The first approach consists of representing the entire particle motion by particles moving in identical periodic trajectories along a flat wall that mimics the bed surface (section~\ref{IdenticalTrajectoryModels}). The second approach consists of deriving general correlations between transport layer-averaged physical quantities and obtain the correlation coefficients from numerical simulations (section~\ref{CorrelationModels}). It will be shown that the latter approach is probably a rough approximation of a variant of the former. Correlation-based model equations elucidate the role that the density ratio $s$ plays for the rebound threshold $\Theta^{\rm Rb}_t$ in a simple manner and therefore provide a simple conceptual explanation for why $\Theta^{\rm Rb}_t$ is smaller in aeolian saltation than in fluvial bedload transport (section~\ref{AeolianvsFluvial}).

\subsubsection*{Open Problem: Effect of Cohesion on Transport Cessation Thresholds}
Most of the sediment transport cessation threshold models reviewed here account for cohesive interparticle forces and do so in a similar manner as transport initiation threshold models. However, \citet{Comolaetal19a} recently revealed that the effects of cohesion on transport cessation and initiation thresholds are actually fundamentally different from one another, which is why this section only considers versions of existing cessation threshold models for cohesionless particles. The effect of cohesion on transport cessation thresholds remains a major open problem.

\subsubsection{Identical Periodic Trajectory Models (IPTMs)} \label{IdenticalTrajectoryModels}
Most studies proposing cessation threshold models start with the assumption that the motion of transported particles can be represented by a system in which all particles hop in the same periodic trajectory, referred to as the average trajectory, driven by the mean turbulent flow along a flat wall, with which they interact according to certain boundary conditions~\citep{ClaudinAndreotti06,Kok10b,Berzietal16,Berzietal17,Pahtzetal20b}. (Note that, although \citet{Kok10b} does not explicitly refer to identical periodic trajectories, his mathematical treatment of the problem is equivalent to IPTMs.) However, the assumption of identical periodic particle trajectories introduces a variety of potentially major weaknesses, which has cast doubt on the reliability of IPTMs~\citep{Andreotti04,LammelKroy17,PahtzDuran17,PahtzDuran18a}:
\begin{enumerate}
 \item In IPTMs, the particle concentration increases with elevation $z$ and jumps to zero when $z$ exceeds the hop height~\citep{AndersonHallet86}. In contrast, in real nonsuspended sediment transport, it monotonously decreases with $z$, often exponentially~\citep[e.g.,][]{Duranetal12}. IPTMs that refer only to the motion of a well-defined species of particles (e.g., continuous rebounders) do not necessarily suffer from this weakness because the concentration profile associated with this species may behave differently from that of the entire ensemble of transported particles.
 \item In IPTMs, the mean square of the vertical particle velocity ($\langle v_z^2\rangle$) decreases with $z$. In contrast, in real nonsuspended sediment transport, it increases with $z$, except far from the bed surface~\citep{PahtzDuran17}. This behavior is a signature of the fact that the transport layer, in general, consists of different species of particles with different characteristic velocities~\citep[e.g.,][Figure~21]{Duranetal11}. That is, IPTMs that refer only to the motion of a well-defined species of particles (e.g., continuous rebounders) do not necessarily suffer from this weakness.
 \item Only particles that take off from the wall with an energy $E_\uparrow$ that is larger than a critical value $E_c$ can continue their motion after the initial few hops (Figure~\ref{CriticalTrajectory}). That is, IPTMs that take into account the motion of entrained particles~\citep{ClaudinAndreotti06,Kok10b} effectively assume that all entrained particles obey $E_\uparrow\geq E_c$ even though most of them do not~\citep{PahtzDuran18a}.
 \item IPTMs neglect particle motion via rolling and sliding, which is significant in bedload transport.
\end{enumerate}
Depending on the boundary conditions, three conceptually different kinds of IPTMs can be distinguished:
\begin{enumerate}
 \item Models of the rebound threshold $\Theta^{\rm Rb}_t$ consider only the dynamics of continuous rebounders. Their rebounds are described, for example, by equations~(\ref{e2Dempirical}) and (\ref{ezempirical}), which link the streamwise ($x$) and normal-wall ($z$) components of the impact velocity $\mathbf{v_i}$ to the streamwise and normal-wall components of the rebound velocity $\mathbf{v_r}$. Such models then look for the smallest Shields number that results in a periodic trajectory under the constraint that the hop height of particles exceeds one particle diameter ($z_h\geq d$). This constraint ensures consistency with the underlying model assumption that continuous rebounders are never captured by the bed surface. The threshold resulting from this constraint is denoted as $\Theta^{\rm Rb\ast\ast}_t$. \citet{Pahtzetal20b} modified this constraint to take into account that the near-surface flow can assist particles in escaping the bed surface and is even predominantly responsible for the escape in the viscous bedload transport regime. These authors' escape criterion reads $\Theta/\Theta_t^{\rm max}\geq\cot\psi/\cot\psi_Y$, where $\Theta_t^{\rm max}=0.12$ is the viscous fluid entrainment threshold (see section~\ref{Yielding}), $\psi_Y=30^\circ$ the pocket angle for particles resting within the deepest pockets of the bed surface, and $\sin\psi=\sin\psi_Y+\mathbf{v_r}^2/(2\tilde gd)$. This criterion means that the rebound kinetic energy only needs to uplift a particle rebounding within the deepest pocket to a point at which the near-surface flow is able to push it out of the pocket. The threshold resulting from this modified constraint is denoted as $\Theta^{\rm Rb\ast}_t$.
 \item Models of the impact entrainment threshold $\Theta^{\rm ImE}_t$~\citep{ClaudinAndreotti06,Kok10b} do not neglect captures of continuous rebounders and therefore take into account the entrainment of bed particles. One possible way to do this is by combining rebound boundary conditions with an additional constraint that describes that one particle leaves the surface per impact on average (e.g., $|\mathbf{v_i}|\propto\sqrt{\tilde gd}$~\citep{ClaudinAndreotti06}). However, the incorporation of entrained particles as part of the average trajectory leads to consistency problems (see third point in the list above).
 \item Hybrids between continuous rebound and impact entrainment models~\citep{Berzietal16,Berzietal17} look for the smallest Shields number (denoted as $\Theta^{\rm Rb|ImE}_t$) that results in a periodic trajectory under the constraint $z_h\geq d$ (like before) and the additional constraint that particle-bed impacts do not lead to entrainment. \citet{Berzietal16,Berzietal17} modeled the latter constraint via $|\mathbf{v_i}|/\sqrt{\tilde gd}\leq\zeta/2\approx20$ (cf. equation~(\ref{Neemp})), which assumes that the fastest particles represented by the average trajectory of continuous rebounders do not exceed the value $\zeta$ of the nondimensionalized impact velocity that is associated with the onset of entrainment (which can be roughly justified by assuming an even impact velocity distribution between $0$ and $\zeta$). However, \citet{PahtzDuran18a} pointed out that this additional constraint is inconsistent with the experimental and numerical evidence that impact entrainment to be effective requires that the transport rate is significantly larger than zero (see section~\ref{ImpactEntrainmentContinuousTransport}), which is never the case at the rebound threshold $\Theta^{\rm Rb}_t$ in the absence of entrainment by turbulent fluctuation events (see section~\ref{TransportCapacity} and equation~(\ref{CapacityM})). Consistently, \citet{Pahtzetal20b} showed that, in the limit $\Theta\rightarrow\Theta^{\rm Rb}_t$, identical periodic trajectories of continuous rebounders are unstable against trajectory fluctuations. That is, the energy that a particle must acquire upon entrainment to become a continuous rebounder is equal to the rebound energy of the continuous rebounder that has entrained it in this limit. This requirement contradicts the fact that the entrainment energy is much smaller than the rebound energy because of energy conservation, which implies that impact entrainment is impossible in this limit (see also discussion in section~\ref{ThoughtExperiment}).
\end{enumerate}
Apart from these conceptual differences, existing IPTMs differ in several details (partly summarized in Table~\ref{SummaryModels}): the form of the fluid drag law, the consideration or neglect of vertical drag forces on the particle motion, the form of the mean flow velocity profile (including the question of whether the viscous sublayer of the turbulent boundary layer is considered; for more details, see Appendix), and the bed boundary conditions (including the incorporation of viscous damping in the rebound laws).
\begin{table}[!htb]
 \begin{tabular}{l | c | c | c | c | c}
  \hline
  Study & Model & Vertical drag & Viscous sublayer & Viscous damping & Boundary Conditions \\ 
  \hline
	CA06 & $\Theta^{\rm ImE}_t$ & yes & yes & no & $e^{2D},\theta^{2D}_r=\mathrm{const}$ \\
	K10 & $\Theta^{\rm ImE}_t$ & yes & no & no & complex \\
	B16/17 & $\Theta^{\rm Rb|ImE}_t$ & no & no & yes & eqs.~(\ref{e2Dempirical}) and (\ref{ezempirical}) \\
	P19 & $\Theta^{\rm Rb\ast}_t$ & yes & yes & no & eqs.~(\ref{e2Dempirical}) and (\ref{ezempirical2}) \\
	\hline
 \end{tabular}
 \caption{Modeling details of the IPTMs by \citet{ClaudinAndreotti06} (CA06), \citet{Kok10b} (K10), \citet{Berzietal16,Berzietal17} (B16/17), and \citet{Pahtzetal20b} (P19).}
 \label{SummaryModels}
\end{table}
In this regard, it is reiterated that the effects of viscous damping on the dynamics of particle-bed rebounds are probably negligible for bedload transport (for which viscous damping is deemed as potentially significant), even for conditions with strongly damped binary particle collisions (see section~\ref{StaticBedCollision}).

In order to facilitate a comparison between the different model types that does not depend on modeling details but focuses only on conceptual differences, the same mean flow velocity profile (equation~(\ref{uxcomplex}), which includes the viscous sublayer), boundary conditions (equations~(\ref{e2D}) and (\ref{ez})), and fluid drag law (the drag law by \citet{Camenen07}) are used for all model types. Following the trajectory calculation by \citet{Pahtzetal20b}, the impact velocity $\mathbf{v_i}$ as a function of the rebound velocity $\mathbf{v_r}$ approximates as
\begin{linenomath*}
\begin{subequations}
\begin{align}
 \hat v_{iz}&=\hat v_{rz}-\hat t_h,\quad\text{with}\quad\hat t_h=1+\hat v_{rz}+W\left[-\left(1+\hat v_{rz}\right)e^{-(1+\hat v_{rz})}\right], \label{viz} \\
 \hat v_{ix}&=\hat v_{rx}e^{-\hat t_h}+V_s^{-1}\sqrt{\Theta}f(\mathrm{Ga}\sqrt{\Theta},V_s^2s\hat z_\ast+Z_\Delta)(1-e^{-\hat t_h}),\quad\text{with}\quad \hat z_\ast\equiv-\hat v_{iz}(\hat v_{rz}+1)-\hat v_{rz}, \label{vix}
\end{align}
\end{subequations}
\end{linenomath*}
where $t_h$ is the hop time, $W$ the principal branch of the Lambert-$W$ function, $V_s\equiv v_s/\sqrt{s\tilde gd}$ the dimensionless value of the settling velocity $v_s$ (defined in equation~(\ref{Settling})), $Z_\Delta d=0.7d$ the average elevation of the particles' center during particle-bed rebounds (obtained from experiments~\citep{Deyetal12,Hongetal15}), and $\sqrt{\Theta}f$ expresses the nonfluctuating wall-bounded flow after \citet{GuoJulien07}, with $f$ the function given in equation~(\ref{uxcomplex}). Furthermore, the hat denotes nondimensionalized quantities using combinations of $\tilde g$ and $v_s$, which is given by
\begin{linenomath*}
\begin{equation}
 v_s=\frac{\sqrt{s\tilde gd}}{\mu_b}\left[\sqrt{\frac{1}{4}\sqrt[m]{\left(\frac{24}{C_d^\infty\mathrm{Ga}}\right)^2}+\sqrt[m]{\frac{4\mu_b}{3C_d^\infty}}}-\frac{1}{2}\sqrt[m]{\frac{24}{C_d^\infty\mathrm{Ga}}}\right]^m,\quad\text{with}\quad\mu_b\equiv\frac{v_{ix}-v_{rx}}{v_{rz}-v_{iz}}, \label{Settling}
\end{equation}
\end{linenomath*}
where $C^\infty_d=1$ and $m=1.5$ are parameter values associated with the drag law for naturally shaped particles. Equations~(\ref{e2D}), (\ref{ez}), (\ref{viz}), and (\ref{vix}) can be iteratively solved for $\Theta(\mathrm{Ga},s,\hat v_{rz})$. Then the thresholds are obtained from
\begin{linenomath*}
\begin{subequations}
\begin{align}
 \Theta^{\rm Rb\ast}_t(\mathrm{Ga},s)&\equiv\min_{\hat v_{rz}}\Theta\left\{\mathrm{Ga},s,\hat v_{rz}\left[\cot^2\psi_Y\frac{\Theta^2}{\Theta_t^{{\rm max}2}}\geq\left(\sin\psi_Y+\frac{\mathbf{v_r}^2}{2\tilde gd}\right)^{-2}-1\right]\right\}, \label{Thetartast} \\
 \Theta^{\rm Rb\ast\ast}_t(\mathrm{Ga},s)&\equiv\min_{\hat v_{rz}}\Theta\left[\mathrm{Ga},s,\hat v_{rz}\left(z_h\geq d\right)\right], \label{Thetart} \\
 \Theta^{\rm ImE}_t(\mathrm{Ga},s)&\equiv\Theta\left[\mathrm{Ga},s,\hat v_{rz}\left(|\mathbf{v_i}|=\frac{1}{2}\zeta\sqrt{\tilde gd}\right)\right], \label{Thetaet} \\
 \Theta^{\rm Rb|ImE}_t(\mathrm{Ga},s)&\equiv\min_{\hat v_{rz}}\Theta\left[\mathrm{Ga},s,\hat v_{rz}\left(z_h\geq d\land|\mathbf{v_i}|\leq\frac{1}{2}\zeta\sqrt{\tilde gd}\right)\right], \label{Thetaret}
\end{align}
\end{subequations}
\end{linenomath*}
where the hop height is given by $z_h=[v_{rz}v_s-v_s^2\ln(1+v_{rz}/v_s)]/\tilde g$ (for small $v_{rz}/v_s$, $z_h\simeq v_{rz}^2/(2\tilde g)$). In equations~(\ref{Thetartast})--(\ref{Thetaret}), the rebound threshold $\Theta^{\rm Rb\ast}_t$ is the only modeled cessation threshold that is linked to the viscous yield stress $\Theta^{\rm max}_t$ and thus to dense granular flow rheology (see section~\ref{Yielding}). In a complete model covering all transport regimes, such a connection must exist because $\Theta^{\rm max}_t$ represents an upper limit to any kind of cohesionless sediment transport threshold. Also, a complete model of any kind of cohesionless transport threshold must reach this maximum value in the limit of vanishing particle inertia (i.e., when typical particle velocities during a trajectory become much smaller than $\sqrt{\tilde gd}$). The characteristic particle velocity scale in IPTMs is given by the settling velocity $v_s$, which scales as $v_s\propto\mathrm{Ga}\sqrt{s\tilde gd}$ in the viscous regime (Eq.~(\ref{Settling}) for small $\mathrm{Ga}$). That is, a complete model of any kind of cohesionless transport threshold must approach $\Theta^{\rm max}_t$ in the limit $v_s/\sqrt{\tilde gd}\propto\mathrm{Ga}\sqrt{s}\rightarrow0$, where $\mathrm{Ga}\sqrt{s}$ can be interpreted as a Stokes-like number~\citep{Berzietal16,Berzietal17,Clarketal17,PahtzDuran18a}.

\begin{figure}[!htb]
 \begin{center}
  \includegraphics[width=1.0\columnwidth]{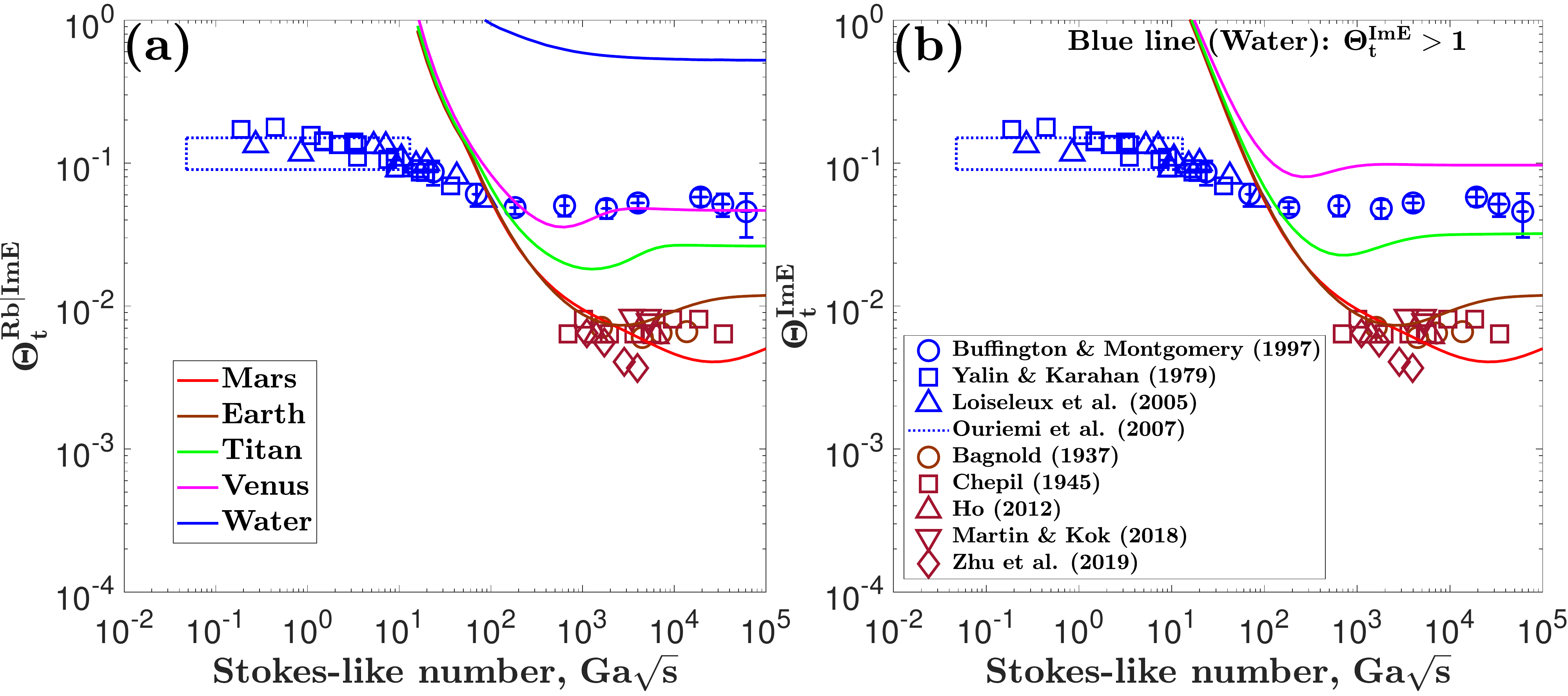}
 \end{center}
 \caption{Predictions of (a) the hybrid between rebound and impact entrainment threshold ($\Theta^{\rm Rb|ImE}_t$) and (b) the impact entrainment threshold $\Theta^{\rm ImE}_t$ from the IPTM as functions of the Stokes-like number $\mathrm{Ga}\sqrt{s}$ (lines) for five different density ratios $s=$~(2.65, 40, 190, 2200, and 250000) corresponding to five different fluvial or aeolian conditions (Water, Venus, Titan, Earth, and Mars). Symbols correspond to threshold measurements (or measurement compilations) from various studies~\citep{Bagnold37,Chepil45,BuffingtonMontgomery97,Loiseleuxetal05,Ouriemietal07,Ho12,MartinKok18,Zhuetal19} and methods (see text). \citet{Ouriemietal07} did not report single measurement values but a constant threshold $0.12\pm0.03$ for a large range of viscous conditions, indicated by the dotted square. Error bars correspond to $95\%$-confidence intervals of the compilation of reference method-based measurements by \citet{BuffingtonMontgomery97}, which make up a large portion of the Shields diagram.}
\label{ThresholdModels1}
\end{figure}

\begin{figure}[!htb]
 \begin{center}
  \includegraphics[width=1.0\columnwidth]{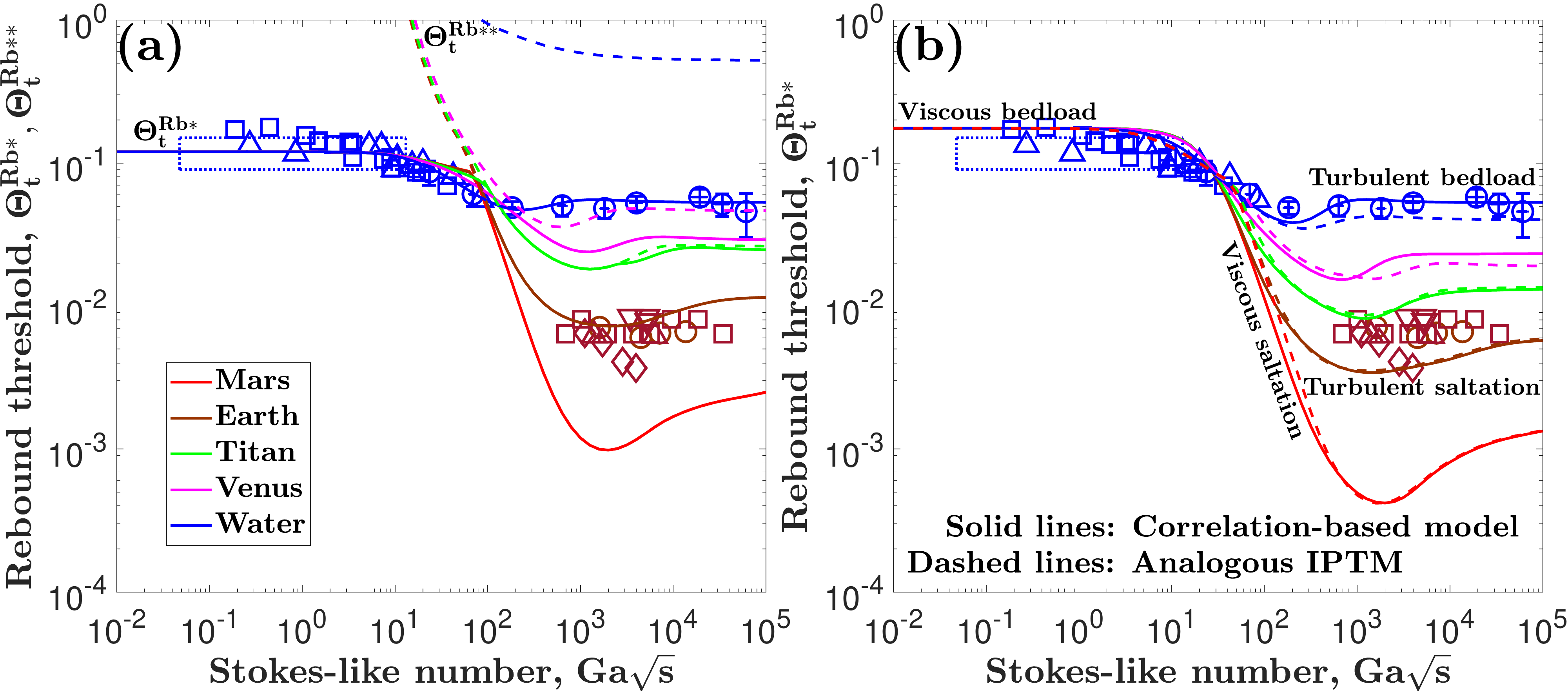}
 \end{center}
 \caption{Predictions of the rebound threshold, $\Theta^{\rm Rb\ast\ast}_t$ (dashed lines in (a)) and $\Theta^{\rm Rb\ast}_t$ (solid lines in (a) and (b) and dashed lines in (b)), from (a) the IPTM and (b) the correlation-based model by \citet{PahtzDuran18a} and its IPTM analogue as a function of the Stokes-like number $\mathrm{Ga}\sqrt{s}$ for five different density ratios $s=$~(2.65, 40, 190, 2200, and 250000) corresponding to five different fluvial or aeolian conditions (Water, Venus, Titan, Earth, and Mars). Symbols correspond to threshold measurements (or measurement compilations) from various studies~\citep{Bagnold37,Chepil45,BuffingtonMontgomery97,Loiseleuxetal05,Ouriemietal07,Ho12,MartinKok18,Zhuetal19} and methods (see text). \citet{Ouriemietal07} did not report single measurement values but a constant threshold $0.12\pm0.03$ for a large range of viscous conditions, indicated by the dotted square. Error bars correspond to $95\%$-confidence intervals of the compilation of reference method-based measurements by \citet{BuffingtonMontgomery97}, which make up a large portion of the Shields diagram. For symbol legend, see Figure~\ref{ThresholdModels1}. The IPTM in (b) uses the modified boundary conditions $\mu_b=0.63$ and $\cot\theta^{2D}_r=\mu_b[1/(\sqrt{3}c_1)-1]\simeq1.4$, and the modified viscous yield stress $\Theta_t^{\rm max}=-\mu_bZ_\Delta/(2c_2)+\sqrt{[\mu_bZ_\Delta/(2c_2)]^2+\mu_b^2/[18(1-c_3)c_2]}\simeq0.175$ to mimic the predictions from the correlation-based model by \citet{PahtzDuran18a}. These modifications are explained in the text.}
\label{ThresholdModels2}
\end{figure}

Figures~\ref{ThresholdModels1}a, \ref{ThresholdModels1}b, and \ref{ThresholdModels2}a show the thresholds calculated by equations~(\ref{Thetartast})-(\ref{Thetaret}) as a function of $\mathrm{Ga}\sqrt{s}$ for five different density ratios $s=$~(2.65, 40, 190, 2200, and 250000) corresponding to five different fluvial or aeolian conditions (Water, Venus, Titan, Earth, and Mars).
These figures also show cessation threshold measurements obtained for nearly cohesionless conditions using different experimental methods. For turbulent bedload transport driven by water, the compilation of reference method-based measurements (measurement mean and its $95\%$ confidence interval) by \citet{BuffingtonMontgomery97}, which make up a large portion of the Shields diagram, is shown. As explained in section~\ref{ShieldsDiagram}, this method yields approximately the rebound threshold $\Theta^{\rm Rb}_t$. For viscous bedload transport driven by water-oil mixtures, the visual incipient motion measurements by \citet{YalinKarahan79} and \citet{Loiseleuxetal05} and cessation threshold measurements by \citet{Ouriemietal07} are shown (for viscous bedload transport, the differences between transport initiation, rebound, and impact entrainment threshold are very small~\citep{PahtzDuran18a}). For aeolian saltation transport, a few studies~\citep[e.g.,][]{Ho12,Zhuetal19} carried out an indirect extrapolation to vanishing transport to obtain $\Theta^{\rm Rb}_t$ using a proxy of $Q$: the surface roughness $z_o$ (see Appendix for its definition in the absence of transport), which undergoes a regime shift when saltation transport ceases. Furthermore, visual measurements of $\Theta^{\rm Rb}_t$ by \citet{Bagnold37} and \citet{Chepil45} are shown, obtained from successively decrementing $\Theta$ until intermittent saltation transport stops. Direct measurements of the intermittent saltation transport threshold (and thus $\Theta^{\rm Rb}_t$), based on the so-called \textit{Time Frequency Equivalence Method} (TFEM)~\citep{Wiggsetal04}, by \citet{MartinKok18} are also shown. Note that, although the evidence that the thresholds obtained from extrapolation to vanishing transport and from direct measurements of the cessation of intermittent saltation transport correspond to the rebound threshold $\Theta^{\rm Rb}_t$ is quite strong (see section~\ref{ImpactEntrainmentContinuousTransport} and \ref{TransportCapacity}), many aeolian researchers believe that they correspond to the impact entrainment threshold $\Theta^{\rm ImE}_t$~\citep[e.g.,][]{MartinKok18}. One of the reasons for this belief can be seen in Figure~\ref{ThresholdModels1}b: the prediction of $\Theta^{\rm ImE}_t$ from equation~(\ref{Thetaet}) is consistent with aeolian saltation transport data on Earth despite not containing fit parameters. In fact, for the range of conditions corresponding to these data, the predictions of $\Theta^{\rm Rb\ast}_t$ and $\Theta^{\rm Rb\ast\ast}_t$ by equations~(\ref{Thetartast}) and (\ref{Thetart}) are equivalent and, coincidentally, very close to the predictions of $\Theta^{\rm ImE}_t$ and $\Theta^{\rm Rb|ImE}_t$ by equations~(\ref{Thetaet}) and (\ref{Thetaret}), which are also equivalent to each other. At this point, it is worth reiterating that differences between the models caused by differences in the modeling details (e.g., those in Table~\ref{SummaryModels}) have been excluded here. Such detail differences cause the predictions of existing models to differ more strongly from one another than shown here.

Figures~\ref{ThresholdModels1}a, \ref{ThresholdModels1}b, and \ref{ThresholdModels2}a show that the predictions of $\Theta^{\rm Rb\ast\ast}_t$, $\Theta^{\rm ImE}_t$, and $\Theta^{\rm Rb|ImE}_t$ from equations~(\ref{Thetart})--(\ref{Thetaret}) overestimate threshold measurements for fluvial bedload transport by at least an order of magnitude. For $\Theta^{\rm Rb\ast\ast}_t$ and $\Theta^{\rm Rb|ImE}_t$, this overestimation is caused by the constraint in the minimization of $\Theta$ that the particle hop height $z_h$ must exceed one particle diameter $d$ to escape the bed surface (equations~(\ref{Thetart}) and (\ref{Thetaret})), preventing solutions with small particle velocities that would have a smaller threshold. However, the prediction of $\Theta^{\rm Rb\ast}_t$ from equation~(\ref{Thetartast}), which is based on a modified escape condition that takes into account the near-surface flow, is consistent with fluvial bedload transport conditions (Figure~\ref{ThresholdModels2}a). The simultaneous agreement of the prediction of $\Theta^{\rm Rb\ast}_t$ from equation~(\ref{Thetartast}) with aeolian and fluvial transport regimes strongly supports modeling nonsuspended sediment transport within the continuous rebound framework.

\subsubsection{Models Based on Correlations Between Transport Layer-Averaged Physical Quantities} \label{CorrelationModels}
Existing correlation-based cessation threshold models start with the assumption of a constant bed friction coefficient $\mu_b$~\citep{Pahtzetal12,PahtzDuran18a} ($\mu_b$ is the inverse of the parameter $\alpha$ in the model by \citet{Pahtzetal12}). As discussed in section~\ref{TransportCapacity}, the approximate constancy of $\mu_b$ has been analytically linked to continuous rebounds~\citep{PahtzDuran18b}. However, in contrast to the purely kinematic meaning of $\mu_b$ in IPTMs (equation~(\ref{Settling})), for realistic nonsuspended sediment transport, $\mu_b$ conveys information about both the particle kinematics and interparticle contacts. Note that $\mu_b\simeq\mathrm{const}$ is also predicted by IPTMs when vertical drag forces are small (i.e., the buoyancy-reduced gravity force dominates the vertical motion) because this fixes $e_z\simeq 1$ and thus $e^{2D}$, $\theta^{2D}_r$, and $\mu_b$ via the rebound laws~\citep{Pahtzetal20b}.

A constant $\mu_b$ links the average horizontal fluid drag acceleration $\overline{a_{\mathrm{d}x}}$ to the buoyancy-reduced gravity $\tilde g$ via $\mu_b\simeq\overline{a_{\mathrm{d}x}}/\tilde g$, where the overbar denotes a particle concentration-weighted height average~\citep{PahtzDuran18a} (which is equal to the average over the hop time for IPTMs). This link subsequently fixes the value of the nondimensionalized average velocity difference $U_x-V_x\equiv(\overline{u_x}-\overline{v_x})/\sqrt{s\tilde gd}=\mu_bv_s/\sqrt{s\tilde gd}$~\citep{Pahtzetal20b} as a function of the Galileo number $\mathrm{Ga}$ via equation~(\ref{Settling}). In fact, equation~(\ref{Settling}) is not limited to IPTMs but actually more general~\citep{PahtzDuran18a}. A further general correlation between $U_x$ and the nondimensionalized transport layer thickness $Z\equiv\overline{z}/d$ can be obtained from approximating $\overline{u_x(z)}\simeq u_x(\overline{z})$~\citep{PahtzDuran18a}. An analogous approximation is also involved in some IPTMs, namely, in the right-hand side of equation~(\ref{vix}), since $\overline{\hat z}\simeq\hat z_\ast\simeq\hat v_{rz}^2/3$ in leading order in $\hat v_{rz}$ (i.e., when vertical drag forces are small). Up to this point, the two existing correlation-based models by \citet{Pahtzetal12} and \citet{PahtzDuran18a} are equivalent. From now on, only the latter model is reviewed as it constitutes a substantial improvement of the former model in many regards. \citet{PahtzDuran18a} derived the further correlation $V_z\equiv\sqrt{\overline{v_z^2}/(s\tilde gd)}=c_1\mu_b^{-1}V_x$, where $c_1$ is a proportionality constant. This correlation with $c_1=[\sqrt{3}(\cot\theta^{2D}_r/\mu_b+1)]^{-1}$ is also predicted by IPTMs in the limit of small vertical drag forces. That is, up to here, the model by \citet{PahtzDuran18a} is effectively an IPTM that neglects vertical drag forces. The main differences between the model by \citet{PahtzDuran18a} and IPTMs lie in the latter two equations of the full set of model equations:
\begin{linenomath*}
\begin{subequations}
\begin{align}
 U_x-V_x&=\left[\sqrt{\frac{1}{4}\sqrt[m]{\left(\frac{24}{C_d^\infty\mathrm{Ga}}\right)^2}+\sqrt[m]{\frac{4\mu_b}{3C_d^\infty}}}-\frac{1}{2}\sqrt[m]{\frac{24}{C_d^\infty\mathrm{Ga}}}\right]^m, \label{mub2} \\
 U_x&=\sqrt{\Theta^{\rm Rb\ast}_t}f[\mathrm{Ga}\sqrt{\Theta_t},(Z+Z_\Delta)], \label{MeanU} \\
 V_z&=c_1\mu_b^{-1}V_x, \label{Vz} \\
 Z&=c_2\mu_b^{-1}\Theta^{\rm Rb\ast}_t+sV_z^2, \label{Meanz} \\
 V_x&=\frac{2\sqrt{\Theta^{\rm Rb\ast}_t}}{\kappa}\sqrt{1-\exp\left[-\frac{1}{4}c_3^2\kappa^2\left(U_x/\sqrt{\Theta^{\rm Rb\ast}_t}\right)^2\right]}, \label{VU}
\end{align}
\end{subequations}
\end{linenomath*}
where $\mu_b=0.63$, $c_1=0.18$, $c_2=0.9$, $c_3=0.79$, and $Z_\Delta=0.7$ are the model parameter values that \citet{PahtzDuran18a} obtained from adjusting equations~(\ref{mub2})--(\ref{VU}) to DEM-based simulations of nonsuspended sediment transport (the same kind of simulations as those by \citet{PahtzDuran17}, see section~\ref{ImpactEntrainmentContinuousTransport}). Equation~(\ref{Meanz}) contains two terms: a term ($sV_z^2$) that is associated with the vertical motion of particles (equivalent to $\overline{\hat z}=\overline{\hat v_z^2}$ in IPTMs) and a term ($\mu_b^{-1}\Theta^{\rm Rb\ast}_t$) that is associated with particle collisions and particle-bed contacts of particles moving above the bed surface level, which occur because of the surface texture~\citep{PahtzDuran18a}. A term analogous to the latter does not appear in existing IPTMs. Equation~(\ref{VU}) empirically merges two extremes. On the one hand, when the transport layer is completely submerged within the viscous sublayer of the turbulent boundary layer (small $U_x/\sqrt{\Theta^{\rm Rb\ast}_t}$), it predicts $V_x=c_3U_x$. For viscous bedload transport (i.e., when the transport layer is small: $Z\ll Z_\Delta$), this correlation with $c_3=1-\mu_b/[18\Theta_t^{\rm max}(c_2\Theta_t^{\rm max}/\mu_b+Z_\Delta)]$ is also predicted by IPTMs that employ the constrained minimization principle in equation~(\ref{Thetartast}) to calculate the rebound threshold $\Theta^{\rm Rb\ast}_t$. For viscous saltation transport (i.e., when the transport layer is large: $Z\gg Z_\Delta$), IPTMs of the rebound threshold that consider vertical drag forces also predict $V_x\propto U_x$~\citep{PahtzDuran18a}. However, the proportionality constant exhibits a different value (but still near unity) that depends on $\mu_b/\cot\theta^{2D}_r$. On the other hand, when most transport occurs within the log-layer of the turbulent boundary layer (large $U_x/\sqrt{\Theta^{\rm Rb\ast}_t}$), equation~(\ref{VU}) predicts $V_x\simeq2\sqrt{\Theta^{\rm Rb\ast}_t}/\kappa$, which also follows from the minimization principle for turbulent saltation transport (i.e., $Z\gg Z_\Delta$)~\citep{PahtzDuran18a}. For these reasons, equation~(\ref{VU}) can be interpreted as a rough approximation of the constrained minimization in equation~(\ref{Thetartast}) yielding $\Theta^{\rm Rb\ast}_t$. In fact, Figure~\ref{ThresholdModels2}b shows that the predictions of $\Theta^{\rm Rb\ast}_t$ from the model by \citet{PahtzDuran18a} are similar to those from an analogous IPTM and that they are also consistent with measurements across aeolian and fluvial environments. The predictions from these two models differ for turbulent bedload transport (large $U_x/\sqrt{\Theta^{\rm Rb\ast}_t}$ and small transport layer: $Z\sim Z_\Delta$), mainly because the scaling $V_x\simeq2\sqrt{\Theta^{\rm Rb\ast}_t}/\kappa$ that equation~(\ref{VU}) predicts for large $U_x/\sqrt{\Theta^{\rm Rb\ast}_t}$ does not capture the outcome of the constrained minimization in equation~(\ref{Thetartast}) for $Z\sim Z_\Delta$. The predictions from these two models also differ for viscous saltation transport because the model by \citet{PahtzDuran18a} neglects vertical drag forces at various instances. However, note that this model does not completely neglect vertical drag forces because the scaling $V_x\propto U_x$ that equation~(\ref{VU}) predicts for viscous saltation transport is associated with vertical drag~\citep{PahtzDuran18a}, which is why deviations between this model and the analogous IPTM are only moderate in this regime.

\subsubsection*{Open Problem: Reliable Models of the Impact Entrainment Threshold and Planetary Saltation Transport}
Existing models of the impact entrainment threshold~\citep{ClaudinAndreotti06,Kok10b,Pahtzetal12}, which is arguably also the continuous transport threshold, do not take into account that the transport rate $Q$ is significantly larger than zero at $\Theta^{\rm ImE}_t$, even in the absence of entrainment by turbulent fluctuation events (see section~\ref{ImpactEntrainmentContinuousTransport}). Instead, $Q$ vanishes at the rebound threshold $\Theta^{\rm Rb}_t$, which is smaller than $\Theta^{\rm ImE}_t$ (see section~\ref{TransportCapacity}). Likewise, as mentioned before, existing models of $\Theta^{\rm ImE}_t$ effectively assume that all entrained particles exhibit a kinetic energy that allows them to participate in the continuous rebound motion even though most of them do not~\citep{PahtzDuran18a}. For these reasons, existing impact entrainment threshold models seem to be missing important physics and need to be improved. This is problematic for modeling and predicting extraterrestrial sediment transport and associated bedform evolution~\citep[e.g.,][]{Almeidaetal08,Bourkeetal10,Kok10a,Ayoubetal14,Lorenz14,Rasmussenetal15,Jiaetal17,Telferetal18,Duranetal19} because most predictions of the aeolian saltation transport rate require that transport is continuous (i.e., at capacity).

\subsubsection{Main Difference Between Aeolian and Fluvial Rebound Threshold} \label{AeolianvsFluvial}
The most important difference between aeolian saltation and fluvial bedload transport is the largely different density ratio $s$, which ranges from close to unity for oil and water to the order of $10^5$ for air on Mars. Equations~(\ref{mub2})-(\ref{VU}) elucidate that $s$ affects the modeled rebound threshold $\Theta^{\rm Rb\ast}_t$ in a relatively simple manner. In fact, it can be seen that $s$ explicitly appears only in Eq.~(\ref{Meanz}), which describes a monotonous increase of the dimensionless transport layer thickness $Z$ with $s$. Subsequently, $Z$ monotonously increases the dimensionless transport layer-averaged flow velocity $U_x$ via equation~(\ref{MeanU}). That is, given a certain solution $\Theta^{\rm Rb\ast}_t(\mathrm{Ga},s)$ of equations~(\ref{mub2})--(\ref{VU}), an increase of $s$ leads to an increase of $U_x$, which must be compensated by a decrease of $U_x$ via a decrease of $\Theta^{\rm Rb\ast}_t$ to achieve a new steady solution for the same Galileo number $\mathrm{Ga}$. This mathematical fact expresses the physical fact that particles that stay longer in the flow can feel a given effective flow forcing at a lower fluid shear stress, which is the ultimate reason for why $\Theta^{\rm Rb\ast}_t$ decreases with $s$ for a given $\mathrm{Ga}$.

\section{Summary and Outlook} \label{Summary}
Section~\ref{Introduction} outlined five old, yet very significant, inconsistencies related to the concept of a threshold shear stress for incipient motion. For the concept of a threshold shear stress to be physically meaningful, these inconsistencies must be addressed and resolved. They can be briefly summarized as follows:
\begin{enumerate}
 \item By design, existing models of incipient motion capture the conditions to which they have been adjusted (aeolian or fluvial transport on Earth). However, the predictions from standard models adjusted to aeolian transport on Earth~\citep{IversenWhite82,ShaoLu00} are in stark disagreement with recent observations of aeolian transport on Mars~\citep[e.g.,][]{SullivanKok17,Bakeretal18}.
 \item Because of turbulent fluctuation events, fluid entrainment gives rise to fluvial bedload transport even for Shields numbers much below the Shields curve~\citep{Paintal71}, which is the curve that is thought to describe incipient motion.
 \item Below the respective shear stress threshold associated with incipient motion, turbulent fluctuation events are able to initiate fluvial bedload transport but not aeolian saltation transport. However, there is no reason to believe that the physics of incipient motion are different in aeolian and fluvial environments.
 \item Old experiments indicate a nonnegligible role of particle inertia in fluvial bedload transport~\citep{Ward69,GrafPazis77}, which is problematic because critical conditions are defined via nonzero transport rates (i.e., particles are in motion at threshold conditions).
 \item In old numerical simulations of turbulent fluvial bedload transport~\citep{NinoGarcia98a}, it was recognized that the threshold shear stress obtained from extrapolating the simulated capacity transport rate to vanishing transport may not be associated with fluid entrainment. This is problematic because many of the threshold data compiled in the Shields diagram have been obtained from such or similar extrapolation methods~\citep[e.g.,][]{Shields36}.
\end{enumerate}
As a result of the latest research reviewed in sections~\ref{DenseGranularFlow}--\ref{ParticleInertia}, a new conceptual picture has emerged (section~\ref{NewConcept}) that resolves these problems. However, it must be emphasized that this conceptual picture represents the authors' synthesis of the current state of the art, and many aeolian and fluvial geomorphologists may disagree. This is because, in some places, it stands in stark contrast to what has been a century old consensus. Likewise, there are still many open problems and controversies, summarized in section~\ref{OpenProblems} (and highlighted in sections~\ref{DenseGranularFlow}--\ref{ParticleInertia}), as well as a number of issues that have not been discussed in this review (e.g., the effects of particle size heterogeneity on transport thresholds and sediment entrainment), into which section~\ref{Outlook} presents a brief outlook.

\subsection{A New and Controversial Conceptual Picture of the Physics of the Thresholds of Nonsuspended Sediment Transport and Bed Sediment Entrainment} \label{NewConcept}
\begin{figure}[!htb]
 \begin{center}
  \includegraphics[width=1.0\columnwidth]{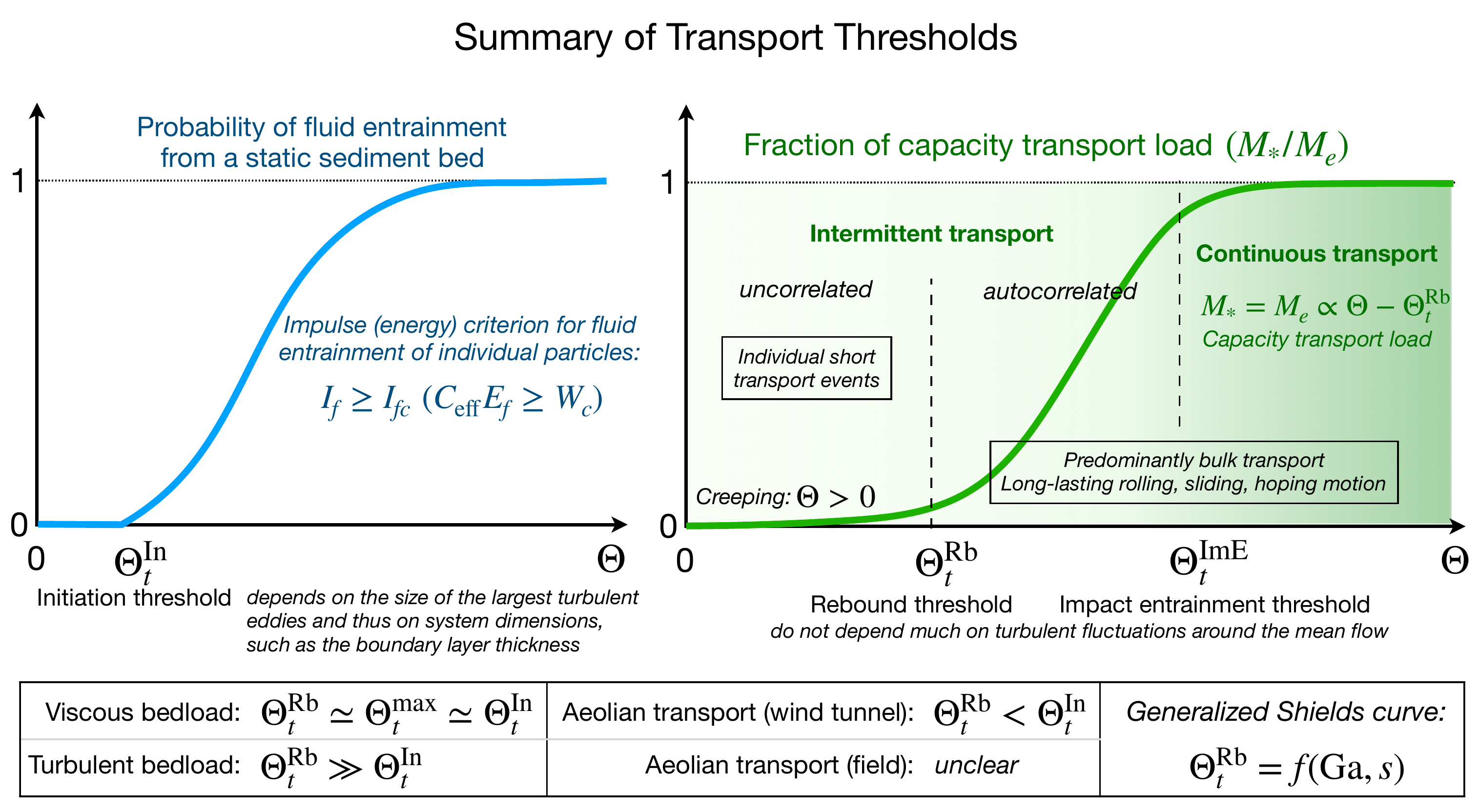}
 \end{center}
 \caption{Sketch summarizing Shields number ($\Theta$) thresholds of nonsuspended sediment transport.}
\label{FigSummary}
\end{figure}

Figure~\ref{FigSummary} summarizes the various shear stress thresholds of nonsuspended sediment transport and their relations to and effects on the transport characteristics.
Details, with references to the research reviewed in sections~\ref{DenseGranularFlow}--\ref{ParticleInertia}, are described below.

\subsubsection{Creeping ($\Theta>0$)}
Creeping (see section~\ref{Creeping}) refers to a superslow granular motion, usually in the form of intermittent local particle rearrangements within the sediment bed (not limited to the bed surface), that occurs below a macroscopic yield criterion (see section~\ref{Yielding}). One form of creeping is triggered by nearby regions above yield, while another form (the origin of which is not fully understood) occurs even in the absence of such regions. The existence of the latter form implies that sediment likely is \textit{always} transported (albeit slowly) for arbitrarily small values of the Shields number $\Theta$, even in the absence of turbulence. Creeping of both kinds is very important in determining the particle motion near transport initiation. It is fundamentally related to the granular material, not a purely fluid-driven effect.

\subsubsection{Viscous Yield Stress $\Theta^{\rm max}_t$}
Apart from creeping, which affects the entire granular bed, bed surface particles can be entrained directly by flow forces. When a sediment bed is subjected to a laminar flow at a sufficiently low shear Reynolds number $\mathrm{Re}_\ast$, there is a critical Shields number, the yield stress $\Theta^{\rm max}_t$, above which motion of bed surface particles is initiated and then never stops, whereas potential transient motion below $\Theta^{\rm max}_t$ will inevitably come to an end (see section~\ref{Yielding}). The viscous yield stress $\Theta^{\rm max}_t$ constitutes the upper limit for any kind of cohesionless sediment transport threshold, including the Shields curve. The values of $\Theta^{\rm max}_t$ reported in the literature are somewhat scattered (between 0.1 and 0.4), but these numbers are within the range of the bulk friction coefficients for granular materials (ranging from low friction spheres to more frictional, rough particles), suggesting that the granular material's yield condition (see section~\ref{Yielding}) is very important in determining the viscous yield stress.

\subsubsection{Initiation Threshold $\Theta^{\rm In}_t$}
While for laminar flows, the entrainment of individual bed surface particles is controlled by a critical Shields number, the entrainment of individual bed surface particles by turbulent flows is better described by an impulse (section~\ref{ImpulseCriterion}) or energy (section~\ref{EnergyCriterion}) criterion. Nonetheless, one can still define an initiation threshold Shields number $\Theta^{\rm In}_t$ (see section~\ref{ShearStressThreshold}) at which the probability of fluid entrainment exceeds zero (i.e., $\Theta^{\rm In}_t\simeq\Theta^{\rm max}_t$ for laminar flows at sufficiently low $\mathrm{Re}_\ast$). Because fluid entrainment is predominantly caused by turbulent fluctuation events, $\Theta^{\rm In}_t$ depends not only on $\mathrm{Re}_\ast$ but also on properties that control the size of the largest turbulent flow eddies, such as the turbulent boundary layer thickness. This may be one of the reasons why aeolian incipient motion models adjusted to wind tunnel measurements fail when applied to atmospheric boundary layers (see first problem outlined at the beginning of section~\ref{Summary}). Further possible reasons include atmospheric instability, topography gradients, surface inhomogeneities, such as obstacles and vegetation, and sublimation of subsurface ice in natural atmospheres (see section~\ref{BoundaryLayerThickness}).

\subsubsection{Rebound Threshold $\Theta^{\rm Rb}_t$ (Generalized Shields Curve)}
The rebound threshold $\Theta^{\rm Rb}_t$ (see section~\ref{ContinuousRebounds}) is largely unrelated to the entrainment of bed sediment (except for viscous bedload transport, for which $\Theta^{\rm Rb}_t\simeq\Theta^{\rm max}_t$) but describes the minimal dimensionless fluid shear stress that is needed for the mean turbulent flow to compensate the average energy loss of rebounding particles by fluid drag acceleration during their trajectories. Hence, for $\Theta\geq\Theta^{\rm Rb}_t$, transported particles rebound for comparably longer periods before they deposit, whereas they deposit very quickly for $\Theta<\Theta^{\rm Rb}_t$. The former transport regime gives rise to transport autocorrelations, while the latter gives rise to individual uncorrelated transport events. Hence, bulk sediment transport vanishes at $\Theta^{\rm Rb}_t$, which is described by a general law for the dimensionless bulk transport load $M_\ast$ at transport capacity (i.e., $M_\ast=M_e\propto\Theta-\Theta^{\rm Rb}_t$, see equation~(\ref{CapacityM})). In fact, fluvial incipient motion measurements compiled in the Shields diagram are actually measurements of $\Theta^{\rm Rb}_t$ (see section~\ref{ShieldsDiagram}), consistent with the fact that turbulent fluvial bedload transport does not vanish even much below the Shields curve because of occasional strong turbulent fluctuation events~\citep{Paintal71}. The notion that $\Theta^{\rm Rb}_t$ is largely unrelated to incipient motion and instead related to particle inertia resolves the second, third, fourth, and fifth problems outlined at the beginning of section~\ref{Summary}. There are relatively simple models (which neglect turbulent fluctuations around the mean turbulent flow) predicting $\Theta^{\rm Rb}_t$ in agreement with measurements across aeolian and fluvial environment without containing fitting parameters (models of $\Theta^{\rm Rb\ast}_t$ in section~\ref{CessationThresholdModels}). Such models predict a generalized Shields curve of the form $\Theta^{\rm Rb}_t(\mathrm{Ga},s)$, where $\mathrm{Ga}\equiv\sqrt{(s-1)gd^3}/\nu_f$ is the Galileo number and $s\equiv\rho_p/\rho_f$ the particle-fluid-density ratio, via modeling steady continuous particle trajectories. In fact, in aeolian environments, comparably large values of $s$ allow the flow to sustain comparably large steady trajectories at a comparably low Shields number $\Theta$, causing $\Theta^{\rm Rb}_t$ to be substantially smaller than in fluvial environments for a given $\mathrm{Ga}$ (see section~\ref{AeolianvsFluvial}).

\subsubsection{Impact Entrainment Threshold $\Theta^{\rm ImE}_t$}
Even for $\Theta>\Theta^{\rm Rb}_t$, randomness introduced by inhomogeneities of the bed and turbulent fluctuations of the flow introduce trajectory fluctuations that can lead to random captures of rebounding particles by the bed. To sustain transport capacity, these captures must be compensated by entrainment of bed sediment into the rebound layer by the action of the fluid (see section~\ref{FluidEntrainment}), by particle-bed impacts (see section~\ref{ImpactEntrainmentContinuousTransport}), or a combination of both (see section~\ref{ImpactEntrainmentBedload}). Because entrainment involving the flow requires strong turbulent fluctuation events (see sections~\ref{FluidEntrainment} and \ref{ImpactEntrainmentBedload}), which occur only at an intermittent basis, transport remains intermittent when impact entrainment alone is insufficient in providing the transport layer with rebounders (i.e., for $\Theta<\Theta^{\rm ImE}_t$). However, once the impact entrainment threshold $\Theta^{\rm ImE}_t$ (see section~\ref{ImpactEntrainmentContinuousTransport}) is exceeded ($\Theta\geq\Theta^{\rm ImE}_t$), impact entrainment is sufficient to do so, even without the assistance of fluid entrainment (i.e., significant fluid entrainment may occur, but is not needed). The impact entrainment threshold is strictly larger than the rebound threshold ($\Theta^{\rm ImE}_t>\Theta^{\rm Rb}_t$), which is associated with a nonzero bulk transport rate ($Q(\Theta^{\rm ImE}_t)>0$). This behavior can be explained within the continuous rebound framework (see section~\ref{ThoughtExperiment}). Nonetheless, reliable models of $\Theta^{\rm ImE}_t$ are currently missing (see section~\ref{CessationThresholdModels}).

\subsubsection{Differences Between Bedload and Saltation Transport}
To avoid confusion, we reiterate that the terms \textit{bedload transport} ($h\sim d$) and \textit{saltation transport} ($h\gg d$) have been defined through the transport layer thickness $h$ relative to the particle diameter $d$ (see notation and section~\ref{Introduction}). Depending on the relationship between the initiation threshold $\Theta^{\rm In}_t$ and the rebound threshold $\Theta^{\rm Rb}_t$, one observes different dynamics. For turbulent fluvial bedload transport, $\Theta^{\rm In}_t\ll\Theta^{\rm Rb}_t$, which means that transport can be initiated much below the Shields curve by occasional turbulent fluctuation events. However, whenever this happens, transport will very rapidly stop again. This is, indeed, the typical situation for gravel-bed rivers, which adjust their shape so that they remain in a low-mobility state~\citep{Parker78,PhillipsJerolmack16}. For aeolian transport in wind tunnels, $\Theta^{\rm In}_t$ is significantly larger than $\Theta^{\rm Rb}_t$. This explains why aeolian bedload transport is usually very short-lived. In fact, even though bed particles are usually entrained into a rolling motion at $\Theta^{\rm In}_t$ (i.e., $h\sim d$), this rolling motion rapidly evolves into saltation transport~\citep{Bagnold41,Iversenetal87,Burretal15} as the flow is sufficiently strong to net accelerate particles moving near the surface. By doing so, their hop height becomes larger and larger (i.e., $h/d$ substantially increases) until a steady state is approached. For aeolian transport in the field, the magnitude of $\Theta^{\rm In}_t$ relative to $\Theta^{\rm Rb}_t$ is unclear as $\Theta^{\rm In}_t$ is smaller than in wind tunnels because of a much larger boundary layer thickness $\delta$, since $\delta$ controls the size of the largest turbulent eddies and thus entrainment by turbulent fluctuation events (see section~\ref{ShearStressThreshold}).

\subsubsection{Implications for Field Phenomenona}
The new conceptual picture described above has been derived nearly entirely from theoretical and laboratory investigations. One may therefore wonder to what degree does the notion of various transport thresholds have implications for natural field conditions, such as bedload transport in rivers and saltation transport driven by planetary winds. There are three major aspects in which the field differs from most laboratory experiments: much broader particle size distributions, much larger and more unstable boundary layers (mainly for aeolian transport), and various kinds of surface inhomogeneities, such as bedforms, obstacles, and vegetation. The effects of particle size heterogeneity have been excluded from this review (they are briefly discussed in the outlook, section~\ref{ParticleSizeDistribution}). The remaining two aspects are both associated with increasing turbulence and thus fluid entrainment (see section~\ref{FluidEntrainment}). In contrast, the rebound threshold $\Theta^{\rm Rb}_t$ and arguably the impact entrainment threshold $\Theta^{\rm ImE}_t$, as well as the transport capacity scaling (which requires $\Theta\geq\Theta^{\rm ImE}_t$), are mainly controlled by the mean turbulent flow and relatively insensitive to turbulent fluctuations around it and should therefore be similar in laboratory and field (provided that the bed particle size distributions are similar). That is, for $\Theta\geq\Theta^{\rm ImE}_t\approx(1.5{-}2)\Theta^{\rm Rb}_t$ (typical for river floods and many aeolian processes), one expects capacity relationships derived from laboratory experiments to reasonably work and laboratory and field to behave similar. This expectation is consistent with observations reported in recent studies~\citep{Recking10,Reckingetal12,MartinKok17}. Even if transport is not at capacity, it is, in principle, possible to separate the turbulence-induced random transport contribution from sediment transport rate data sets~\citep{Salevanetal17} (see section~\ref{TransportCapacity}) and to modify capacity relationships to account for noncapacity transport~\citep{Comolaetal19b}.

\subsection{Summary of Important Open Problems and Controversies} \label{OpenProblems}
Sections~\ref{DenseGranularFlow}--\ref{ParticleInertia} have highlighted several important open questions and controversies that need to be addressed in future studies, which are summarized below. Section~\ref{DenseGranularFlow}:
\begin{enumerate}
 \item Why do fluid-sheared surfaces creep below a macroscopic yield criterion? And why do they do so even for seemingly arbitrarily small values of the Shields number $\Theta$ and in the absence of turbulence?
 \item What is responsible for the large spread of experimentally measured values of the viscous yield stress $\Theta^{\rm max}_t$?
 \item Is flow-induced bed failure (i.e., yielding) a critical phenomenon?
 \item What is the rheology of nonsuspended sediment transport?
\end{enumerate}
Section~\ref{FluidEntrainment} (although this section concerns both fluvial and aeolian transport conditions, open questions and controversies in this section regard mainly aeolian transport):
\begin{enumerate}
 \item Why do different experimental designs for measuring the initiation threshold $\Theta^{\rm In}_t$ of aeolian rolling and saltation transport cause qualitative differences in the scaling of $\Theta^{\rm In}_t$ with the particle diameter $d$?
 \item Is the measured dependency of $\Theta^{\rm In}_t$ on the density ratio $s$ for constant Galileo number $\mathrm{Ga}$ real or an artifact of differences in the boundary layer thickness of the wind tunnels used to carry out the experiments?
 \item Is the measured strong increase of $\Theta^{\rm In}_t$ with $d$ for very large $s$ in a wind tunnel with Martian pressure conditions real or an artifact of a limited boundary layer thickness of this wind tunnel?
 \item Is aeolian transport in the field on Earth and other planetary bodies, in contrast to wind tunnels with similar atmospheric pressure conditions, always being initiated close to the rebound threshold $\Theta^{\rm Rb}_t$ because of thick boundary layers, atmospheric instability, topography inhomogeneities, and subsurface ice sublimation? The answer to this question is probably the most important one, since a positive answer would imply that a reliable model for $\Theta^{\rm In}_t$ (i.e., answers to the previous three questions) is not required for predicting aeolian processes on such bodies.
 \item Does equilibrium aeolian bedload transport (i.e., $h\sim d$) exist in the field because of thick boundary layers?
 \item Direct measurements of aeolian sediment transport initiation, which are currently missing, can help answering the questions above.
\end{enumerate}
Section~\ref{ParticleInertia}:
\begin{enumerate}
 \item For a particle collision with a static sediment bed: how does the rebound probability $P_r$ depend on impact velocity and angle?
 \item How do particle shape and size distribution affect particle-bed collisions?
 \item Does viscous damping truly not much affect particle-bed collisions, as suggested by the insensitivity of DEM-based sediment transport simulations to the normal restitution coefficient $\epsilon$ of binary collisions? And if so, what is the physical reason?
 \item How do cohesive interparticle forces affect the collision process and thus sediment transport cessation?
 \item How do the laws describing a particle collision with a static bed change for a particle collision with a mobile bed?
 \item It is straightforward to define intermittent and continuous sediment transport for the absence of fluid entrainment because the sediment transport rate exhibits a discontinuous jump from nearly zero to a finite value at the continuous transport threshold. However, how does one universally define intermittent and continuous transport if fluid entrainment does occur?
 \item Is the transition from intermittent to continuous aeolian saltation transport associated with fluid entrainment (the current consensus) or with impact entrainment (the authors' opinion, based on recent developments in the field)?
 \item What controls the impact entrainment threshold $\Theta^{\rm ImE}_t$ and how does one model it?
\end{enumerate}

\subsection{Outlook} \label{Outlook}
To limit the scope of this review, several important topics have been excluded. Two of them are briefly discussed below.

\subsubsection{Effects of Particle Size Heterogeneity on Sediment Transport Initiation, Cessation, and Entrainment} \label{ParticleSizeDistribution}
Perhaps the most important topic that has been excluded from this review is the effects of the heterogeneity of the size of bed surface particles on sediment transport initiation, cessation, and entrainment. Naturally, sediment transport initiation and entrainment are size-selective. However, it is unclear whether this is also true for sediment transport cessation. While the continuous rebound mechanism (see section~\ref{ContinuousRebounds}) is clearly a size-selective process (coarser particles are less accelerated during their trajectories), impact entrainment may not be~\citep{MartinKok19,Zhuetal19}. Furthermore, in heterogeneous sediment beds, relatively fine particles tend to be surrounded by coarser ones and their protrusion (i.e., the particle height above surrounding sediment) is thus smaller than on average, whereas relatively coarse particles tend to have a larger-than-average protrusion. Because driving forces decrease and resisting forces increase with decreasing protrusion~\citep{Yageretal18}, relatively fine particles are more difficult to be entrained when compared with a bed made only of such fine particles. The ability of fine particles to continuously rebound is also suppressed by the presence of coarse particles~\citep{Zhuetal19}. All these effects can make heterogeneous sediment beds much less mobile than homogeneous ones of the same median particle size. For example, for both fluvial bedload~\citep{MacKenzieEaton17,MacKenzieetal18} and aeolian saltation transport~\citep{Zhuetal19}, it was found for certain heterogeneous beds that the largest particles of the particle size distribution (larger than the 90th percentile) have a very strong control on overall mobility. However, the manner and degree of the heterogeneousness seem to play an important role as not all kinds of heterogeneous beds are so strongly affected by the presence of large particles~\citep[e.g.,][]{Wilcock93,MartinKok19}. In particular, in the early stages of bed armoring, the sediment transport rate can increase because collisions between transported fine particles and coarse bed particles are more elastic than collisions between particles of the same size~\citep{Bagnold73}.

\subsubsection{Effects of Steep Bed Slope on Sediment Transport Initiation, Cessation, and Entrainment}
Another important topic that has been excluded from this review is the effects of steep bed slope angles on sediment transport initiation, cessation, and entrainment. For example, horizontal downslopes should, if everything else stays the same, increase bed mobility because of the additional horizontal gravity force acting on particles~\citep{Maurinetal18}. However, in fluvial environments, steep slopes are usually accompanied by a very small water depth of the order of one particle diameter (or even lower), which strongly suppresses the magnitude of hydrodynamic forces acting on particles, thus decreasing rather than increasing bed mobility~\citep{PrancevicLamb15}. Then again, an increasing downslope angle $\alpha$ increases the bulk friction coefficient $\mu$ within the sediment bed (for turbulent flows, $\mu\simeq\tan\alpha[1+[(\rho_p/\rho_f-1)\phi_b]^{-1}]$~\citep{Maurinetal18}, where $\phi_b$ is the bed volume fraction). Once $\mu$ exceeds the static friction coefficient $\mu_s$ associated with the yielding transition (see section~\ref{Yielding}), the entire bed fails and a debris flow forms~\citep{Takahashi78,Prancevicetal14,Chengetal18b}.

\setcounter{equation}{0}
\setcounter{figure}{0}
\renewcommand{\theequation}{A\arabic{equation}}
\renewcommand{\thefigure}{A\arabic{figure}}

\section*{Appendix: Mean Flow Velocity Profile (Law of the Wall)}
The mean flow velocity profile within the inner turbulent boundary layer above a flat wall (the \textit{law of the wall}) exhibits three regions: a log-layer for large nondimensionalized elevations (\textit{wall units}) $\mathrm{Re}_\ast z/d$, a viscous sublayer for small $\mathrm{Re}_\ast z/d$, and a buffer layer for transitional $\mathrm{Re}_\ast z/d$. For more details on turbulent wall-bounded flows, see the review by \citet{Smitsetal11}. In section~\ref{CessationThresholdModels}, the following form of the law of the wall is used~\citep{GuoJulien07}:
\begin{linenomath*}
\begin{align}
 \frac{u_x}{\sqrt{(s-1)gd}} & = \sqrt{\Theta}f\left( \mathrm{Re}_\ast,z/d \right), \nonumber \\
 f\left(\mathrm{Re}_\ast,z/d\right) & =  7\arctan\left( \frac{\mathrm{Re}_\ast}{7}\frac{z}{d} \right)+\frac{7}{3}\arctan^3 \left( \frac{\mathrm{Re}_\ast}{7}\frac{z}{d} \right) \nonumber \\ 
 & - 0.52\arctan^4\left( \frac{\mathrm{Re}_\ast}{7}\frac{z}{d} \right) + \ln\left[1+\left(\frac{\mathrm{Re}_\ast}{B}\frac{z}{d}\right)^{(1/\kappa)}\right] \nonumber \\
 &-\frac{1}{\kappa}\ln\left\{1+0.3\mathrm{Re}_\ast \left[1-\exp\left(-\frac{\mathrm{Re}_\ast}{26}\right)\right]\right\}, \label{uxcomplex}
\end{align}
\end{linenomath*}
where $\kappa=0.4$ and $B=\exp(16.873\kappa-\ln9)$. Within the viscous sublayer of the turbulent boundary layer, $u_x/\sqrt{(s-1)gd} \rightarrow \Theta\mathrm{Ga}z/d$, whereas in the log-layer, $u_x/\sqrt{(s-1)gd} \rightarrow \kappa^{-1}\sqrt{\Theta}\ln(z/z_o)$. The roughness length $z_o$ equals $d/(9\mathrm{Re}_\ast)$ in the hydraulically smooth and $d/30$ in the hydraulically rough regime~\citep{GuoJulien07}.

\acknowledgments
All data shown in the figures of this review can be found in the following references: \citet{PeyneauRoux08a}, \citet{DaCruzetal05}, \citet{KamrinKoval14}, \citet{Houssaisetal15}, \citet{AllenKudrolli17,AllenKudrolli18}, \citet{Diplasetal08}, \citet{Valyrakis13}, \citet{Williamsetal94}, \citet{IversenWhite82}, \citet{Burretal15}, \citet{Swannetal20}, \citet{Beladjineetal07}, \citet{Vowinckeletal16}, \citet{Salevanetal17}, \citet{BuffingtonMontgomery97}, \citet{YalinKarahan79}, \citet{Loiseleuxetal05}, \citet{Ouriemietal07}, \citet{Bagnold37}, \citet{Chepil45}, \citet{Ho12}, \citet{MartinKok18}, and \citet{Zhuetal19}. We thank Michael Church and three anonymous reviewers for their critical reviews and numerous insightful comments and suggestions. T.P. acknowledges support from grant National Natural Science Foundation of China (No.~11750410687).

\begin{notation}
 \notation{$\tau$} Fluid shear stress [Pa]
 \notation{$\tau_p$} Particle shear stress [Pa]
 \notation{$P$} Particle pressure [Pa]
 \notation{$\rho_p$} Particle density [kg/m$^3$]
 \notation{$\rho_f$} Fluid density [kg/m$^3$]
 \notation{$m_p$} Particle mass [kg]
 \notation{$u$} Instantaneous local flow velocity [m/s]
 \notation{$U_b$} Bulk flow velocity [m/s]
 \notation{$u_\ast\equiv\sqrt{\tau/\rho_f}$} Fluid shear velocity [m/s]
 \notation{$\nu_f$} Kinematic fluid viscosity [m$^2$/s]
 \notation{$\delta$} Boundary layer thickness [m]
 \notation{$H$} Flow thickness [m]
 \notation{$W$} Flow width [m]
 \notation{$d$} Characteristic particle diameter [m]
 \notation{$h$} Transport layer thickness [m]
 \notation{$\dot\gamma$} Particle shear rate (strain rate) [1/s]
 \notation{$T$} Granular temperature [m$^2$/s$^2$]
 \notation{$g$} Gravitational constant [m/s$^2$]
 \notation{$\tilde g\equiv(1-\rho_f/\rho_p)g$} Buoyancy-reduced gravitational constant [m/s$^2$]
 \notation{$M$} Sediment transport load [kg/m$^2$]
 \notation{$Q$} Sediment transport rate [kg/(ms)]
 \notation{$\Theta\equiv\tau/((\rho_p-\rho_f)gd)$} Shields number or Shields parameter
 \notation{$s\equiv\rho_p/\rho_f$} Particle-fluid-density ratio
 \notation{$\mathrm{Re}\equiv U_bH/\nu_f$} Reynolds number
 \notation{$\mathrm{Re}_\ast\equiv u_\ast d/\nu_f$} Shear Reynolds number
 \notation{$\mathrm{Ga}\equiv\sqrt{(s-1)gd^3}/\nu_f$} Galileo number (also called Yalin parameter)
 \notation{$\mathrm{St}\equiv s|\mathbf{v_r}|d/(9\nu_f)$} Stokes number, where $|\mathbf{v_r}|$ is the relative velocity between two particles just before they collide
 \notation{$M_\ast\equiv M/(\rho_pd)$} Nondimensionalized sediment transport load
 \notation{$Q_\ast\equiv Q/(\rho_pd\sqrt{(\rho_p/\rho_f-1)gd})$} Nondimensionalized sediment transport rate
 \notation{$I\equiv\dot\gamma d/\sqrt{P/\rho_p}$} Inertial number
 \notation{$J\equiv\rho_f\nu_f\dot\gamma/P$} Viscous number
 \notation{$K\equiv J+c_KI^2$} Viscoinertial number, where $c_K$ is an order-unity fit parameter
 \notation{$\mathrm{Pe}\equiv\dot\gamma d/\sqrt{T}$} P\'eclet number
 \notation{$C_m=1/2$} Added mass coefficient
 \notation{$\kappa=0.4$} von K\'arm\'an constant
 \notation{$\psi$} Pocket angle
 \notation{$\psi_Y$} Pocket angle for particles resting within the deepest pockets of the bed surface
 \notation{$L_{\rm arm}$} Lever arm length [m]
 \notation{$\alpha$} Bed slope angle
 \notation{$\Delta Z$} Critical dimensionless vertical particle displacement required for entrainment
 \notation{$\Delta X$} Critical dimensionless horizontal particle displacement required for entrainment
 \notation{$\mu_C$} Effective Coulomb friction coefficient encoding the combined effects of sliding and rolling friction in entrainment
 \notation{$\mu\equiv-\tau_p/P$} Ratio between particle shear stress and particle pressure (bulk friction coefficient)
 \notation{$\mu_g$} Surface friction coefficient of granular particle
 \notation{$\mu_s$} Static friction coefficient of granular bulk (yield stress ratio)
 \notation{$\mu_b$} Bulk friction coefficient at the interface between bed and transport layer. In contrast to $\mu$ and $\mu_s$, $\mu_b$ includes contributions from stresses associated with the particle fluctuation motion in addition to contributions from intergranular contacts.
 \notation{$\xi\propto|\mu-\mu_s|^{-\nu}$} Correlation length associated with the yielding transition, where $\nu=0.5$ is the critical exponent
 \notation{$F,F_D,F_L,F_t,F_n,F_e$} Instantaneous force applied by the fluid on a particle [kgm/s$^2$]. Subscript ($D$, $L$, $t$, $n$, $e$) refers to nature of force (drag, lift, tangential, normal, effective).
 \notation{$T,T_D,T_L,T_t,T_n,T_e$} Duration of turbulent fluctuation event [s]. Subscript ($D$, $L$, $t$, $n$, $e$) refers to nature of applied fluid force (drag, lift, tangential, normal, effective).
 \notation{$I_f$} Impulse of turbulent fluctuation event [kgm/s]
 \notation{$E_f$} Energy of turbulent fluctuation event [kgm$^2$/s$^2$]
 \notation{$F_c$} Force resisting initial particle motion [kgm/s$^2$]
 \notation{$u_c$} Critical instantaneous local flow velocity associated with resisting forces [m/s]
 \notation{$I_{fc}$} Critical impulse required for fluid entrainment [kgm/s]
 \notation{$W_c$} Critical work done by flow event required for fluid entrainment [kgm$^2$/s$^2$]
 \notation{$C_{\rm eff}$} Energy transfer coefficient, describing the fraction of energy transferred from flow to target particle during turbulent fluctuation event
 \notation{$C\equiv\alpha_f^{-1}f(G)\sqrt{s}d/\delta$} Inverse dimensionless boundary layer thickness
 \notation{$\alpha_f\equiv u_m/\overline{u}$} Ratio between the characteristic flow velocity $u_m$ associated with the largest turbulent fluctuations and the local mean flow velocity $\overline{u}$
 \notation{$T_{\rm max}$} Maximal duration of turbulent fluctuation events [s]
 \notation{$f(G)$} Factor that encodes information about particle shape, orientation, and the pocket geometry
 \notation{$\mathbf{v_i}$} Impact velocity [m/s]
 \notation{$\mathbf{v_r}$ ($\mathbf{v^{2D}_r}$)} Rebound velocity (projected into incident plane) [m/s]
 \notation{$\mathbf{v_e}$ ($\mathbf{v^{2D}_e}$)} Ejection velocity (projected into incident plane) [m/s]
 \notation{$\theta_i$} Impact angle
 \notation{$\theta_r$ ($\theta^{2D}_r$)} Rebound angle (projected into incident plane)
 \notation{$\theta_e$ ($\theta^{2D}_e$)} Ejection angle (projected into incident plane)
 \notation{$\mathbf{E_i}$} Impact energy [kgm$^2$/s$^2$]
 \notation{$\mathbf{E_e}$ ($\mathbf{E^{2D}_e}$)} Ejection energy (projected into incident plane) [kgm$^2$/s$^2$]
 \notation{$\mathbf{N_e}$} Average number of ejected particles
 \notation{$\mathbf{P_r}$} Rebound probability
 \notation{($e^{2D}\equiv|\mathbf{v^{2D}_r}|/|\mathbf{v_i}|$) $e\equiv|\mathbf{v_r}|/|\mathbf{v_i}|$} (Projected) rebound restitution coefficient
 \notation{$e_z\equiv-v_{rz}/v_{iz}$} Vertical rebound restitution coefficient
 \notation{$A,B,A^{2D},B^{2D},\chi,r,r^{2D},n_0,\zeta$} Dimensionless parameters appearing in empirical or semi-empirical relations describing the collision process between an incident bead and a granular packing 
 \notation{$\alpha_r$} Normal rebound restitution coefficient in the impact plane
 \notation{$\beta_r$} Tangential rebound restitution coefficient in the impact plane
 \notation{$\epsilon$} Restitution coefficient for binary particle collision
 \notation{$V_b$} Effective value of the local particle velocity averaged over elevations near the bed surface [m/s]
 \notation{$f_{\rm in}$} Particle feeding frequency at flume entrance [1/s]
 \notation{$f_{\rm out}$} Frequency of particles passing an illuminated window near the flume exit [1/s]
 \notation{$f_Q$} Fraction of active aeolian saltation transport
 \notation{$v_\uparrow$} Initial particle velocity in thought experiment in section~\ref{ThoughtExperiment} [m/s]
 \notation{$E_\uparrow$} Initial particle energy in thought experiment in section~\ref{ThoughtExperiment} [kgm$^2$/s$^2$]
 \notation{$E_c$} Critical energy that $E_\uparrow$ must exceed for particle to continuously rebound along the surface [kgm$^2$/s$^2$]
 \notation{$n_{\mathrm{tr}}/n_{\mathrm{tot}}$} Number of transported particles relative to the total number of bed surface particles
 \notation{$n_{v_t}$} Number of particles that are faster than a certain velocity threshold $v_t$
 \notation{$z_h$} Hop height [m]
 \notation{$t_h$} Hop time [s]
 \notation{$f(\mathrm{Re}_\ast,z/d)$} Function given by equation~(\ref{uxcomplex})
 \notation{$v_s$} Settling velocity [m/s]
 \notation{$U_x\equiv\overline{u_x}/\sqrt{s\tilde gd}$} Dimensionless transport layer-averaged fluid velocity
 \notation{$V_x\equiv\overline{v_x}/\sqrt{s\tilde gd}$} Dimensionless transport layer-averaged horizontal particle velocity
 \notation{$V_z\equiv\sqrt{\overline{v_z^2}}/\sqrt{s\tilde gd}$} Dimensionless transport layer-averaged vertical particle velocity
 \notation{$Z\equiv\overline{z}/d$} Dimensionless transport layer thickness
 \notation{$Z_\Delta=0.7$} Dimensionless average elevation of the particles' center during particle-bed rebounds
 \notation{$z_o$} Surface roughness [m]
 \notation{$c_1,c_2,c_3$} Model constants in equations~(\ref{Vz})-(\ref{VU})
 \notation{Bed sediment entrainment} Mobilization of bed sediment
 \notation{Fluid entrainment} Entrainment caused by the action of flow forces
 \notation{Incipient motion} Initiation of sediment transport by fluid entrainment
 \notation{Impact entrainment} Entrainment caused by the impacts of transported particles onto the bed
 \notation{Sediment transport} Sediment motion caused by the shearing of an erodible sediment bed by flow of a Newtonian fluid
 \notation{Aeolian sediment transport} Wind-driven sediment transport
 \notation{Fluvial sediment transport} Liquid-driven sediment transport (despite its name, not limited to fluvial environments)
 \notation{Nonsuspended sediment transport} Sediment transport in which the fluid turbulence is unable to support the submerged particle weight
 \notation{Saltation transport} Nonsuspended sediment transport with comparably large transport layers ($h\gg d$)
 \notation{Bedload transport} Nonsuspended sediment transport with comparably small transport layers ($h\sim d$)
 \notation{Transport capacity (or saturation)} Loosely, the maximum amount of sediment a given flow can carry without causing net sediment deposition at the bed. More precisely, in the context of nonsuspended transport of nearly monodisperse sediment, it is defined as a steady transport state at which any further net entrainment of bed sediment into the transport layer would weaken the mean turbulent flow to a degree at which it is no longer able to compensate the average energy loss of particles rebounding with the bed by their energy gain during their trajectories via fluid drag acceleration. The so defined transport capacity obeys equation~(\ref{CapacityM}).
 \notation{Creeping} A superslow granular motion, usually in the form of intermittent local particle rearrangements within the sediment bed (not limited to the bed surface), that occurs below a macroscopic yield criterion
 \notation{$\Theta_t$ ($\tau_t$)} Shields number (fluid shear stress) at a nonspecified transport threshold. For specifications, see below.
 \notation{$t_{\rm conv}\propto|\Theta-\Theta_t|^{-\beta}$} Time scale for transport property to converge in the steady state near $\Theta_t$, where $\beta$ is a positive exponent
 \notation{Shields diagram (Shields curve)} Diagram compiling measurements of $\Theta_t$ as a function of $\mathrm{Re}_\ast$ (the Shields curve $\Theta_t(\mathrm{Re}_\ast)$) for fluvial bedload transport conditions
 \notation{$\Theta^{\rm max}_t$ ($\tau^{\rm max}_t$)} Viscous yield stress. The upper limit of the threshold Shields number (fluid shear stress) in the Shields diagram, which is associated with viscous bedload transport. For $\Theta\lesssim\Theta^{\rm max}_t$, a sediment bed subjected to a laminar flow at low Shear Reynolds number $\mathrm{Re}_\ast$ may temporarily fail but will eventually rearrange itself into a more stable packing that resists the applied fluid shear stress. For $\Theta\gtrsim\Theta^{\rm max}_t$, a sediment bed subjected to a laminar flow can no longer find packing geometries that are able to resist the applied fluid shear stress.  
 \notation{$\Theta^{\rm In}_t$ ($\tau^{\rm In}_t$)} Initiation threshold. Shields number (fluid shear stress) at which the probability of fluid entrainment of bed particles exceeds zero (which, for turbulent fluvial bedload transport, occurs much below the Shields curve). For sediment beds subjected to turbulent flows, a critical fluid shear stress does no longer describe the fluid entrainment of individual particles. However, one can still define a Shields number ($\Theta^{\rm In}_t$) below which fluid entrainment does never occur. Like for $\Theta^{\rm max}_t$, transient behavior associated with the flow temporarily pushing particles from less stable to more stable pockets is excluded in the definition of $\Theta^{\rm In}_t$.
 \notation{$\Theta^{\rm In\prime}_t$ ($\tau^{\rm In\prime}_t$)} Rocking initiation threshold. Shields number (fluid shear stress) above (below) which there is a nonzero (zero) probability that peaks of flow forces associated with turbulent fluctuation events acting on bed particles exceed resisting forces. That is, there is a nonzero probability that particles rock (or wobble or oscillate) within their bed pockets. Rocking may ($\Theta^{\rm In\prime}_t=\Theta^{\rm In}_t$) or may not ($\Theta^{\rm In\prime}_t<\Theta^{\rm In}_t$) lead to complete entrainment depending on the maximal duration of the strongest possible turbulent fluctuation events.
 \notation{$\Theta^{\rm Rb}_t$ ($\tau^{\rm Rb}_t$)} Rebound threshold. Shields number (fluid shear stress) above which the mean turbulent flow is able to compensate the average energy loss of transported particles rebounding with the bed by their energy gain during their trajectories via fluid drag acceleration, giving rise to a long-lasting rebound motion. In general, this threshold is unrelated to the entrainment of bed sediment. It is also the threshold that appears in most threshold shear stress-based sediment transport expressions.
 \notation{$\Theta^{\rm Rb\ast}_t$ ($\Theta^{\rm Rb\ast\ast}_t$)} Modeled rebound threshold. Values of $\Theta^{\rm Rb}_t$ from models that consider (neglect) that the near-surface flow can assist rebounding particles in escaping the bed surface.
 \notation{$\Theta^{\rm ImE}_t$ ($\tau^{\rm ImE}_t$)} Impact entrainment threshold. Shields number (fluid shear stress) above which entrainment of bed sediment by impacts of transported particles onto the bed is able to compensate captures of long-lasting rebounders (see $\Theta^{\rm Rb}_t$ above) by the bed. This threshold is arguably also the threshold of continuous nonsuspended sediment transport.
 \notation{$\Theta^{\rm Rb|ImE}_t$ ($\tau^{\rm Rb|ImE}_t$)} Modeled hybrid between rebound and impact entrainment threshold
\end{notation}


\end{document}